\documentclass[12pt,a4paper, notitlepage]{article}
\usepackage[utf8]{inputenc}
\usepackage[english]{babel}
\usepackage[T1]{fontenc}
\usepackage{hyperref}
\usepackage{amsmath}
\usepackage{stmaryrd}
\usepackage{amsfonts}
\usepackage{amssymb}
\usepackage{xcolor}
\usepackage{caption}
\usepackage{amsmath}
\usepackage[nottoc, notlof, notlot]{tocbibind}
\usepackage{relsize,exscale}
\usepackage{array,multirow,makecell}
\usepackage[toc,page]{appendix}
\DeclareSymbolFont{largesymbols}{OMX}{cmex}{m}{n}

\usepackage{lineno}
\newcommand*\patchAmsMathEnvironmentForLineno[1]{%
  \expandafter\let\csname old#1\expandafter\endcsname\csname #1\endcsname
  \expandafter\let\csname oldend#1\expandafter\endcsname\csname end#1\endcsname
  \renewenvironment{#1}%
     {\linenomath\csname old#1\endcsname}%
     {\csname oldend#1\endcsname\endlinenomath}}%
\newcommand*\patchBothAmsMathEnvironmentsForLineno[1]{%
  \patchAmsMathEnvironmentForLineno{#1}%
  \patchAmsMathEnvironmentForLineno{#1*}}%
\AtBeginDocument{%
\patchBothAmsMathEnvironmentsForLineno{equation}%
\patchBothAmsMathEnvironmentsForLineno{align}%
\patchBothAmsMathEnvironmentsForLineno{flalign}%
\patchBothAmsMathEnvironmentsForLineno{alignat}%
\patchBothAmsMathEnvironmentsForLineno{gather}%
\patchBothAmsMathEnvironmentsForLineno{multline}%
}

\setcellgapes{1pt}
\makegapedcells
\newcolumntype{R}[1]{>{\raggedleft\arraybackslash }b{#1}}
\newcolumntype{L}[1]{>{\raggedright\arraybackslash }b{#1}}
\newcolumntype{C}[1]{>{\centering\arraybackslash }b{#1}}
\newcommand{\Tr}{\mathrm{Tr}}
\newcommand{\tr}{\mathrm{tr}}

\newtheorem{definition}{Definition}

\newtheorem{remark}{Remark}

\usepackage{enumerate}
\usepackage{subfig}
\usepackage{graphicx}

\newcommand{\STr}{\mathrm{STr}}

\usepackage{graphicx}

\newcommand{\beq}{\begin{equation}}
\newcommand{\eeq}{\end{equation}}
\newcommand{\bea}{\begin{eqnarray}}
\newcommand{\eea}{\end{eqnarray}}
\definecolor{mygray}{gray}{0.3}

\newcommand{\bes}{\begin{eqnarray}}
\newcommand{\ees}{\end{eqnarray}}

\newcommand\restr[2]{{
  \left.\kern-\nulldelimiterspace 
  #1 
  \vphantom{\big|} 
  \right|_{#2} 
  }}

\usepackage{multicol}
\usepackage{amssymb}

\makeatletter

\begin{document}

\begin{center}
\textbf{\Large{Functional renormalization group for multilinear disordered Langevin dynamics I}}\\

\bigskip
\textbf{{\large Formalism and first numerical investigations at equilibrium}}

\medskip
\vspace{15pt}

{\large Vincent Lahoche$^a$\footnote{vincent.lahoche@cea.fr}  \,\,
Dine Ousmane Samary$^{a,b}$\footnote{dine.ousmanesamary@cipma.uac.bj}
\,\,and Mohamed Ouerfelli$^a$\footnote{mohamed-oumar.ouerfelli@cea.fr} }
\vspace{15pt}

a)\,  Commissariat à l'\'Energie Atomique (CEA, LIST),
8 Avenue de la Vauve, 91120 Palaiseau, France

b)\, Facult\'e des Sciences et Techniques (ICMPA-UNESCO Chair),
Universit\'e d'Abomey-
Calavi, 072 BP 50, B\'enin
\vspace{0.5cm}
\begin{abstract}
\noindent 

This paper aims at using  the functional renormalization group formalism to study the equilibrium states of a stochastic process described by a quench--disordered multilinear Langevin equation. Such an equation characterizes the evolution of a time-dependent $N$-vector $q(t)=\{q_1(t),\cdots q_N(t)\}$  and is traditionally encountered in the dynamical description of glassy systems at and out of equilibrium through the so-called Glauber model. From the connection between Langevin dynamics and quantum mechanics in imaginary time, we are able to coarse-grain the path integral  of the problem in the Fourier modes, and to construct a renormalization group flow for effective  Euclidean action.  In the large $N$-limit we are able to solve the flow equations for both matrix and tensor disorder. The numerical solutions of the resulting exact flow equations are then investigated using standard local potential approximation, taking into account the quench disorder. In the case where the interaction is taken to be matricial, for  finite $N$ the flow equations are also solved. However, the case of finite $N$ and taking into account the non-equilibrum process  will be considered in a companion investigation.

\medskip

\end{abstract}
\end{center}
\pagebreak

\tableofcontents

\pagebreak

\section{Introduction}
Multilinear Langevin dynamics are characterized by a specific notion of disorder, said to be quenched, and referring to the existence of two fluctuations times: the time $\tau$ of some random variables $\{q_i\}$ and the time $T$ of the couplings between these random variables. For a glassy system, we have $T\gg \tau$, meaning that the couplings are essentially constant during the typical fluctuation time of the random variables $\{q_i\}$. Initially introduced as a mathematical description of the magnetic field, it allows to obtain strange thermodynamics properties through for instance the so-called Glauber model \cite{Castellani1,Cugliandolo1,FischerHertz,Bouchaud1,MezardAl}.  Glassy system have acquired an independent existence, essentially due to their mathematical complexity, and are encountered in a large area of research fields, among which we find the condensed matter physics, chemistry, genetics, computer science (as a description of neural networks behavior), collective processes in markets, economy,  tensorial principal component analysis, etc. (see \cite{Sherrington2,Bouchaud2,BryngelsonAl, Coolen1,BenArous1, BenArous2,Sherrington} for more details).

\medskip
The existence of a freezing temperature $T_{f}$, below which the system may be trapped into an equilibrium non-ergodic region of the phase space, and where the system is enforced to fluctuate around some states but cannot visit every other possible equilibrium states, is the fundamental behavior of the disordered system. The physical meaning of this phenomenon leads to the existence of many deep minima of the energy landscape, with large energy barriers between the different minima, such that the system requires a very long time to escape from them. The existence of such an ergodicity breaking is one of the standard signatures of the glassy transition, occurring generally in dynamical investigations based on a stochastic, Langevin equation. Other standard approaches for glass transition use replica method, the cavity method through a generalized mean-field approach, or the so-called ‘‘TAP'' (Thouless-Anderson-Palmer \cite{thouless1977solution}) approach focusing on complexity (i.e. configuration entropy, counting the number of energy minima) \cite{Castellani1, DeDominicisbook,Contuccibook, FischerHertz,Nishimori,MezardAl}. All of them provide the characterization of the transition with different signatures, replica symmetry breaking for the first one and the divergence of the magnetic susceptibility for the second one and then the divergence of the complexity for the third one, which highlights their complementarity.
\medskip

The Langevin equation formulation has the advantage to be able to describe systems from a dynamical point of view both near and far from equilibrium, therefore we particularly focus on this formulation in the following work because it corresponds well to the system that we want to study. This kind of equation has been widely studied recently, and we can in particular mention some relevant results \cite{ Kim2001, Caiazzo1, Cugliandolo4,Cugliandolo2, Cugliandolo3,Sommers1}. A large part of these investigations combine numerical and analytic methods; but analytic solutions were still found only in some limited cases, in particular the references \cite{Cugliandolo2, Cugliandolo3} studied the large $N$ limit where self-averaging property holds. Another specific case is matrices disorder where an exact analytic solution can be found due to the well-known large-$N$ properties of the random matrices, whose spectrum converges toward the Wigner semi-circle law \cite{Benaych1, Krajewski1,DiFrancesco1, Potters1, Wigner1}. However, such a simplification does not occur for the multilinear case, which is the main topic of this paper, essentially because no exact analogue of the Wigner law exists for tensors\footnote{Note however that the recent contribution: arXiv:2004.02660, may probably change this situation in the future}. In this paper, we propose an alternative approach using coarse-graning in frequencies modes to construct effective states of the system for large times. More precisely, the degrees of freedom can be labelled with frequency modes $\omega$, and we are aiming to use renormalization group (RG) and nonperturbative techniques to construct an RG flow for effective quantities, integrating out large frequencies modes (i.e. rapid modes effects) $1/\omega \gg T$, for some running time-scale $T$.
\medskip

In particular, we deal with the large-time behavior of the system, describing equilibrium states from initial, out-of-equilibrium conditions. To this end, we consider a coarse-graining approach in time, through an RG description allowing us to investigate physics at different temporal scales integrating out large frequencies modes.  Let us recall that RG realizes an interpolation between two physical descriptions of the same system, through a progressive integration of degrees of freedom; which provides an effective description valid at a large scale from a microscopic description \cite{Delamotte1, Pol1, Riv2, Riv1,ZinnJustinBook1,ZinnJustinBook3,ZinnJustinBook2}. This coarse-graining procedure dilutes information from each step, and the effective description discards some irrelevant effects which decrease along with the RG flow. Hence, RG is usually considered as a powerful tool to discuss universality in quantum and classical field theories; and generally in systems involving intrinsic variability. There is a recent literature on the link between RG and information theory, see for instance \cite{BENY2015,2018Entropie,erdmenger2021quantifying}. In \cite{2018Entropie} for instance, authors propose a point of view where RG looks like a maximum entropy dynamics where coarse-graining is generated by a decrease of the relative entropy (with respect to a fixed background probability). For recent connections with data analysis see \cite{B_ny_2018,lahoche2020field2,lahoche2020field,lahoche2020generalized,lahoche2021signal,lahoche2022field}.
\medskip

In the  present manuscript we use the functional renormalization group (FRG) formalism due to Wetterich \& al. \cite{Berges_2002,Wett1,Wett2,Delamotte1,Dupuis_2021,Morris_19942,MORRIS_1994}. The main interest of this method is that it avoids some difficulties occurring in other mathematical incarnation of the RG idea, especially concerning the strong coupling regimes; where instabilities may occur with methods based on other functional relation as Polchinski equation \cite{Dupuis_2021,pawlowski2015physics,Synatschke_2009}. Thus, its ability to support crude truncation is especially relevant in a case where the coupling's magnitude fluctuates and can become arbitrarily large. Coarse-graining in time has been considered through the FRG formalism for quantum mechanical systems \cite{Kapoyannis_2000,Synatschke_2009,Zappal__2001}; especially regarding supersymmetric aspects of quantum systems. Exploiting the link between Langevin equation and Schrödinger equation in imaginary time, the same techniques have been used for stochastic systems \cite{Duclut_2017,Prokopec_2018,wilkins2021functional,wilkins2021functional2}; focusing on the effective  Euclidean action. In the quantum description, stochastic systems exhibit elementary supersymmetry which simplifies the choice of the truncation; explicit supersymmetric truncation strongly improves the approximate RG flow for non-supersymmetric ones \cite{Canet_2007,Canet_2011,Prokopec_2018,wilkins2021functional2}. 
\medskip

RG has been considered as a powerful tool of investigation for such a kind of problem, especially in regard to the behavior of spin glass \cite{castellana2013renormalization,Castellana_20152,Castellana_2015,Castellana3,Pimentel_2002} or the so-called random-field Ising problem \cite{Balog_2018,Balog_2020,DeDominicisbook,kaviraj2021random,Tarjus_20082,tarjus2015avalanches,
Tarjus_2020,Tissier_2008,Tissier_2011,Tissier_20122,
Tissier_2012}. In this paper we propose to investigate dynamics of a zero-dimensional  \textit{p-spin} like system, through FRG formalism, mathematically incarnated as a Langevin equation with quenched disorder, describing dynamics of a $N$ dimensional random vector $q_i(t)$. A large part of this paper is devoted to the multilinear case, with tensorial disorder. Our aim in  this paper is essential to establish the formalism at equilibrium, meaning that we are only able to discuss vitreous transitions above $T_c$. Beyond physical applications, such a formalism can be helpful for other areas of sciences, like data analysis, where such a kind of equations appears, and we plan to investigate these aspects in a  future work.
\medskip

The outline is the following. In section \ref{sec1} we define the models and review some aspects about the standard connection between Langevin equation, Fokker-Planck, imaginary time Schrödinger equations and supersymmetry. In section \ref{sec2} we introduce the FRG formalism able to deal with time-reversal and quenched disorder. In section \ref{sec3}, we use standard local potential approximation to construct solutions of the exact RG equation compatible with the supersymmetry. We introduce a graphical formalism allowing us to draw flow equations in a very convenient way, especially relevant for multilinear models where the number of terms in the truncations become large. Finally,  we investigate  a similar analysis for  the same model with the spherical constraint. In section \ref{sec4} we provide a numerical analysis of the flow equations, and compare them with numerical simulations for the Langevin equation, especially in the large $N$ limit. In section \ref{sec5} we consider the special case of matricial disorders. Finally, section \ref{sec6} is devoted to the conclusions on which we provide some  remarks and  ingredients which can help to investigate the non-symmetric phase and the non-equilibrium process  which will be addressed taking taken into account in forthcoming works.

\section{Preliminaries} \label{sec1}
The  theory  of  Brownian  motion  is  perhaps  the  simplest  approximate  way  to  treat  the dynamics of nonequilibrium systems. The fundamental equation that governs this dynamic is called the Langevin equation. This equation is a stochastic differential equation describing the time evolution of a subset of the degrees of freedom. These degrees of freedom typically are collective (macroscopic) variables changing only slowly in comparison to the other (microscopic) variables of the system.
In this section, we provide some definitions and basics for the disordered Langevin equation, required for the rest of this paper. For more detail concerning this notion, the  reader can consult the standard references as \cite{ZinnJustinBook1,ZinnJustinBook2}.

\subsection{The model}\label{sec11}

We introduce the functional renormalization group methods to investigate the behavior of a disordered system described by a random dynamical vector $Q(t)=\{q_1(t),\cdots q_N(t) \}\in \mathbb{V}_N$ depending on time $t \in \mathbb{R}$ and where $\mathbb{V}_N$ designates the configuration space. We assume that the dynamics obeys to the non-linear Langevin equation:
\begin{equation}
\frac{dq_i}{dt}=-\frac{\partial}{\partial q_i} H_0[J, Q]+\eta_i(t)\,, \label{eq1}
\end{equation}
where $\eta(t)=(\eta_1(t),\cdots,\eta_N(t))\in \mathbb{R}^N$ is a random vector assumed to be Gaussian distributed, with zero mean value  and the variance given by:
\begin{equation}
\langle \eta_{i_1}(t) \eta_{i_2}(t^\prime) \rangle=2D\delta_{i_1i_2} \delta(t-t^\prime),\,\,D\in \mathbb{R}^+.\label{deltanoise}
\end{equation}
Two specific cases are relevant:
\begin{enumerate}
\item \textit{The unconstrained model} (or \textit{soft p-spin model}), for which $\mathbb{V}_N\equiv \mathbb{R}^N$, and the stochastic Hamiltonian $H_0(J,Q)$ is defined such that:
\begin{equation}
H_0[J, Q]:= \frac{1}{p}J_{i_1 i_2\cdots i_p} q_{i_1}(t)q_{i_2}(t)\cdots q_{i_p}(t)+\sum_{m=1}^M\frac{h_m}{2N^{m-1}m} (Q^2)^m\,.
\end{equation}
This Hamiltonian content two parts. The first one is  a stochastic expression linear on the (symmetric) tensor $J$. The second one  is the deterministic part with $M\in \mathbb{N}$ and $Q^2:=\sum_i q_i^2$.
\item \textit{The spherical model} $\mathbb{V}_N\equiv \mathbb{S}^{(N)}(\sqrt{N})$, the $N$ dimensional sphere of radius $\sqrt{N}$, for which the stochastic Hamiltonian must be modified from a Lagrange parameter $\mu$ to ensure the spherical constraint $Q^2(t)=N$:
\begin{equation}
H_0[J, Q] \rightarrow H_0^\prime[J, Q]=H_0[J, Q]-\frac{1}{2}\mu (Q^2-N)\,,
\end{equation}
with the initial condition which  satisfies  the spherical constraint.
\end{enumerate}
Note that we assumed the Einstein convention for summation of repeated indices. The time independent symmetric rank-$p$ tensor $J\in \mathbb{R}$ is assumed to be randomly distributed, following to a Gaussian distribution $p(J)$ such that:
\begin{equation}
\overline{ J_{i_1\cdots i_p} J_{j_1\cdots j_p }} = \frac{\lambda}{N^{p-1}} C_{i_1\cdots i_p;j_1 \cdots j_p}\,,\quad \overline{ J_{i_1\cdots i_p} } =\frac{J_0}{N^{p-1}}\,, \label{distributionJ}
\end{equation}
where $\lambda \in \mathbb{R}^+$,  and:
\begin{equation}
C_{i_1\cdots i_p;j_1 \cdots j_p}:=\frac{2}{p!}\sum_{\pi} \delta_{\pi(i_1) j_1}\cdots \delta_{\pi(i_p) j_p}\,,
\end{equation}
the sum running over permutations $\pi$ of the first $p$ positives integers. Note that since we are aiming to take the average with respect to two sources of fluctuation, we have to fix the notation to designate the mean quantities. Thus, we use the notation $\langle x \rangle$ for averaging for the noise $\eta$, and we use the notation $\overline{x}$ for averaging for $J$.
The power of $N$ is chosen to ensure that the  extensive quantities like $H_0$ are proportional to $N$. $\sqrt{\lambda}$ define the size of the fluctuating tensorial coupling $J$ which interpolates between two regimes:
\begin{itemize}
\item For $\lambda=0$, the fluctuations are frozen and $J$ goes toward a fixed value $J_0$. When $J_0$ becomes large, the influence of the noise fades out, and the dynamics reach a \textit{deterministic regime}. For $J_0\to 0$ on the other hand, the noise dominates, and the dynamics is essentially \textit{Brownian}.
\item For $\lambda\to +\infty$, the fluctuations of the coupling dominate, and the system reach a \textit{glassy regime}. We note that the nature of the dynamics could be characterized using the Hurst Exponent \cite{hurst1951long}. Indeed, this statistical measure allows to investigate the variability of a time series and thus identify a random walk from deterministic dynamics.
\end{itemize}
The spherical model is the one generally considered in the spin-glass literature. The first model however is as well considered as a simplification of the constrained case, which it reduces in some limits \cite{Crisanti1993,Janssen1,Kirkpatrick1987,Bausch1976,Sommers1}. For instance, with $J_0=0$, and for arbitrary size $N$, the spherical model can be recovered as a limit case, with quartic deterministic part such that $h_1\to -\infty$, $h_2\to +\infty$ but $h_1/h_2= \mathrm{const.}$. In the large $N$ limit, the self-averaging of $O(N)$ invariant quantities as $Q^2:=\sum_i q_i^2$ (stemming as for the central limit theorem from the assumption that random variables $q_i$ are weakly correlated) ensures that quantity as $\langle Q^2 Q^2\rangle$ satisfies the cluster properties \cite{Moshe_2003,ZinnJustin4,ZinnJustinBook1}:
\begin{equation}
\langle Q^2 Q^2\rangle \underset{N\to +\infty}{\sim} \langle Q^2\rangle \langle Q^2\rangle\,.
\end{equation}
Thus, by projecting equation \eqref{eq1} along $Q$, we get for the quartic potential ($h_m=0\,\forall m>2$):
\begin{equation}
\overline{\langle \sum_i q_i \dot{q}_i \rangle }\approx h_1 \overline {\langle Q^2\rangle }+\frac{h_2}{N}\, \overline {\langle Q^2\rangle }^2\,.
\end{equation}
In addition to $\overline {\langle Q^2\rangle }=0$, the potential admits another minimum when $h_1<0$, for the value:
\begin{equation}
\overline {\langle Q^2\rangle }= -\frac{h_1}{h_2}N\,.
\end{equation}
The value of the mean potential for this value being $\overline {\langle V(Q)\rangle }=-h_1^2N/4h_2$. Thus, in the large $N$ limit, the stable minimum of the potential becomes infinitely deep, and all the trajectories are trapped in effective spherical dynamics. We also note that the non-linearities present in the local potential could induce interesting non-gaussian effects such as described in \cite{carreras1996fluctuation,kimanderson2008nonperturbative,kim2009probability}.
\medskip

\subsection{Associated Euclidean path integral and supersymmetry}\label{sec12}

The stochastic Langevin equation can be suitably reformulated as an  Euclidean quantum mechanical issue involving a real wave function that obeys an imaginary time Schrödinger equation. This representation  admits a path integral formulation, expressing the transition probability between two configurations as a sum over histories bounding them, weighted by the exponential of the classical action $e^{-\mathcal{S}[Q]}$. This representation moreover admits interesting supersymmetry, which is of great interest to construct an approximation of the exact RG flow in section \ref{sec3}.
\medskip

The randomness of the trajectory $Q(t)$ induced by the Gaussian noise $\eta(t)$ can be equivalently described in terms of a functional probability distribution $P(Q_f,t)$ giving the probability to reach the final point $Q_f$ in the lapse $t$ from the initial condition $Q(t=0):=c$. Fixing disorder the probability distribution, suitably normalized, can be read as: $P(Q_f,t)= \langle \delta(Q(t)-Q_f) \rangle$, or explicitly:
\begin{equation}
P(Q_f,t) \propto \int [d\eta(t)] \, e^{-\int dt\frac{\eta^2(t)}{4D}} \delta(Q(t)-Q_f) \,, \label{integral1}
\end{equation}
$[d\eta(t)]$ denoting the standard path integral measure. This probability distribution satisfies a Fokker-Planck type equation. Moreover, defining $\Psi(Q,t)=:e^{H_0[Q]/2D} P(Q,t)$, we find that $\Psi(Q,t)$ satisfies the  Euclidean Schrödinger equation $\dot{\Psi}=-\hat{\mathcal{H}} \Psi$, with positive definite Hamiltonian:
\begin{equation}
\hat{\mathcal{H}}:=\sum_i D\left(-\frac{\partial}{\partial q_i}+\frac{1}{2D}\frac{\partial H_0}{\partial q_i}\right)\left(\frac{\partial}{\partial q_i}+\frac{1}{2D}\frac{\partial H_0}{\partial q_i}\right)\,.
\end{equation}
The ground state $\Psi_0$ such that $\hat{\mathcal{H}}\Psi_0=0$ is formally solved by $\Psi_0\propto e^{-H_0(q)/2D}$ (if it is normalizable). This has an important consequence in regard to the large time behavior of the distribution. Indeed, we expect that the transition operator $\hat{U}(t):=e^{-t \hat{\mathcal{H}}}$ projects into the ground state for $t\to \infty$, and thus $P(Q,t)$ reach its equilibrium solution $\lim_{t\to \infty} P(q,t) \to e^{-H_0(q)/D}$. The quantum transition probability can be formally deduced from the standard Feynman prescription. Alternatively it can be defined from the integral \eqref{integral1} from the formal identity\footnote{Note that we removed the absolute value of the determinant, due to the fact that the Langevin equation is a first order differential equation, admitting a single causal solution see \cite{ZinnJustinBook1}. Moreover, note that out of equilibrium, equation \eqref{eq1} having several solutions (i.e. the action having several minima), equation \eqref{identity1} would no longer be valid. }:
\begin{equation}
1\equiv \int_{c} [dq(t)] \delta\left(\dot{Q}+\nabla_Q H_0-\eta \right) \det \mathcal{M}\,,\label{identity1}
\end{equation}
where $\nabla_Q:= \{\partial/\partial q_i\}$ and $\mathcal{M}$ is the matrix with entries:
\begin{equation}
\mathcal{M}_{ij}(t^\prime,t):=\frac{d}{dt}\delta_{ij}\delta(t-t^\prime) +\frac{\partial^2 H_0}{\partial q_i(t) \partial q_j(t^\prime)}\,. \label{matM}
\end{equation}
Inserting the identity \eqref{identity1} into the integral \eqref{integral1}, we get:
\begin{align}
P(Q_f,t) \propto \int [d\eta]&\int_c^{Q_f}[dq] \, e^{-\int dt\frac{\eta^2}{4D}} \times \delta\left(\dot{Q}+\nabla_Q H_0-\eta \right) \det \mathcal{M}\,. \label{integral2}
\end{align}
The integration over the noise $\eta(t)$ can be straightforwardly performed. Moreover, the determinant can be computed from the identity $\det \mathcal{M}= e^{\Tr \ln \mathcal{M}}$. Taking into account the normalization of the Gaussian integral over $\eta$ and expanding in power of the trace, we get:
\begin{equation}
\frac{e^{\Tr \ln \mathcal{M}}}{\det \mathcal{M} \big\vert_{H_0=0}}= \exp \left(\theta(0) \int dt \sum_i\frac{\partial^2H_0}{\partial q_i(t) \partial q_i(t)} \right)\,.\label{Stratonovich}
\end{equation}
The standard Stratonovich prescription imposes $\theta(0)=1/2$ for the value of the Heaviside function at origin. We thus obtain:
\begin{equation}
P(Q_f,t) \propto e^{-\frac{1}{2D}(H_0(Q_f)-H_0(c))}\int_c^{Q_f}[dq] \, e^{-\mathcal{S}[Q]} \,, \label{integral3}
\end{equation}
where we used the identity, valid only for the Stratonovich prescription,  see \cite{Nishimori} for more detail:
\begin{equation}
\int_A^B dt \dot{Q} \nabla_Q H_0=H_0(B)-H_0(A)\,.
\end{equation}
Remark that the integral in the expression \eqref{integral3} must be identified with the path integral associated to the quantum states $\Psi$. The corresponding classical action $\mathcal{S}$ is defined as:
\begin{equation}
\mathcal{S}_{QM}[Q]:=\int dt \, \left[ \frac{1}{4D} \left( \dot{Q}^2+(\nabla_Q H_0)^2 \right)-\frac{1}{2} \nabla^2_Q H_0\right]\,.\label{classic1}
\end{equation}
Also let us specify that this dynamic  action associated to a stochastic process admits a natural BRS (Becchi-Rouet-Stora) symmetry \cite{becchi1974abelian} that can be embedded in a quantum mechanical supersymmetric formulation. This supersymmetry is known to have important consequences on the RG flow, especially relevant in the construction of nonperturbative approximate solutions of the exact RG flow \cite{Canet_2007,Canet_2011}. A simple way to reveal this supersymmetry is to rewrite the determinant $\det \mathcal{M}$ in \eqref{integral1} as a Grassmann integral:
\begin{equation}
\det \mathcal{M}= \int [d\bar{c}dc] \exp \left[-\int dt \bar{c}_i \left(-\delta_{ij}\partial_t+\frac{\partial^2 H_0}{\partial q_i(t) \partial q_j(t)}\right)c_j \right]\,.
\end{equation}
Now by imposing the following redefinition  of the  fields as:
\begin{equation}
\bar{c}_i c_i =: i \bar{\psi}_i \psi_i \quad q_i(t)=:\sqrt{2D}\phi_i(t)\quad H_0=: 2D W_0\,,
\end{equation}
and introducing the auxiliary fields $\bar{\varphi}_i\in \mathbb{R}$ in order to break the square ${\frac{1}{2}}(\nabla_\phi W_0)^2$ as $\bar{\varphi}^2/2+i \bar{\varphi}_i \partial W_0/\partial \phi_i$,  the classical action \eqref{classic1} takes the form:
\begin{align}
\mathcal{S}_{BRS}= \int dt \Bigg( &\frac{\dot{\phi}^2}{2} +\frac{\bar{\varphi}^2}{2} + i \bar{\varphi}_i \frac{\partial W_0}{\partial \phi_i} +i \bar{\psi}_i \left(-\delta_{ij}\partial_t+\frac{\partial^2 W_0}{\partial \phi_i(t) \partial \phi_j(t)}\right)\psi_j\Bigg)\,. \label{classic2}
\end{align}
The supersymmetry can be made even more explicit by introducing the superfield:
\begin{equation}
\Phi_i(t;\bar{\theta},\theta)=\phi_i(t)+\theta \bar{\psi}_i(t)+\psi_i(t) \bar{\theta}+ \theta \bar{\theta} \bar{\varphi}_i(t)\,, \label{superfield}
\end{equation}
where $\theta$ and $\bar{\theta}$ are Grassmann valued, as well as the operators:
\begin{equation}
\mathcal{D}=i \frac{\partial}{\partial \bar{\theta}}-\theta \frac{\partial }{\partial t}\,,\quad \bar{\mathcal{D}}=i \frac{\partial}{\partial {\theta}}-\bar{\theta} \frac{\partial }{\partial t}\,,
\end{equation}
which satisfies  the  anti-commutator relations $\{\mathcal{D},\mathcal{D}\}=\{\bar{\mathcal{D}},\bar{\mathcal{D}}\}=0$, and $\{{\mathcal{D}},\bar{\mathcal{D}}\}=-2i\partial_t$. Then the following relation holds:
\begin{equation}
\frac{1}{2}\Phi (\mathcal{D}\bar{\mathcal{D}}-\bar{\mathcal{D}} \mathcal{D}) \Phi =\phi \ddot{\phi} -\bar{\varphi}^2-2i \bar{\psi} \dot{\psi}\,.
\end{equation}
Moreover, expanding $H_0(\Phi)$ to the power of $\theta$ and $\bar{\theta}$, and using the properties of Grassmann algebra, it is not hard to see that the superpath integral:
\begin{equation}
S[\Phi]=\int dt\, d\bar{\theta} d\theta \, \left(\frac{1}{2}\Phi \tilde{K}\Phi+ iW_0(\Phi) \right)\,,\label{supersym}
\end{equation}
is reduced to the \textit{off-shell} classical action \eqref{classic1}, if $\tilde{K}:= -[{\mathcal{D}},\bar{\mathcal{D}}]/2$. Therefore, the supersymmetry properties may be explicitly checked by introducing the nilpotent operators $Q:=i\partial_{\bar{\theta}}+\theta \partial_t$ and $\bar{Q}:=i\partial_{{\theta}}+\bar{\theta} \partial_t$ such that $\{Q,\bar{Q} \}=2i \partial_t$, and then the infinitesimal supersymmetric transformations acts on the superfield $\Phi$ through the operator $\delta_\epsilon=\bar{\epsilon} Q-\epsilon \bar{Q}$ as $\Phi\to \Phi+\delta_\epsilon \Phi$. A few algebraic computation show that the action \eqref{supersym} is invariant under such a transformation at first order in $\epsilon$. More details on supersymmetry can be found in standard references \cite{Kurchan1992,Synatschke_2009,ZinnJustinBook1}. 

\subsection{Generating functional}\label{sectiongen}

 A generating functional of the  correlation functions can be defined through a standard method known as Martin-Siggia-Rose-Janssen-de
Dominicis formalism  which is largely discussed in the literature (see \cite{DeDominicisbook,Takayama1986}, and references therein). From a solution of the Langevin equation by fixing $J$, $\phi_i(t)$, we may define formally the generating functional:
\begin{equation}
Z[j]= \left \langle \exp \left[ \int dt \sum_{i=1}^N j_i(t) \phi_i(t) \right] \right \rangle \,.\label{generatingfunctional}
\end{equation}
Because of the normalization of the Gaussian noise, the averaging over $\eta$ conserves the normalization and $Z[j=0]=1$. This condition allows us to perform an averaging over disorder without introducing the replica method \cite{DeDominicisbook}. The computation of the averaging follows the same strategy as for the previous section. The end conditions may be enforced with a suitable delta function, and the trick \eqref{identity1} can be used in the same way. The determinant can be rewritten using Grassmann fields, and after integrating out the delta function $\delta\left(\dot{Q}+\nabla_Q H_0-\eta \right)$, we may break the square $(\dot{Q}+\nabla_Q H_0)^2$ introducing an auxiliary vectorial field $\bar{\varphi}_i$. We obtain:
\begin{equation}
Z[j]= \int [d\phi] [d\bar{\psi}] [d\psi] [d\bar{\varphi}] e^{-\tilde{S}[\phi,\bar{\psi},\psi,\bar{\varphi}]+\int dt j(t)\phi(t) }\,,
\end{equation}
where we introduced the short notation $j(t)\phi(t):= \sum_{i=1}^N j_i(t) \phi_i(t)$ and:
\begin{align}
\tilde{S}[\phi,\bar{\psi},\psi,\bar{\varphi}]:=\int dt &\Bigg( \frac{\bar{\varphi}^2}{2} + i \bar{\varphi}_i\left( \dot{\phi}_i+ \frac{\partial W_0}{\partial \phi_i} \right)+i \bar{\psi}_i \left(-\delta_{ij}\partial_t+\frac{\partial^2 W_0}{\partial \phi_i(t) \partial \phi_j(t)}\right)\psi_j\Bigg)\,.\label{classicalaction1}
\end{align}
It is suitable to introduce a second generating functional $\tilde{Z}[j,\tilde{j}]$ both for  the fields $\phi$ and $\bar{\varphi}$,
\begin{equation}
\tilde{Z}[j,\tilde{j}]:= \int [d\phi] [d\bar{\psi}] [d\psi] [d\bar{\varphi}] e^{-\tilde{S}[\phi,\bar{\psi},\psi,\bar{\varphi}]+\int dt( j\phi + \tilde{j} \bar{\varphi})}\,, \label{prelim}
\end{equation}
with the normalization condition $\tilde{Z}[j=0,\tilde{j}=0]=1$. As explained above,  this property allows us to perform an integral over quenched disorder $J$ without introducing replica:
\begin{equation}
\overline{\tilde{Z}[j,\tilde{j}]}\equiv \int [d\Xi] [dJ] p(J) e^{-\tilde{S}[\phi,\bar{\psi},\psi,\bar{\varphi}]+\int dt( j\phi + \tilde{j} \bar{\varphi})} \label{partitionzero}
\end{equation}
where $\Xi:=(\phi, \psi, \bar{\psi}, \bar{\varphi})$ denotes collectively all the fields involved in the path integral.

\section{Functional renormalization group formalism}\label{sec2}

\subsection{Basics}

The RG aims to interpolate between a microscopic description and a macroscopic description, taking into account fluctuations at all scales. RG provides a progressive description of effective physics at different scales, integrating out firstly the modes having a small wavelength and ending by the ones having a large wavelength. This coarse-graining is the most operational incarnation of the block-spin idea of Kadanoff. For stochastic and non-equilibrium systems, temporal fluctuations cannot be ignored and can be included in the coarse-graining through a progressive integration in the frequencies space. This has been considered in some recent works, among them \cite{Duclut_2017}. In this paper, we follow the same strategy, and consider a coarse-graining in time that interpolate between two regimes:
\begin{enumerate}
\item The UV regime, where fluctuations are frozen and fields configurations are determined by stationary points of the classical action $\tilde{S}$.
\item The IR regime, where fluctuations are all integrated out and field configurations described through the effective action $\Gamma$, Legendre transform of the Gibbs free energy.
\end{enumerate}

\noindent
The standard procedure consists in adding to the classical action $\tilde{S}$ a suitable non-local functional $$\Delta S_k:= \int \Xi_a(t)(R_k)_{ab}(t-t^\prime)\Xi_b(t^\prime)\,,$$ (for some energy scale $k$), assuming $R_k$, the \textit{regulator}, to be differentiable in $k$ and $t$. It plays the role of a momentum dependent mass, such that high energy modes concerning $k$ are integrated out whereas low energy modes are frozen. In such a way we are aiming to construct a smooth interpolation $\Gamma_k$ of $\tilde{S}$ and $\Gamma$ i.e. between some UV scale $k=\Lambda$ (for some $\Lambda$) where all fluctuations are frozen and the IR scale $k=0$ where all fluctuations are integrated out. The last condition generally imposes that $R_k$ vanish for $k=0$, and the symmetry must be formally restored in the deep IR for the exact flow equations. However, this equation in itself cannot be solved exactly except for very special cases; and the approximations used to solve which work generally into a reduced dimensional phase space may introduce a spurious dependence on the regulator \cite{pawlowski2015physics}. To avoid these difficulties, we have to be able to construct an RG that preserve as many symmetries as possiblly allowed from equilibrium condition. Among them, we demand that  the RG flow will be such that it preserves \textit{causality}, \textit{time reversal symmetry} and \textit{supersymmetry}. Note that time-reversal symmetry and supersymmetry are not truly independents, in the sense that all the Ward-Takahashi identities coming from supersymmetry can be used to derive equilibrium relations as a fluctuation-dissipation theorem, which are the consequence of the time-reversal symmetry (as the Onsager relation). However, as pointed out in \cite{Aron_2010} the inverse is not true and supersymmetry fails to provide relations where time-reversal appears explicitly. The origin of this non-reciprocity can be traced back to the fact that time-reversal symmetry cannot be decomposed as a sum of supersymmetry generators. From this observation, one expects that time-reversal symmetry implies supersymmetry, but that the reverse is false. Thus, we essentially focus on the time-reversal, and we will check at the end that supersymmetry and the corresponding Ward-Takahashi identities hold. Finally, the last requirement concerns the averaging over the disorder. Indeed, if we aim to coarse-grain before averaging over the disorder, we show from \eqref{prelim} and \eqref{partitionzero} that such a program requires $\tilde{Z}[j=0,\tilde{j}=0]=1$ before averaging, and the introduction of the regulator must not change this normalization.

\subsection{Physical constraints on the regulator at equilibrium}

\textit{1. Coarse-grained partition function.} We start our analysis by constructing a coarse-grained partition function, labeled with a scale index $k$, in a way compatible with the normalization condition. Following the analyze in \cite{Duclut_2017} we add to the force $\partial H_0/\partial q_i$ on the right-hand side of equation \eqref{eq1} a non-local driving force $f_i[q(t)]$ such that:
\begin{equation}
\dot{q}_i(t)=-\frac{\partial H_0}{\partial q_i(t)}- \int dt^\prime R_k^{(1)}(t-t^\prime) q_i(t^\prime)+\eta_i(t)\,.\label{modified1}
\end{equation}
The additional driving force is constructed as a non-local functional of the random trajectory $q(t)$, depending on the history of the trajectory:
\begin{equation}
f_i[q(t)]:=\int dt^\prime \, R_k^{(1)}(t-t^\prime) q_i(t^\prime)\,.\label{driv1}
\end{equation}
In addition, we modify the noise correlation in a non-local manner, adding a non-local function to the Dirac delta. In such a way, the inverse propagator for $\eta$ becomes:
\begin{equation}
2D\delta (t-t^\prime) \to 2D \left(\delta (t-t^\prime)+R_k^{(2)}(t-t^\prime)\right)\,,
\end{equation}
which can be understood as the introduction of a short memory in the dynamics of the system. Note that the averaging over disorder also produces such an explicit memory effect which couples different times. The non-local functions $R_k^{(1)}$ and $R_k^{(2)}$ aims to cut-off small frequencies modes $\omega \lesssim k$, for some infrared cut-off $k$. In such a way, the formal derivation of the previous section leads to the one family parameter of models described by the generating functional:
\begin{equation}
\overline{\tilde{Z}_k[j,\tilde{j}]}\equiv \int [d\Xi] [dJ] p(J) e^{-\tilde{S}[\Xi]-\Delta\tilde{S}_k[\Xi]+\int dt( j\phi + \tilde{j} \bar{\varphi})}\,,\label{partitionmean1}
\end{equation}
the additional term $\Delta\tilde{S}_k[\Xi]$ being defined as:
\begin{align}
\nonumber \Delta\tilde{S}_k:=\int dt dt^\prime& \sum_i \Big(i \bar{\varphi}_i(t) R_k^{(1)}(t-t^\prime) \phi_i(t^\prime)\\
&+\frac{1}{2}\bar{\varphi}_i(t)R_k^{(2)}(t-t^\prime)\bar{\varphi}_i(t^\prime)+i\bar{\psi}_i(t) R_k^{(1)}(t^\prime-t)\psi_i(t^\prime)
\Big)\,, \label{regulatordef}
\end{align}
Finally, we can perform the integral over $J$ in equation \eqref{partitionmean1}. The integration is easier using the supersymmetric formalism, the integral that we need to compute reads as:
\begin{equation}
I[\Xi]:=\int [dJ] p(J)\, \exp{ \bigg(i \int dt d\bar{\theta} d\theta\, \frac{(2D)^{\frac{p}{2}-1}}{p} \sum_{\{i_\ell\}} J_{i_1i_2\cdots i_p} \Phi_{i_1}\Phi_{i_2}\cdots \Phi_{i_p}\bigg)}\,,
\end{equation}
where $\Phi_{i}\equiv \Phi_{i}[\Xi]$ is defined from the one to one mapping \eqref{superfield}. From the properties \eqref{distributionJ} defining the Gaussian distribution $p(J)$, we get the effective field theory non-local in time \cite{DeDominicisbook,DeDominicis1978}:
\begin{equation}
\overline{\tilde{Z}_k[j,\tilde{j}]}\equiv \int [d\Xi] e^{-\overline{S}[\Xi]-\Delta\tilde{S}_k[\Xi]+\int dt( j\phi + \tilde{j} \bar{\varphi})}\,, \label{partitionfinale}
\end{equation}
with:
\begin{align}
\overline{S}[\Xi]:=\int dt \sum_i&\Bigg( \frac{\bar{\varphi}^2_i}{2} + i \bar{\varphi}_i\dot{\phi}_i+i\bar{\psi}_i\dot{\psi}_i\Bigg)+i\int dz\overline{W_0}[\Phi(z)]\,,\label{classicalactionbis}
\end{align}
the super-coordinate $z=(t,\theta,\bar{\theta})$ having measure $dz:= dt d\bar{\theta} d\theta$ and the effective (non-local) stochastic potential $\overline{W_0}$ being given by:
\begin{align}
\overline{W_0}&:=-i\lambda\frac{(2D)^{p-2}}{p^2N^{p-1}} \int dz^\prime\big(\Phi(z)\cdot \Phi(z^\prime)\big)^p+\frac{(2D)^{\frac{p}{2}-1}}{p} \frac{J_0}{N^{p-1}} \big(\Phi_{\bullet}(z)\big)^p+\sum_m \frac{h_m}{2m}\bigg(\frac{2D}{N}\bigg)^{m-1} (\Phi^2)^m\,.
\end{align}
where we introduced the notations:
\begin{equation}
\Phi\cdot \Phi^\prime:= \sum_i \Phi_i \Phi_i^\prime\,, \quad \Phi_{\bullet}:= \sum_i \Phi_{i}\,.
\end{equation}
Taking derivative with respect to $k$ of the partition function \eqref{partitionfinale}, we deduce an equation describing the evolution of the free energy $\overline{\mathcal{W}}_k:= \ln \overline{\tilde{Z}}_k$, which  can be translated as an equation describing the evolution of the effective averaged action $\Gamma_k$, which interpolates between the microscopic action \eqref{classicalactionbis} at $k=\Lambda$ (for some cut-off $\Lambda$ in high frequency modes) and the full effective action $\Gamma$ for $k=0$. It is defined as:
\begin{equation}
\Gamma_k[\mathcal{M}]+\overline{\mathcal{W}}_k[\mathcal{J}]=\int dt (\mathcal{J},\mathcal{M})- \Delta\tilde{S}_k[\mathcal{M}]\,, \label{Gammak}
\end{equation}
where we introduced the inner product between superfields:
\begin{equation*}
(\mathcal{J},{\Xi})=\sum_i (j_i(t) \phi_i(t)+\tilde{j}_i(t) \bar{\varphi}_i(t)+ \bar{\psi}_i(t) \chi_i(t)+\bar{\chi}_i \psi_i(t))\,,
\end{equation*}
including external sources $(\chi_i, \bar{\chi}_i)$ for fermions; and the notation $\mathcal{M}_{\alpha}\equiv \partial \overline{\mathcal{W}}_k/\partial \Xi_{\alpha}$ denoting collectively the classical fields -- the extra index $\alpha$ running over all fermionic and bosonic fields involved into the set $\Xi$. The resulting flow equation for $\Gamma_k$, describes the move through the full theory space from UV scales ($k\approx \Lambda$) to IR scales ($k\approx 0$) is \cite{Wett1,Wett2}:
\begin{align}
\partial_k{\Gamma}_k:=\frac{1}{2}\Tr_B& \left([\Gamma_k^{(2)}+\textbf{R}_k]^{-1} \partial_k{\textbf{R}}_k\right)-\frac{1}{2} \Tr_F \left([\Gamma_k^{(2)}+{R}_k^{(1)}]^{-1} \partial_k{{R}}_k^{(1)}\right)\,, \label{Wetterich}
\end{align}
the traces $\Tr_B$ and $\Tr_F$ denoting respectively traces over all relevant bosonic and fermionic indices; and the matrix ${\textbf{R}}_k$ is defined as:
\begin{equation*}
\Delta\tilde{S}_k^{(\mathcal{B})}:= \frac{1}{2}\sum_i \int dt dt^\prime (\bar{\varphi}_i(t), \phi_i(t)) {\textbf{R}}_k(t-t^\prime) \begin{pmatrix}
\bar{\varphi}_i(t^\prime)\\
\phi_i(t^\prime)
\end{pmatrix}\,,
\end{equation*}
and:
\begin{equation}
\Gamma_{k,\alpha \beta}^{(2)}:= \frac{\overrightarrow\partial}{\partial \mathcal{M}_\alpha} \Gamma_k \frac{\overleftarrow\partial}{\partial \mathcal{M}_\beta}\,.
\end{equation}
Note that in the rest of this paper we will denote as $\Gamma^{(n)}_k$ the nth functional derivative with respect to the classical superfield.
\medskip

\textit{2. Time reversal symmetry and causality.} We now move on to the main physical constraints, namely the time reversal symmetry and the causality, and following \cite{Duclut_2017} we will investigate the conditions on $R_k^{(1)}$ and $R_k^{(2)}$ which preserve these physical requirements. Indeed, from the hypothesis that the system relaxes toward equilibrium for sufficiently large time, one infers time reversal invariance of the move equations, and thus of the classical action \eqref{classicalaction1}. The fermionic part of the action has to be invariant from time reversal by construction. Indeed, because all the prescriptions to express the determinant must be physically true, it is sufficient to check invariance from Stratonovich prescription \eqref{Stratonovich}. The remaining part of the action is invariant in the transformation, $(\phi,\bar{\varphi})\to (\phi^\prime,\bar{\varphi}^\prime)$ defined as:
\begin{equation}
\phi^\prime_i(t)=\phi_i(-t)\,,\quad \bar{\varphi}^\prime_i(t)=\bar{\varphi}_i(-t)+2i\dot{\phi}_i(-t)\,, \label{timerevers}
\end{equation}
up to a total derivative. These transformations acting on the bosonic sector of $\Delta\tilde{S}_k[\Xi]$ leads to:
\begin{align*}
\nonumber\Delta\tilde{S}_k^{(\mathcal{B})}\to \int dt dt^\prime& \sum_i \Big[i \bar{\varphi}_i(t) R_k^{(1)}(t^\prime-t) \phi_i(t^\prime)-2 \dot{\phi}_i(t) R_k^{(1)}(t^\prime-t) \phi_i(t^\prime)+\frac{1}{2}\bar{\varphi}_i(t)R_k^{(2)}(t^\prime-t)\bar{\varphi}_i(t^\prime)\\
& +i\dot{\phi}_i(t)\big(R_k^{(2)}(t^\prime-t)+R_k^{(2)}(t-t^\prime)\big)\bar{\varphi}_i(t^\prime)-2\dot{\phi}_i(t)R_k^{(2)}(t^\prime-t)\dot{\phi}_i(t^\prime) \Big]\,,
\end{align*}
or, from the symmetries of $R_k^{(2)}$:
\begin{align*}
\nonumber\Delta\tilde{S}_k^{(\mathcal{B})}\to& \int dt dt^\prime \sum_i \Big[i \bar{\varphi}_i(t) R_k^{(1)}(t^\prime-t) \phi_i(t^\prime)+2 \dot{\phi}_i(t) R_k^{(1)}(t^\prime-t) \phi_i(t^\prime)+\frac{1}{2}\bar{\varphi}_i(t)R_k^{(2)}(t^\prime-t)\bar{\varphi}_i(t^\prime) \\
&+i\dot{\phi}_i(t)\big(R_k^{(2)}(t^\prime-t)+R_k^{(2)}(t-t^\prime)\big)\bar{\varphi}_i(t^\prime)-\dot{\phi}_i(t)\big(R_k^{(2)}(t^\prime-t)+R_k^{(2)}(t-t^\prime)\big)\dot{\phi}_i(t^\prime) \Big]\,.
\end{align*}
Thus, integrating by part, the requirement that the global action is invariant from time reversal thus writes as, up to a total derivative:
\begin{align*}
0\equiv \int dt dt^\prime& \sum_i \Big[i \bar{\varphi}_i(t) \big(R_k^{(1)}(t^\prime-t)-R_k^{(1)}(t-t^\prime)+\dot{R}_k^{(2)}(t-t^\prime)-\dot{R}_k^{(2)}(t^\prime-t)\big){\phi}_i(t^\prime)\\
&+ \dot{\phi}_i(t) \big( -2 R_k^{(1)}(t^\prime-t) +\dot{R}_k^{(2)}(t^\prime-t)-\dot{R}_k^{(2)}(t-t^\prime)\big){\phi}_i(t^\prime) \Big]\,.
\end{align*}
This constraint must be satisfied by all fields $\bar{\varphi}_i$ and $\phi_i$. Moreover, an integration by part of the term $\int dtdt^\prime\sum_i \dot{\phi}_i(t)R_k^{(1)}(t^\prime-t) \phi_i(t^\prime)$ show that a sufficient condition of $R_k^{(1)}$ and $R_k^{(2)}$ is the following:
\begin{equation}\begin{boxed}{
R_k^{(1)}(t)-R_k^{(1)}(-t)+\dot{R}_k^{(2)}(-t)-\dot{R}_k^{(2)}(t)=0\,. \label{condition1}}\end{boxed}
\end{equation}
Equation \eqref{condition1} is the first physical requirement that we impose on the regulators. The second requirement comes from causality. Indeed, because the additional driving force $f_i[q(t)]= \int dt^\prime R_k^{(1)}(t-t^\prime) q_i(t^\prime)$ depends itself on the history of the trajectory, it could have feedback on itself, thus breaking causality. To avoid such an effect, we have to impose $R_k^{(1)}(t-t^\prime)=0$ for $t<t^\prime$, i.e.
\begin{equation}
R_k^{(1)}(t-t^\prime)\propto\theta(t-t^\prime)\,,
\end{equation}
where $\theta(t)$ denotes the standard Heaviside function. \\

\textit{3. Supersymmetry and Ward-Takahashi identities.} The original model \eqref{partitionzero} exhibits a non-trivial supersymmetry which can be easily checked from the discussion of the previous sections. Indeed, the kinetic part of the classical action \eqref{classicalactionbis}
\begin{equation}
\overline{S}_{\text{kin}}[\Xi]:=\int dt \sum_i\Bigg( \frac{\bar{\varphi}^2_i}{2} + i \bar{\varphi}_i\dot{\phi}_i-i\bar{\psi}_i\dot{\psi}_i\Bigg)\,,\label{kinetic}
\end{equation}
can be rewritten explicitly in a supersymmetric form \cite{ZinnJustinBook1}. Introducing the differential operators $\bar{D}$ and $D$ as:
\begin{equation}
\bar{D}:=i \frac{\partial}{\partial \bar{\theta}}\,,\qquad D= i\frac{\partial}{\partial \theta}- 2 \bar{\theta}\frac{\partial}{\partial t}\,,
\end{equation}
satisfying $D^2=\bar{D}^2=0$ and $\{D,\bar{D}\}=-2i\partial_t$, we must have:
\begin{equation}
\overline{S}_{\text{kin}}[\Xi]=\frac{1}{2}\sum_i\int dz\, \bar{D} \Xi_i D\Xi_i\,.
\end{equation}
Thus, we introduce the nilpotent operators $\mathcal{Q}$ and $\bar{\mathcal{Q}}$,
\begin{equation}
\mathcal{Q}= i \frac{\partial}{\partial \theta}\,,\quad \bar{\mathcal{Q}}=i \frac{\partial}{\partial \bar{\theta}}+2\theta \frac{\partial}{\partial t}\,.
\end{equation}
which anticommute with $D$ and $\bar{D}$ and such that $\{\mathcal{Q},\bar{\mathcal{Q}} \}=2i\partial/\partial t$, it is a simple exercise to prove that the classical action \eqref{classicalaction1} is invariant at first order in $\bar{\epsilon}$ under the field transformation $\delta \Phi_i(t,\theta,\bar{\theta})=\bar{\epsilon} \bar{\mathcal{Q}} \Phi_i$; the variation of the kinetic action $\overline{S}_{\text{kin}}[\Xi]$ taking the form of a total derivative. At the quantum level, this invariance translates as exact (i.e. valid to all orders in the perturbative expansion) relations between correlations functions, summarized through the Ward-Takahashi identities. Let us consider the following partition function, including sources for fermionic fields:
\begin{equation}
\overline{\tilde{Z}[\mathcal{J}]}:=\int [d\Xi] e^{-\bar{S}[\Xi]+\int dt (\mathcal{J}(t),{\Xi}(t))}\,. \label{partitionmean2}
\end{equation}
From an infinitesimal supersymmetric transformation, the partition function becomes:
\begin{equation}
\delta \ln \overline{\tilde{Z}}[\mathcal{J}]= \int dt (\mathcal{J}(t), \delta_*\mathcal{M})\,, \label{varprime}
\end{equation}
where $\delta_*$ is defined from linearity of $\delta$ as the image of the transformation $\delta$ on the effective field $\mathcal{M}(t)$. Assuming the integration measure $[d\Xi]$ is translation invariant $[d\Xi]=[d(\Xi+\delta\Xi)]$ (without anomaly), we must have $\delta \overline{\tilde{Z}}_k[\mathcal{J}]=0$. Therefore, taking into account the explicit expression of the supersource:
\begin{equation}
\mathcal{J}_i:= \tilde{j}_i + \theta \bar{\chi}_i+ \chi_i \bar{\theta}+\theta\bar{\theta} j_i \,,
\end{equation}
we get in terms of the first derivative of the classical action the compact relation:
\begin{equation}
\int dz \sum_i \Phi_i(z) \bar{\mathcal{Q}}\frac{\delta \Gamma}{\delta \Phi_i(z)} =0\,.\label{WT}
\end{equation}
Which is the standard expression of Ward-Takahashi identity, from which all the relations between correlation functions coming from supersymmetry can be deduced. One expects that such a relations have to hold along all physical relevant RG candidates. However, the presence of the regulator must break them, if it is not explicitly supersymmetric, adding to the variation \eqref{varprime} the contribution $\langle \delta \Delta \tilde{S}_k \rangle$. It is however easy to check that supersymemtry is ensured by time reversal symmetry, and such that $ \delta \Delta \tilde{S}_k=0$. To check this, note that the infinitesimal transformation $\delta \Phi_i(t,\theta,\bar{\theta})=\bar{\epsilon} \bar{\mathcal{Q}} \Phi_i$, reads explicitly as:
\begin{align}
\nonumber \delta \phi_i= i\psi_i \bar{\epsilon}\,, \qquad &\delta \psi_i=0\,, \\
\delta \bar{\psi}_i=\left(i\bar{\varphi}_i-2\dot{\phi}_i\right)\bar{\epsilon}\,, \qquad&\delta \bar{\varphi}_i=2\dot{\psi}_i \bar{\epsilon}\,. \label{SUSYtransform}
\end{align}
Thus, computing the variation $ \Delta \tilde{S}_k$ we get:
\begin{align}
\nonumber\delta \Delta \tilde{S}_k= &\int dt dt^\prime \sum_i \Big(2i \dot{\psi}_i(t) R^{(1)}_k(t-t^\prime) \phi_i(t^\prime)-\bar{\varphi}_i(t)R^{(1)}_k(t-t^\prime)\psi_i(t^\prime)\\
&+\bar{\varphi}_i(t)R^{(1)}_k(t^\prime-t)\psi_i(t^\prime)+2\dot{\psi}_i(t) R^{(2)}_k(t-t^\prime) \bar{\varphi}(t^\prime) +2i \dot{\phi}_i(t) R^{(1)}_k(t^\prime-t) \psi(t^\prime)\Big) \bar{\epsilon}\,.
\end{align}
The first and the last terms cancel exactly. Moreover, taking into account the physical condition \eqref{condition1} for the second and the third terms we get:
\begin{align*}
\delta \Delta \tilde{S}_k= \int dt dt^\prime \sum_i &\Big[2\dot{\psi}_i(t) R^{(2)}_k(t-t^\prime) \bar{\varphi}_i(t^\prime) - \bar{\varphi}(t)[\dot{R}^{(2)}_k(t-t^\prime)-\dot{R}^{(2)}_k(t^\prime-t)]\psi(t^\prime)\Big]\bar{\epsilon}\,.
\end{align*}
Finally, integrating by parts the first term, we obtain:
\begin{align*}
\delta \Delta \tilde{S}_k= - \int dt dt^\prime \sum_i & \bar{\varphi}(t)[\dot{R}^{(2)}_k(t-t^\prime)+\dot{R}^{(2)}_k(t^\prime-t)]\psi(t^\prime)\bar{\epsilon}\,.
\end{align*}
Because the regulator $R^{(2)}_k(t)$ is expected to be an even function, $\dot{R}^{(2)}_k(t)$ must be even, and the bracket vanish\footnote{This can be specially checked for the regulator \eqref{r2}, which reads explicitly as: $${R}^{(2)}_k(t)=\alpha k \big(\theta(t) e^{-kt}+\theta(-t) e^{kt} \big)\,.$$}. Therefore, $\delta \Delta \tilde{S}_k=0$ and supersymmetry is ensured from time reversal symmetry. Moreover, from the way we derive the relation \eqref{WT}, but working with the coarse-grained partition function, we get the identity:
\begin{equation}
\int dz \sum_i \Phi_i(z) \bar{\mathcal{Q}}\frac{\delta \Gamma_k}{\delta \Phi_i(z)} =0\,.\label{WT2}
\end{equation}
ensuring that the relations between correlations functions coming from supersymmetry, and valid for the IR theory, remains true along the RG flow, at least as long as no singularities are encountered. \\

\textit{4. Choice of the regulator.} The derivation of the flow equation proceeds in two steps. Following \cite{Duclut_2017} we choose the one-parameter family of regulators:
\begin{equation}
R_k^{(1)}(t):= \alpha Z_k k^2 e^{- k t}\,\theta(t)\,,
\end{equation}
which is compatible with causality and whose Fourier transform can be computed analytically \footnote{Note that the fact that the pole is localized in the half inferior part of the complex plane, as a consequence of the locality prescription.}:
\begin{equation}
R_k^{(1)}(\omega)=\alpha \frac{Z_k k^2}{k-i\omega}=: Z_k \rho_k^{(1)}(\omega)\,. \label{r1}
\end{equation}
The Fourier transform (using the same symbol to designate the function and its Fourier transform) being defined as in the rest of this paper as:
\begin{equation}
f(\omega):= \int dt\, f(t) e^{i\omega t}\,.
\end{equation}
The parameter $\alpha\in \mathbb{R}$ can be adjusted numerically to optimize the truncation. Moreover, the condition \eqref{condition1} allows computing $R_k^{(2)}$:
\begin{equation}
R_k^{(2)}(\omega)=\frac{R_k^{(1)}(-\omega)-R_k^{(1)}(\omega)}{2i\omega}\,, \label{relationR}
\end{equation}
where we assumed $R_k^{(2)}(t)$ even in time from its definition below \eqref{partitionmean1}. Explicitly:
\begin{equation}
R_k^{(2)}(\omega)=-\alpha \frac{Z_k k^2}{k^2+\omega^2}=:Z_k\rho_k^{(2)}(\omega)\,, \label{r2}
\end{equation}
Let us examine the boundary conditions. For $k\to \Lambda \gg 1$, we designed the first regulator $R_k^{(1)}(\omega)$ such that $R_{k=\Lambda}^{(1)}(\omega)\sim \Lambda$, ensuring that fluctuations are frozen in the deep UV limit. In the same way, however, $R_{k\to \Lambda}^{(2)}(\omega)\to -Z_\Lambda \alpha$; and the initial correlation noise is not exactly recovered. Nevertheless, we expect that such an initial condition, which has only for effect to modify the initial value of  $D\to D(1-\alpha Z_\Lambda)$ does not affect the universal behaviors of the system, at least as long as $\alpha Z_\Lambda \neq 1$. Finally, for $k\to 0$, the two regulators both vanish ensuring that all fluctuations are integrated out. \\

\section{Multi-local truncations for $p=3$ models}\label{sec3}

We denote this section to the construction of approximate solutions of the exact flow equation \eqref{Wetterich} which cannot be solved exactly excepts for very special cases. We are aiming to construct approximate solutions able to deal both with the nonperturbative regime and non-locality of the interactions. The difficulty to solve the exact RG flow equation can be traced back to the fact that it describes a trajectory into an infinite-dimensional functional space, thus we resort to approximations that consist in truncating it to form a finite-dimensional subspace. Here, the non-local structure of the interactions provides a non-conventional difficulty. To deal with it, we introduce a method inspired by the replicated construction considered by \cite{Tarjus_20082,Tissier_2008,Tissier_2011,Tissier_20122,Tissier_2012} for the random field Ising model, organizing truncation as a hierarchical multi-local expansion but keeping in mind two differences with their construction:
\begin{enumerate}
\item The role of the discrete replica index is now played by the continuous super-coordinate $z$.
\item The coarse-grained dimension is time rather than space.
\end{enumerate}
We introduce the methodology for the unconstrained model and close this section by considering the spherical model.

\subsection{Parametrization of the theory space: minimal bi-local truncation}

As the perturbative expansion shows straightforwardly, the non-local structure of the interaction term in \eqref{classicalactionbis} contaminates the free energy $\overline{\mathcal{W}}:= \ln \overline{\tilde{Z}}$, which becomes a non-local function of the external sources $\mathcal{J}$ as well. In such a way, we expect an expansion of the form:
\begin{equation}
\overline{\mathcal{W}}[\mathcal{J}]= \sum_{n=1}^\infty\frac{1}{n!} \int \prod_{\ell=1}^n dz_\ell\, \varpi_n(\mathcal{J}(z_1),\cdots,\mathcal{J}(z_n))\,.\label{hierarchy}
\end{equation}
The origin of this expansion can be easily understood as follows. The potential in the classical action \eqref{classicalactionbis} writes schematically as a sum of two terms:
\begin{align}
\nonumber \overline{W_0}= A\,\vcenter{\hbox{\includegraphics[scale=1]{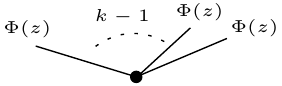} }}+B\,\vcenter{\hbox{\includegraphics[scale=0.8]{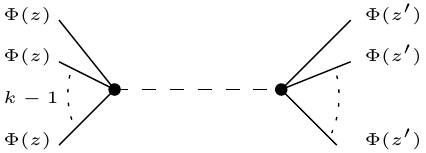} }}\,,
\end{align}
the dotted edge materializing the contraction of internal indices $i$. Thus, from standard perturbation theory and properties of Gaussian integrations, the loop expansion in power of $A$ and $B$ organizes schematically as:
\begin{align*}
\overline{\mathcal{W}}= & \underbrace{A \,\vcenter{\hbox{\includegraphics[scale=0.8]{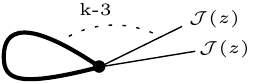} }}+\cdots + B \, \vcenter{\hbox{\includegraphics[scale=0.8]{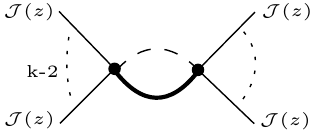} }}+\cdots }_{\varpi_1^{(1)}}\\
&+\underbrace{A^2 \vcenter{\hbox{\includegraphics[scale=0.8]{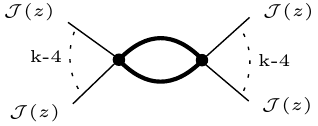} }} + AB \vcenter{\hbox{\includegraphics[scale=0.8]{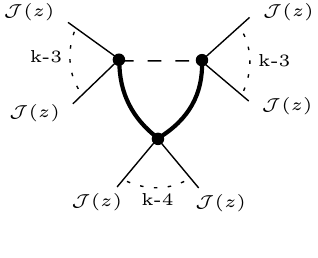} }}+B^2\vcenter{\hbox{\includegraphics[scale=0.8]{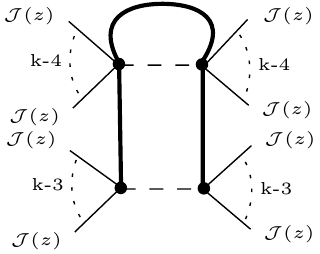} }}+\cdots }_{\varpi_1^{(2)}}\\
&+\underbrace{A^2\,\vcenter{\hbox{\includegraphics[scale=0.8]{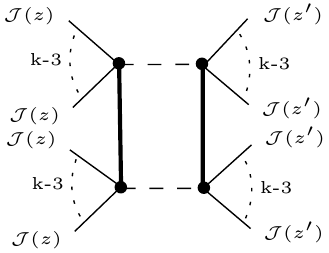} }}+\cdots}_{\varpi_2^{(2)}}+\underbrace{A^3\,\vcenter{\hbox{\includegraphics[scale=0.8]{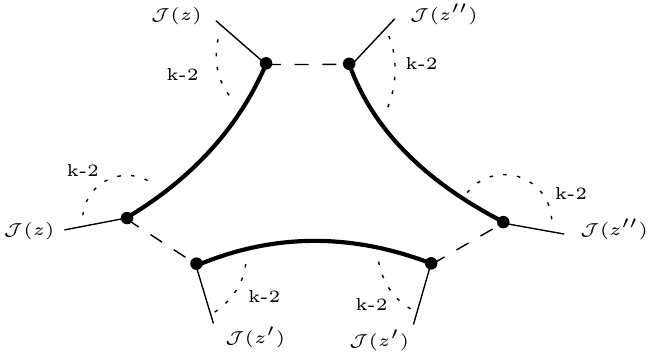} }}}_{\varpi_3^{(3)}}+\cdots
\end{align*}
where the thick edges materialize the Wick contraction with propagator defined by the kinetic part of the classical action \eqref{classicalactionbis}. Note that we voluntarily omitted the combinatorial factors, signs and additional numerical factors as well as classical (i.e. tree) contributions, irrelevant for our purpose. Finally, the upper index $p$ in $\varpi_n^{(p)}$ refers to the order of the perturbative expansion. Furthermore, the hierarchical structure of the expansion \eqref{hierarchy} contaminates the Legendre transform $\Gamma[\mathcal{M}]$ of $\overline{\mathcal{W}}[\mathcal{J}]$ which have to be expands as:
\begin{equation}
\Gamma[\mathcal{M}]= \sum_{n=1}^\infty\frac{1}{n!} \int \prod_{\ell=1}^n dz_\ell\, \gamma_n(\mathcal{M}(z_1),\cdots,\mathcal{M}(z_n))\,.\label{hierarchy2}
\end{equation}
Now, we are aiming to construct an approximation of the exact equation \eqref{Wetterich} through a suitable parametrization of the full theory space able to deal with the non-local structure of the interactions. Moreover, we can only consider a finite-dimensional region of this theory space. This can be achieved through an ansatz for the effective averaged action $\Gamma_k[\mathcal{M}]$ retaining only a finite set of couplings. This ansatz moreover must be able to deal with some exact constraint coming from the partition function \eqref{partitionfinale}. In particular, because we focus on equilibrium, we consider a truncation compatible with the time-reversal symmetry, i.e. with the transformation \eqref{timerevers}. Moreover, assuming that supersymmetry holds, we have only to retain linear interactions in $\bar{\varphi}$ (this can be viewed as a requirement imposed by the fluctuation-dissipation theorem as well, see \cite{Duclut_2017}).
\medskip

We choose to work with notations that make the supersymmetry explicit. To simplify, we denote the classical fields using the same symbols corresponding to the random variables, and view $\mathcal{M}=(\phi, \psi, \bar{\psi}, \bar{\varphi})$ as a superfield. Moreover, we define the operators $D=\partial_{\theta}-2i\bar{\theta}\partial_t$ and $\bar{D}=\partial_{\bar{\theta}}$, such that $D^2=\bar{D}^2$ and $\{ D,\bar{D}\}=-2i \partial_t$. We thus consider the following ansatz for $\Gamma_k$, keeping only first order derivatives in time and bi-local contributions for the potential:
\begin{align}
\Gamma_k[\mathcal{M}]= \Gamma_k^{(I)}[\mathcal{M}]+\Gamma_k^{(II)}[\mathcal{M}]\,,
\label{truncation1}
\end{align}
with:
\begin{align}
\nonumber \Gamma_k^{(I)}[\mathcal{M}]:= \int dz \Big( &- \frac{1}{2} \bar{D}Y_k(\mathcal{M}(z)) DY_k(\mathcal{M}(z)) +i U_k^{(1)}(\mathcal{M}(z)) \Big)\,,
\end{align}
and
\begin{equation}
\Gamma_k^{(II)}[\mathcal{M}]:=i \int dz dz^\prime\, U_k^{(2)}(\mathcal{M}(z),\mathcal{M}(z^\prime\,))\,.
\end{equation}
To return to the original variables, we expand the functions $Y_k(\mathcal{M}(z))$, and use the standard properties of the Grassmann variables. After a tedious calculation, the kinetic part of the truncation \eqref{truncation1} (keeping only the quadratic terms involving a first derivative with respect to time) writes as:
\begin{equation}
\Gamma_{k,\text{kin}}= \int dt \sum_{i} (Z_k)^{\frac{1}{2}}_{i} (Z_k)^{\frac{1}{2}}_{i}\left( \frac{1}{2} \bar{\varphi}_i\bar{\varphi}_i+i\bar{\varphi}_i \dot{\phi}_i-i \bar{\psi}_i \dot{\psi}_i \right)\,, \label{truncationkin}
\end{equation}
where $(Z_k)_{i}^{\frac{1}{2}}$ must be a function of $\phi$ only. Explicitly:
\begin{equation}
(Z_k)_{i}^{\frac{1}{2}}(\phi):= \frac{dY_k}{d\phi_i}(\phi)\,.
\end{equation}
This equation means that the own wave function renormalization factors $Z_{k,\bar{\varphi}_i}^{\frac{1}{2}}$, $Z_{k,\phi_i}^{\frac{1}{2}}$ and $Z_{k,\bar{\psi}_i \psi_i}^{\frac{1}{2}}$ respectively associated with the fields $\bar{\varphi}_i$, $\phi_i$ and the product $(\bar{\psi}_i\psi_i)^{\frac{1}{2}}$ (without summation) must be equal:
\begin{equation}
\begin{boxed}{
Z_{k,\bar{\varphi}_i}^{\frac{1}{2}}=Z_{k,\phi_i}^{\frac{1}{2}}=Z_{k,\bar{\psi}_i \psi_i}^{\frac{1}{2}}\equiv (Z_k)_i^{\frac{1}{2}}\,.}
\end{boxed} \label{Ward1}
\end{equation}
This nonperturbative constraint comes from time reversal symmetry, and drastically reduces the number of independent parameters in the truncation. The remaining part of the truncation, that we call “interaction" writes as:
\begin{align*}
\nonumber\Gamma_{k,\text{int}}= -2i\int dt \sum_{ijl}& (X_k)_{ijl}\Bigg( i\bar{\varphi}_l\bar{\psi}_i \psi_j +\dot{\phi}_l \bar{\psi}_i \psi_j+\dot{\phi}_j \bar{\psi}_i \psi_l \Bigg) \\
&\qquad+i\int dz U_k^{(1)}(\mathcal{M}(z))+i \int dz dz^\prime\, U_k^{(2)}(\mathcal{M}(z),\mathcal{M}(z^\prime\,))\,,
\end{align*}
where the symbols $(X_k)_{ijl}$ depend only on $\phi$ and are defined as:
\begin{equation}
\begin{boxed}{
(X_k)_{ijl}(\phi):= (Z_k)^{\frac{1}{2}}_{l}(\phi) \frac{d}{d\phi_i}(Z_k)^{\frac{1}{2}}_{j}(\phi)\,.}
\end{boxed}
\end{equation}
This is the second constraint coming from time reversal symmetry. In the rest of this paper we focus on the simpler approximation which consist in neglecting the dependence of the wave function renormalization on the classical field $\phi$. In such a way $(X_k)_{ijl}(\phi)=0$. Finally, rescaling the fields as $\mathcal{M}_i \to (Z_{k})^{\frac{1}{2}}_{i}\mathcal{M}_i/Z^\frac{1}{2}_k$, with $Z_k:= \sum_i (Z_k)^{\frac{1}{2}}_{i} (Z_k)^{\frac{1}{2}}_{i}$; and using the same symbols to designate the rescaled fields, we get the simplified form:
\begin{equation}
\begin{boxed}{\Gamma_{k}[\mathcal{M}]=\int dz \Big\{ \frac{Z_k}{2} \bar{D} \mathcal{M}\cdot D \mathcal{M} +i U_k^{(1)}(\mathcal{M}(z))+i \int dz^\prime\, U_k^{(2)}(\mathcal{M}(z),\mathcal{M}(z^\prime\,))\Big\}\,.}\end{boxed}\label{truncation3}
\end{equation}
We focus on rank three tensors ($p=3$) and quartic deterministic forces ($M=2$, see \eqref{eq1} and below). The bare potential $\overline{W_0}$ writes as:
\begin{align}
\overline{W_0}[\mathcal{M}]:=&-i\lambda\frac{(2D)}{9N^2} \int dz^\prime \big(\Phi(z)\cdot \Phi(z^\prime)\big)^3+\frac{(2D)^{\frac{1}{2}}}{3N^2} J_0 \big(\Phi_{\bullet}(z)\big)^3+V[\Phi]\,.\label{classicalpotential}
\end{align}
where the deterministic part $V$ is:
\begin{equation}
V[\Phi]=\frac{h_1}{2} \Phi^2(z)+\frac{1}{2D}\frac{h_2}{4N} (\Phi^2)^2(z)\,.
\end{equation}
In the rest of this paper, we will use the notation $h\equiv Z_k h_1$. Note that due to the integration over the disorder, the higher powers in fields are always even, and thus have non-vanishing Witten index \cite{Synatschke_2009,Witten1981}, meaning that supersymmetry is expected to be unbroken dynamically. Moreover, we focus on the large-$N$ limit, discarding irrelevant contributions in $1/N$. Due to the non-local nature of the interaction, we have to choose a suitable parametrization for the running effective potential $U_k[\mathcal{M}]$. As a first approximation we consider sixtic truncation; imposing:
\begin{equation}
\prod_{i=1}^n \frac{\partial}{\partial \Phi[z_i]} U_k[\Phi]=0 \quad \text{for}\,\, n>6\,.
\end{equation}
As a first step we choose a parametrization build as a sum of monomial interactions reproducing essentially the structure of the first effective loop effects deduced from the potential \eqref{classicalpotential}. Explicitly:
\begin{align}
\nonumber &U_k[\mathcal{M}]=u_0N\int dt+u_1 \int dz \mathcal{M}_{\bullet}(z)+\frac{u_3^{(1)}}{N^2} \int dz \mathcal{M}_{\bullet}^3(z)+\frac{u_3^{(2)}}{N} \int dz \mathcal{M}_{\bullet}(z)\big(\mathcal{M}(z)\cdot \mathcal{M}(z)\big)\\\nonumber
&-\frac{iZ_k\Delta^{(1)}(k)}{2N} \int dz dz^\prime \mathcal{M}_{\bullet}(z)\mathcal{M}_{\bullet}(z^\prime)\delta(t-t^\prime)-\frac{iZ_k\Delta^{(2)}(k)}{2} \int dz dz^\prime \big(\mathcal{M}(z)\cdot \mathcal{M}(z^\prime)\big) \\\nonumber
&-\frac{iZ_k\Delta^{(3)}(k)}{2N} \int dz dz^\prime \mathcal{M}_{\bullet}(z)\mathcal{M}_{\bullet}(z^\prime)+\frac{Z_k h}{2} \int dz\big(\mathcal{M}(z)\cdot \mathcal{M}(z)\big) +\frac{u_4^{(1)}}{N} \int dz \big(\mathcal{M}(z)\cdot \mathcal{M}(z)\big)^2\\\nonumber
&+i\frac{u_4^{(2)}}{N} \int dz dz^\prime \big(\mathcal{M}(z)\cdot \mathcal{M}(z)\big)\big(\mathcal{M}(z)\cdot \mathcal{M}(z^\prime)\big)+ i\frac{u_5^{(1)}}{N^3} \int dz dz^\prime \mathcal{M}_{\bullet}^2(z) \big(\mathcal{M}(z)\cdot \mathcal{M}(z^\prime)\big) \mathcal{M}_{\bullet}(z^\prime)\\\nonumber
&+ \frac{u_5^{(2)}}{N^2} \int dz \mathcal{M}_{\bullet}(z) \big(\mathcal{M}(z)\cdot \mathcal{M}(z)\big)^2+ \frac{u_5^{(3)}}{N^3} \int dz \mathcal{M}_{\bullet}^3(z) \big(\mathcal{M}(z)\cdot \mathcal{M}(z)\big)\\
&+i \frac{u_6}{N^2} \int dzdz^\prime \big(\mathcal{M}(z)\cdot \mathcal{M}(z^\prime)\big)^3\,. \label{truncation4}
\end{align}
The  initial conditions are given by the potential \eqref{classicalpotential}. In particular, for some UV cut-off $\Lambda$ in frequencies (assuming $\Lambda \gg 1$),
\begin{equation}
u_4^{(1)}(\Lambda)=\frac{h_2}{4}\,.
\end{equation}
Physically, the bi-local terms must include $\bar{\varphi}$-$\bar{\varphi}$ products. These terms renormalize the Gaussian measure for the response field, which is nothing but the effective noise correlation. In contrast, the local terms must include only a single field $\bar{\varphi}$. They correct the effective driving force. Note moreover that we chose the power on $N$ in front of each interaction accordingly with the extensive nature of the effective action $\Gamma_k$. Finally, note that we discarded $2$-point couplings of the form:
\begin{equation}
\delta U_k \propto \int dz \mathcal{M}_{\bullet}(z)\mathcal{M}_{\bullet}(z)\,.
\end{equation}
With this approximation the $\bar{\varphi}-\phi$ component of the effective propagator remains diagonal in the internal indices (see \eqref{compbarphiphi} below). It can be justified from the observation that the corresponding coupling must be of order $(u_3^{(1)})^2 u_4^{(1)}$, a sub-leading order correction with respect to the other contributions. Moreover, even though we provide a formalism including odd vertices (which come from the fact that $J_0\neq 0$), a large part of our numerical analysis focus on the centered distribution $J_0=0$, discarding in that way all the odd contributions as $u_3^{(1)}$. This will be especially the case for our investigations about the spherical model, see Section \ref{sec4}. The same observation holds for terms such that:
\begin{equation}
\delta U_k\propto \int dz dz^\prime \mathcal{M}_{\bullet}(z^\prime)(\mathcal{M}\cdot \mathcal{M})(z)\,, \quad \delta U_k\propto \int dz \mathcal{M}_{\bullet}^2(z)(\mathcal{M}\cdot \mathcal{M})(z)\,,
\end{equation}
and the truncation \eqref{truncation4} can be justified semi-perturbatively in $J_0$\footnote{One expect that this approximation is especially justified in the large $N$ limit, as the following explicit calculations show.}. It reduces the investigations into the interior of a (despite everything vast) subregion of dimension $15$ of the full theory space spanned by the different couplings involved in it (including $Z_k$). In the rest of this section, we will derive and solve numerically the resulting flow equations.

\subsection{Solving RG flow equation: $0$ and $1$-point functions}

In this section, we illustrate the computations of flow equations for the couplings $u_1$ and $u_0$, before introducing a more systematic procedure in the next subsection.

\subsubsection{Computation of the effective propagator}\label{CEP}

As a first step we derive the explicit expression of the effective propagator at scale $k$, ${G}_k(\omega)$. Denoting as ${\Phi}=(\bar{\varphi},\phi)$ the field with $2N$ components bringing together the entire bosonic sector, the corresponding components of the effective propagator $\textbf{G}_k$ must read in Fourier space:
\begin{equation}
\textbf{G}_k(\omega)=\begin{pmatrix}
0 & G_{k, \bar{\varphi} \phi }(\omega)\\
G_{k, \bar{\varphi}\phi }(-\omega) & G_{k, \phi \phi}(\omega) \label{formG}
\end{pmatrix}\,,
\end{equation}
each component $(\textbf{G}_{k,\Phi_\alpha \Phi_\beta}(\omega))$ being a $N\times N$ matrix. Note that it is suitable to work in the frequency space, and integrate over frequencies $\omega$ rather than on time $t$. To compute each component, we start by computing the inverse matrix from the truncation \eqref{truncation4}. We have:
\begin{equation}
\textbf{G}_k^{-1}(\omega)=\begin{pmatrix}
A_{k,\bar{\varphi}\bar{\varphi}}(\omega) & B_{k, \bar{\varphi} \phi }(\omega)\\
B_{k, \bar{\varphi}\phi }(-\omega) & 0
\end{pmatrix}\,,
\end{equation}
where $A$ and $B$ are $N\times N$ matrices given explicitly by:
\begin{equation}
A_{k,\bar{\varphi}_i\bar{\varphi}_j}(\omega)=Z_k\big[ ( 1 +\rho_k^{(2)}+2\pi \Delta^{(2)}_k \delta (\omega)) \delta_{ij}+ \frac{1}{N}(\Delta^{(1)}_k+2\pi \Delta^{(3)}_k \delta (\omega))\big]\,,\label{A}
\end{equation}
and:
\begin{equation}
B_{k,\bar{\varphi}_i{\phi}_j}(\omega)= Z_k(\omega+ih+i \rho^{(1)}_k(\omega)) \delta_{ij}\,,\label{B}
\end{equation}
where $R^{(q)}_k=:Z_k \rho^{(q)}_k$, $q=1,2$. The matrix can be inverted as a $2\times 2$ block matrix;
\begin{equation*}
\textbf{G}_k(\omega)=\begin{pmatrix}
0 & B^{-1}_{k,\bar{\varphi}{\phi}}(-\omega)\\
B^{-1}_{k,\bar{\varphi}{\phi}}(\omega) & -B^{-1}_{k,\bar{\varphi}{\phi}}(\omega)A_{k,\bar{\varphi}\bar{\varphi}}(\omega)B^{-1}_{k,\bar{\varphi}{\phi}}(-\omega)
\end{pmatrix}\,.
\end{equation*}
Explicitly, the component $ G_{k,\bar{\varphi} \phi}(\omega)$ and $ G_{k,\bar{\varphi} \bar{\varphi}}(\omega)$ reads as:
\begin{equation}
\Big(G_{k,\bar{\varphi} \phi}(\omega)\Big)_{ij}=-\frac{i}{Z_k} \frac{\delta_{ij}}{i \omega+h+(\rho_k^{(1)})^*(\omega)} \,,\label{compbarphiphi}
\end{equation}
and:
\begin{align}
\Big(G_{k,{\phi} ,\phi}(\omega)\Big)_{ij}=\frac{( 1 +\rho_k^{(2)}+2\pi\Delta^{(2)}_k \delta (\omega)) \delta_{ij}+\frac{1}{N}(\Delta^{(1)}_k+2\pi\Delta^{(3)}_k \delta (\omega))}{Z_k\vert -i\omega+h +\rho^{(1)}(\omega)\vert^2}\,, \label{compphihpi}
\end{align}
where we assumed $(\rho_k^{(1)})^*(\omega)=\rho_k^{(1)}(-\omega)$ (see equation \eqref{r1}), the star $*$ denoting standard complex conjugation.
\medskip

\begin{remark}
\textbf{Two-points function for the response field.} The fact that $G_{\bar{\varphi}\bar{\varphi}}=0$ in \eqref{formG} (or equivalently that $\Gamma_{\phi\phi}=0 $) is not a simplification but an exact relation \cite{DeDominicis1978,Aron_2010}. It can be traced back to the observation that adding a linear driving force: $W_0\to W_0^{\prime}= W_0+ \int dt \sum_i k_i(t) \phi_i(t)$ is equivalent to translating by $-ik_i(t)$ the $i$-th component of the source $\tilde{j}$ for the response field $\bar{\varphi}$ into the partition function \eqref{partitionzero} before averaging over disorder:
\begin{equation}
\tilde{Z}_{W_0^{\prime}}[j,\tilde{j}]=\tilde{Z}_{W_0}[j,\tilde{j}-ik]\,.
\end{equation}
Therefore, because $\tilde{Z}_{W_0}[j=0,\tilde{j}=0]=1$ we must have $\tilde{Z}_{W_0}[0,-ik]=1$ $\forall\, k$. Thus, for vanishing source $j$:
\begin{equation}
G_{\bar{\varphi}\bar{\varphi}}(t,t^\prime)=-\int dJ p(J) \frac{\delta^2 \tilde{Z}_{W_0}[0,-ik] }{\delta k(t)\delta k(t^\prime) }= -\int dJ p(J)\frac{\delta^2 1}{\delta k(t)\delta k(t^\prime) }=0\,.
\end{equation}
\end{remark}

\subsubsection{Flow equations for $u_1$ and $u_0$}

We denote as $\tilde{\textbf{R}}_k$ the super-matrix in the boson-fermion space:
\begin{equation}
\tilde{\textbf{R}}_k=\begin{pmatrix}
\textbf{R}_k & 0\\
0 &R_k^{(f)}
\end{pmatrix}\,,
\end{equation}
the regulator for fermions $R_k^{(f)}$ being defined as:
\begin{equation}
R_k^{(f)}(t-t^\prime)=i R_k^{(1)}(t^\prime-t)\,.\label{rf}
\end{equation}
Note that from this point we reserve the dot for derivative with respect to $\ln(k)$, $\dot{X}=k dX/dk$ and we denote with a ‘‘$\prime$\,'' the time derivatives, $X^\prime:= dX/dt$. The flow equation for $\big(\Gamma_k^{(1)}\big)_{\alpha}$ can be deduced from the Wetterich equation \eqref{Wetterich} taking the first derivative with respect to $\mathcal{M}_{\alpha}$. We get (note that from this point, $\dot{X}=k dX/dk$ and $X^\prime:= dX/dt$):
\begin{equation}
\big(\dot{\Gamma}_k^{(1)}\big)_{\alpha} = -\frac{1}{2}\STr \, \dot{\tilde{\textbf{R}}}_k \frac{1}{\Gamma_k^{(2)}+\tilde{\textbf{R}}_k} \Gamma_{k,\cdots \alpha}^{(3)} \frac{1}{{\Gamma_k^{(2)}+\tilde{\textbf{R}}_k}}\,, \label{floweq1}
\end{equation}
the notation $\STr$ for ‘‘supertrace" meaning that we trace over all indices, with a global minus sign in front of fermionic contributions, and $\Gamma_{k,\cdots \alpha}^{(3)}$ denotes the $3$-point function meaning that the third index is fixed to be $\alpha$. To make the projection on both sides, we notice that:
\begin{equation}
\int dz \mathcal{M}_\bullet(z)= \int dt {\bar{\varphi}}_\bullet(t)\,,
\end{equation}
and\footnote{We recall the definition of the standard functional derivative in the superspace: $\frac{\partial \Phi_i(z)}{\partial \Phi_j(z^\prime)}=\delta_{ij} \delta(z-z^\prime)$, with $\delta(z-z^\prime):=\delta(t-t^\prime)\delta(\theta-\theta^\prime)\delta(\bar{\theta}-\bar{\theta}^\prime)$.}:
\begin{equation}
\int dz \mathcal{M}_\bullet^3(z)=3\int dt \bar{\varphi}_{\bullet}(t)\phi_\bullet^2(t)+6\int dt \phi_\bullet \bar{\psi}_{\bullet}(t) \psi_{\bullet}(t)\,. \label{eq3pts}
\end{equation}
In such a way, $\Gamma^{(1)}_k=iu_1 +\mathcal{O}(\mathcal{M})$. Therefore, to deduce the flow equation for $u_1$, we have to expand the right-hand side of the equation \eqref{floweq1} at the zeroth order in $\mathcal{M}$, and identify the terms of order zero on both sides. Note that $\alpha$ must belong to the auxiliary field $\bar{\varphi}$; so that from \eqref{eq3pts} the two remaining free indices of $\Gamma_{k,\cdots \alpha}^{(3)}$ have to  belong to the bosonic field $\phi$. Thus, because the supermatrix $\tilde{\textbf{R}}_k$ is diagonal in the boson-fermion superspace, and that we have to keep only the zeroth order term on fields in the expansion of $1/(\Gamma_k^{(2)}+\tilde{\textbf{R}}_k)$, the supertrace reduces to a trace over the bosonic sector. Straightforwardly, we get:

\begin{align}
\Gamma_{k,\phi_i(t^\prime)\phi_j(t^{\prime\prime})\bar{\varphi}_k(t^{\prime\prime\prime})}^{(3)}\bigg\vert_{\mathcal{M}=0}=6i \delta(t^\prime-t^{\prime\prime})\delta(t^\prime-t^{\prime\prime\prime}) \bigg[ \frac{u_3^{(1)}}{N^2}+\frac{u_3^{(2)}}{3N}\big( \delta_{ij}+\delta_{ik}+\delta_{jk}\big) \bigg]\,,
\end{align}

and denoting as $G_{k,\alpha\beta}(t)$ the (symmetric) inverse of the matrix $(\Gamma_k^{(2)}+\tilde{\textbf{R}}_k)_{\alpha\beta}$ evaluated for $\mathcal{M}=0$, we thus obtain:
\begin{align}
\nonumber \left( \int_{-\infty}^{+\infty} dt \right) \dot{u}_1& = \frac{1}{2} \tr \int dt dt_1 dt_2 \mathcal{V}^{(3)}G_{k,\bar{\varphi} \phi}(t-t_1)\\
&\times \Big( \dot{R}_k^{(2)}(t_2-t_1) G_{k, \phi \bar{\varphi}} (t_2-t)+ 2i\dot{R}_k^{(1)}(t_2-t_1) G_{k, \phi {\phi}} (t_2-t) \Big)\,, \label{floweq10}
\end{align}
the remaining trace $\tr$ running over the internal Latin indices, and $\mathcal{V}^{(3)}$ denotes the ‘‘vertex matrices" with elements:
\begin{equation}
\big(\mathcal{V}^{(3)}\big)_{ij}:=6\bigg[ \frac{u_3^{(1)}}{N^2}+\frac{u_3^{(2)}}{3N^2}\big(2+N \delta_{ij}\big) \bigg]\,.
\end{equation}
To go beyond, we have to specify the choice of the regulator. Note that the undefined integral over all times $\int dt$ in \eqref{floweq10} must be formally identified with $\delta(\omega=0)$; $\delta(\omega)$ being the standard Dirac function. Such an undefined factor appears generally on both sides of the flow equations, and are formally canceled as an artifact of the momentum conservation. Thus, accordingly with the definitions \eqref{r1} and \eqref{r2}, the first integral on the right-hand side of equation \eqref{floweq10} reads as follows:
\begin{align}
\nonumber \tr \int dt dt_1 dt_2 \mathcal{V}^{(3)}G_{k,\bar{\varphi} \phi}(t-t_1) \dot{R}_k^{(2)}(t_2-t_1) G_{k, \phi \bar{\varphi}} (t_2-t) = \frac{6\pi\delta(0)}{NZ_k} \bigg[ u_3^{(1)}+\frac{u_3^{(2)}}{3} (2+N)\bigg]I_2\,,\label{floweq11}
\end{align}
where:
\begin{equation}
I_2:=-\int \frac{d\omega}{2\pi} \frac{\eta_k \rho_k^{(2)}(\omega)+\dot{\rho}_k^{(2)}(\omega) }{\big\vert -i\omega+h+ \rho_k^{(1)}(\omega) \big\vert^2}\,,
\end{equation}
and following its standard definition we introduced the running anomalous dimension $\eta_k$ as:
\begin{equation}
\eta_k:= k \frac{d}{dk} \ln \big( Z_k \big)\,.
\end{equation}
The derivative of $\rho_k^{(1)}$ and $\rho_k^{(2)}$ with respect to the flow parameter $\ln(k)$ can be easily computed and leads to:
\begin{equation}
\dot{\rho}_k^{(1)}(\omega)= \frac{\alpha k^2}{k-i\omega} \left( 2- \frac{k}{k-i\omega} \right)\,,
\end{equation}
and:
\begin{equation}
\dot{\rho}_k^{(2)}(\omega)=-2\alpha \frac{k^2\omega^2}{(k^2+\omega^2)^2}
\end{equation}
From dimensional analysis, it is suitable to introduce a dimensionless version of $I_2$, say $\bar{I}_2$, such that $I_2= \bar{I}_2/k$, with:
\begin{equation}
\bar{I}_2:=-\int \frac{dx}{2\pi} \frac{\eta_k \rho^{(2)}(x)+\dot{\rho}^{(2)}(x) }{\big\vert -ix+\bar{h}+ \rho^{(1)}(x) \big\vert^2}\,,
\end{equation}
where we introduced the notation $\rho^{(i)}(x):= \rho^{(i)}_{k=1}(x)$. In the same way the explicit expression for $G_{k,\phi\phi}(\omega)$ can be straightforwardly computed, and it reads as:
\begin{equation}
\big(G_{k,\phi\phi}(\omega)\big)_{ij}=g_{k,\phi\phi}^{(1)}(\omega)\delta_{ij}+\frac{1}{N}g_{k,\phi\phi}^{(2)}(\omega)+2\pi( l^{(1)}_{k,\phi\phi} \delta_{ij}+\frac{1}{N}l^{(2)}_{k,\phi\phi})\delta(\omega) \,,
\end{equation}
with:
\begin{equation}
g_{k,\phi\phi}^{(1)}(\omega):= \frac{1}{Z_k} \frac{1-\frac{\alpha k^2}{k^2+\omega^2}}{\big\vert -i\omega+h+ \rho_k^{(1)}(\omega) \big\vert^2} \,, \quad l^{(1)}_{k,\phi\phi}=\frac{1}{Z_k}\frac{\Delta^{(2)}}{\vert h+\alpha k\vert^2} \,,
\end{equation}
and
\begin{equation}
g_{k,\phi\phi}^{(2)}(\omega):= \frac{1}{Z_k} \frac{\Delta^{(1)} }{\big\vert -i\omega+h+ \rho_k^{(1)}(\omega) \big\vert^2} \,, \quad l^{(2)}_{k,\phi\phi}=\frac{1}{Z_k}\frac{ \Delta^{(3)}}{\vert h+\alpha k\vert^2} \,.
\end{equation}
The second integral on the right-hand side of the equation \eqref{floweq10} reads as $4\pi \delta(0) (Z_kkN)^{-1} {\bar{L}_2}$, with:
\begin{align}
\bar{L}_2=: \bigg[ \bar{u}_3^{(1)}+\frac{\bar{u}_3^{(2)}}{3} (2+N)\bigg]\bar{J}_2^{(1)}+\bigg[ \bar{u}_3^{(1)}+\bar{u}_3^{(2)}\bigg]\bar{J}_2^{(2)}\,,\label{L2}
\end{align}
the kernels $\bar{J}_2^{(1)}$ and $\bar{J}_2^{(2)}$ being defined as follows:
\begin{equation}
\bar{J}_2^{(1)}:= Z_k^2 \int \frac{dx}{2\pi}(\eta_k \rho^{(1)}(x)+ \dot{\rho}^{(1)}(x) )  i G_{1,\bar{\varphi} \phi}(-x) (g_{1,\phi\phi}^{(1)}(x)+2\pi\bar{l}^{(1)}_{1,\phi\phi}\delta(x))\,,\label{J21}
\end{equation}
and:
\begin{equation}
\bar{J}_2^{(2)}:= Z_k^2 \int \frac{dx}{2\pi}(\eta_k \rho^{(1)}(x)+ \dot{\rho}^{(1)}(x) )  i G_{1,\bar{\varphi} \phi}(-x) (g_{1,\phi\phi}^{(2)}(x)+2\pi\bar{l}^{(2)}_{1,\phi\phi}\delta(x))\,.\label{J22}
\end{equation}
In this equation we abusively used of the notation $G_{1,\bar{\varphi} \phi}(x)$ to denote the diagonal component of the $2$-point function, namely:
\begin{equation}
G_{1,\bar{\varphi} \phi}(x)=-\frac{i}{Z_k} \frac{1}{i x+\bar{h}+(\rho_1^{(1)})^*(x)}\,,
\end{equation}
where $h=:k \bar{h}$. Moreover, note that $\bar{l}^{(1,2)}_{k,\phi\phi}\delta(x)$ express in terms of the dimensionless couplings $\bar{\Delta}^{(2,3)}=\Delta^{(2,3)}/k$; because from dimensional analysis $[ \bar{\varphi}]= 1/2$, $[\phi]=-1/2$ and $[\psi]=[\bar{\psi}]=0$, the bracket being defined such that $[\omega]=1$. As a result we must have $[h]=1$, $[\Delta^{(1)}]=0$, $[\Delta^{(2,3)}]=1$, $[u_1]=1/2$ and $[u_3]=3/2$. For this reason, we introduced the dimensionless couplings $\bar{u}_1$ and $\bar{u}_3$ in \eqref{L2} as:
\begin{equation}
u_1= Z_k^{\frac{1}{2}} k^{\frac{1}{2}} \bar{u}_1\,,\qquad u_3= Z_k^{\frac{3}{2}}k^{\frac{3}{2}} \bar{u}_3\,.
\end{equation}
In terms of these dimensionless couplings, the flow equation for $\bar{u}_1$ reads as follows:

\begin{equation}
\boxed{
\dot{\bar{u}}_1=-\frac{1}{2} (1+\eta_k)\bar{u}_1-\frac{3 }{N} \bigg[ \bar{u}_3^{(1)}+\frac{\bar{u}_3^{(2)}}{3} (2+N)\bigg] ( \bar{I}_2+2\bar{J}_2^{(1)})-\frac{6}{N}\big[ \bar{u}_3^{(1)}+\bar{u}_3^{(2)}\big]\bar{J}_2^{(2)} \,. }
\end{equation}

In the same way, we get for $u_0$:
\begin{align*}
\left( \int dt \right)N \dot{{u}}_0&= \int dt dt^\prime \tr\Big[ \frac{1}{2} G_{k,\bar{\varphi}\bar{\varphi}}(t-t^\prime) \dot{R}^{(2)}_k(t^\prime-t)\\
&\qquad \qquad+ \left(i\dot{R}_k^{(1)}(t^\prime-t) G_{k, \bar{\varphi} \phi}(t-t^\prime) +\dot{R}_k^{(f)}(t^\prime-t) G_{k,\bar{\psi}\psi}(t-t^\prime) \right)\Big] \,.
\end{align*}
The effective $\bar{\phi}$-loop vanishes because $G_{k,\bar{\varphi}\bar{\varphi}}=0$. However, the stability of supersymmetry ensured by W-T identities requires $u_0=0$, and we expect that the two remaining contributions cancel exactly. We will check briefly this point. First, the bosonic contribution writes explicitly as:
\begin{equation}
L_{\text{B}}:=\int \frac{d\omega}{2\pi}\tr\Big(i\dot{R}_k^{(1)}(\omega) G_{k,\bar{\varphi}\phi}(-\omega) \Big)=\frac{1}{Z_k}\int \frac{d\omega}{2\pi}\tr\Big(\dot{R}_k^{(1)}(\omega) \Big) \frac{1}{-i\omega+h+\rho^{(1)}_k(\omega)}\,. \label{LB}
\end{equation}
In the same way, because of $G_{k,\bar{\psi}\psi}(t-t^\prime)=-G_{k,\psi\bar{\psi}}(t^\prime-t)$ and taking into account that from \eqref{rf} ${R}_k^{(f)}(\omega)=iR^{(1)}_k(-\omega)$, we get for the fermionic loop:
\begin{equation}
L_{\text{F}}:= -\int \frac{d\omega}{2\pi}\tr\Big(i\dot{R}_k^{(f)}(\omega) G_{k,\psi\bar{\psi}}(\omega) \Big)=-\frac{1}{Z_k}\int \frac{d\omega}{2\pi}\tr\Big(\dot{R}_k^{(1)}(-\omega) \Big) \frac{1}{i\omega+h+\rho^{(1)}_k(-\omega)}\,.\label{LF}
\end{equation}
thus, up to the change of variable $\omega\to -\omega$ into the last integral, we explicitly showed that $L_{\text{B}}+L_{\text{F}}=0$, ensuring $\dot{u}_0=0$. \\

Finally, note that we can easily make contact with the choice of the regulator done by some authors who studied stochastic equations from nonperturbative RG, for instance, \cite{Synatschke_2009}. The apparent contradiction seems to come from the fermionic regulator, but the relation between our choice and the choice done in the literature can be stressed from the relation \eqref{relationR} coming from time-reversal symmetry:
\begin{equation}
R^{(f)}(\omega)=R^{(1)}(\omega)-2i\omega R^{(2)}(\omega)\,.
\end{equation}

\subsection{‘‘$\bar{\varphi}\bar{\varphi}$\,'' $2$-points couplings}

Taking the second derivative on both sides of equation \eqref{Wetterich} with respect to the classical superfield, we get:
\begin{equation}
\big(\dot{\Gamma}_k^{(2)}\big)_{\alpha\beta} =\big(\dot{\Gamma}_k^{(2)}\big)_{\alpha\beta}^{(4)}+\big(\dot{\Gamma}_k^{(2)}\big)_{\alpha\beta}^{(3,3)}\,,
\end{equation}
with:
\begin{align}
&\big(\dot{\Gamma}_k^{(2)}\big)_{\alpha\beta}^{(4)}:= -\frac{1}{2}\STr \, \dot{\tilde{\textbf{R}}}_k G_k \Gamma_{k,\cdots \alpha\beta}^{(4)} G_k\\
&\big(\dot{\Gamma}_k^{(2)}\big)_{\alpha\beta}^{(3,3)}:=\STr \, \dot{\tilde{\textbf{R}}}_k G_k \Gamma_{k,\cdots \alpha}^{(3)} G_k \Gamma_{k,\cdots \beta}^{(3)} G_k\,, \label{floweq2pts1}
\end{align}
and
\begin{equation}
G_k:=\frac{1}{\Gamma_k^{(2)}+\tilde{\textbf{R}}_k}\,.
\end{equation}
Because of the constraints coming from relations \eqref{Ward1}, it is suitable to set $\alpha=\beta=\bar{\varphi}$. The relevant component for the $4$-point function is $\Gamma^{(4)}_{k,\phi_i(t)\phi_j(t^\prime)\bar{\varphi}_k(t^{\prime\prime})\bar{\varphi}_l(t^{\prime\prime\prime})}$, which can be explicitly computed from the truncation \eqref{truncation4} as:
\begin{align}
\nonumber &\Gamma^{(4)}_{k,\phi_i(t)\phi_j(t^\prime)\bar{\varphi}_k(t^{\prime\prime})\bar{\varphi}_l(t^{\prime\prime\prime})}=- \frac{2u_4^{(2)}}{N} \Big( \delta_{ik} \delta_{jl}+\delta_{jk}\delta_{il}+\delta_{ij}\delta_{kl}\Big) \times \big( \delta(t^{\prime\prime}-t)+\delta(t^{\prime\prime\prime}-t) \big) \delta(t^{\prime}-t)\,.
\end{align}
From this expression, we deduce that $\big(\dot{\Gamma}_k^{(2)}\big)_{\bar{\varphi}_k \bar{\varphi}_l}^{(4)}$ must have the following structure:
\begin{equation}
\big(\dot{\Gamma}_k^{(2)}\big)_{\bar{\varphi}_k \bar{\varphi}_l}^{(4)}=\vcenter{\hbox{\includegraphics[scale=1]{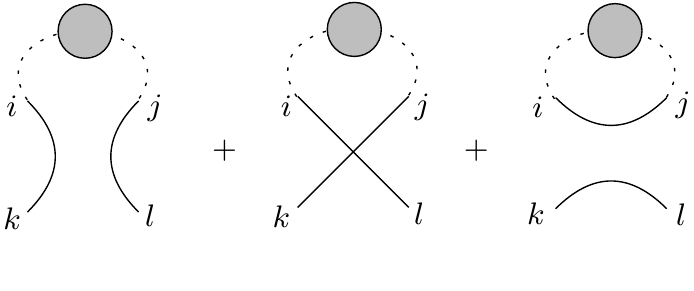} }}\,, \label{graph1}
\end{equation}
where the solid edges materialize the Kronecker deltas, and the dotted edges weighted with a grey bubble materialize the effective propagator $G_k\dot{\tilde{\textbf{R}}}_k G_k$. Because of the $1/N$ arising from $u_4^{(2)}$; it is not hard to convince that only the third term, which create a closed trace in the Latin indices, survives in the large $N$ limit. Computing it, we get:
\begin{align}
\big(\dot{\Gamma}_k^{(2)}\big)_{\bar{\varphi}_k \bar{\varphi}_l}^{(4)}&= \delta(\Omega)\delta(\Omega^\prime) \frac{2(2\pi)^2u_4^{(2)}}{Z_k} \Big\{\Big[\Big(1+\frac{2}{N} \Big) ({I}_2+2{J}_2^{(1)})+\frac{2}{N}{J}_2^{(2)}\Big]\delta_{kl}+\frac{4}{N^2}{J}_2^{(2)} \Big\}
\end{align}
In the same way, we can represent the contribution $\big(\dot{\Gamma}_k^{(2)}\big)_{\bar{\varphi}_k \bar{\varphi}_l}^{(3,3)}$ as:
\begin{align}
\nonumber\big(\dot{\Gamma}_k^{(2)}\big)_{\bar{\varphi}_k \bar{\varphi}_l}^{(3,3)}&=\vcenter{\hbox{\includegraphics[scale=0.8]{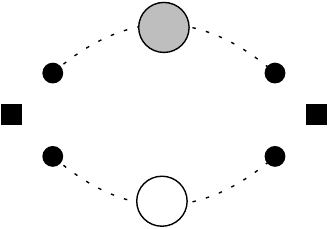} }}+\vcenter{\hbox{\includegraphics[scale=0.8]{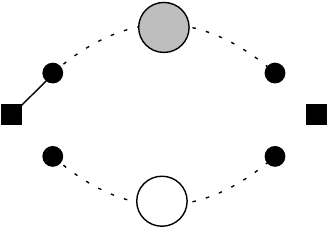} }}+\vcenter{\hbox{\includegraphics[scale=0.8]{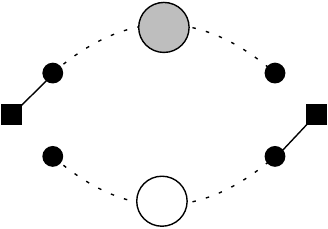} }}\\
&+\vcenter{\hbox{\includegraphics[scale=0.8]{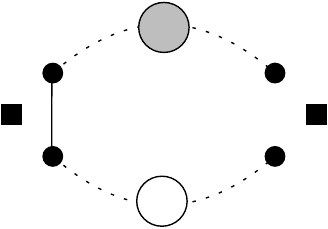} }}+\vcenter{\hbox{\includegraphics[scale=0.8]{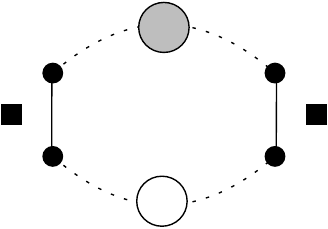} }}+\vcenter{\hbox{\includegraphics[scale=0.8]{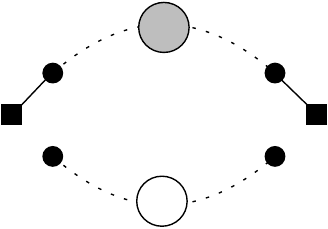} }}\,,\label{graph2}
\end{align}
where the white circle materializes the propagator $G_k$, the black dots and square correspond respectively to the fields $\phi_{\bullet}$ and $\bar{\varphi}_{\bullet}$ and the solid edge materialize the scalar product between fields. In such a way, groups of three isolated dots must correspond to interaction with coupling $u_3^{(1)}$ and groups with an edge to interaction with coupling $u^{(2)}_3$. Because of the way the internal indices are contracted between the remaining free fields $\bar{\varphi}_{\bullet}$, the first five diagrams must contribute to the flow of $\Delta^{(3)}$ whereas the last diagram must contribute to the flow of the field strength $Z_k$. Let us consider the first diagram. Explicitly we get:
\begin{align}
\nonumber\vcenter{\hbox{\includegraphics[scale=0.8]{oneloop2.pdf} }}&=\frac{(6iu_3^{(1)})^2}{N^4} \tr\int \frac{d\omega d\Omega}{(2\pi)^2} G_{k, \bar{\varphi} \phi} (-\omega)\times \Big(\tr G_{k,\phi\phi}(\Omega-\omega)\Big)\\
&\qquad \times \big[ \dot{R}^{(2)}_k(\omega) G_{k,\bar{\varphi}\phi}(\omega)+2i\dot{R}^{(1)}_k(\omega) G_{k,\phi\phi}(-\omega) \big]e^{-i\Omega(t-t^\prime)}\,,
\end{align}\label{gamma33}
\noindent
where the traces ‘‘$\tr$" run over internal indices and $\Omega$ denotes to the external momenta. However, because we truncate around local couplings, we have to keep only the first term in the ‘‘local" expansion in power of $\Omega$:
\begin{equation}
\int \frac{d\Omega}{2\pi} \, f(\Omega) e^{-i\Omega(t-t^\prime)} = f(0) \delta(t-t^\prime)+\text{(derivative interactions)}\,,
\end{equation}
where \textit{derivative interactions} includes derivatives of the Dirac delta. Our truncation scheme \eqref{truncation4} focuses on products of local terms; however, we have to be careful with the component $G_{k,\phi\phi}$ of the effective propagator \eqref{compphihpi}. Indeed, omitting the internal index structure, this component reads schematically as:
\begin{equation}
G_{k,\phi\phi}(\omega,\omega^\prime)= A\delta(\omega+\omega^\prime)+B\delta(\omega)\delta(\omega^\prime). \label{decomplocal}
\end{equation}
Two local contributions linked by the component denoted as $A$ provide an effective local contribution; when the same local contributions linked together by the component $B$ provides us with a product of local contributions again.
\medskip

Focusing on the local approximation, from the explicit expressions for the components of the propagator, we get for instance in Fourier variables:
\begin{align}
\nonumber \vcenter{\hbox{\includegraphics[scale=0.8]{oneloop2.pdf} }}= \frac{2\pi(6i u_3^{(1)})^2}{Z_k^2 N^2 k^3}&\Big\{\delta(\Omega+\Omega^\prime) \big[\bar{I}_3^{(1)}+\bar{I}_3^{(2)}+2(\bar{J}_3^{(11)}+\bar{J}_3^{(12)}+\bar{J}_3^{(22)})\big] \\
&+4\pi\delta(\Omega)\delta(\Omega^\prime) (\bar{K}_3^{(11)}+\bar{K}_3^{(12)}+\bar{K}_3^{(22)}) \Big\}\,.
\end{align}
the numerical coefficients $I_3^{(i)}$ and $J^{(ij)}_3$ (for $(i,j)\in \{1,2\}$) arising from loop integrals being defined as:
\begin{equation}
\bar{I}_3^{(i)}:=\eta_k a_3^{(i)}+ b_3^{(i)} \,, \quad \bar{J}_3^{(ij)}:=\eta_k (a_3^{(ij)}+a_3^{\prime(ij)})+ b_3^{(ij)}+ b_3^{\prime(ij)}\, \quad \bar{K}_3^{(ij)}=\eta_k a_3^{\prime\prime(ij)}+b_3^{\prime\prime(ij)} \,,
\end{equation}
where:
\begin{align}
a_3^{(i)}&:=(Z_k)^3\int \frac{dx}{2\pi} \rho^{(2)}(x) G_{1,\bar{\varphi} \phi}(-x) G_{1,\bar{\varphi} \phi}(x) (g_{1,\phi\phi}^{(i)}(x)+2\pi\bar{l}^{(i)}_{1,\phi\phi}\delta(x))\,,\\
b_3^{(i)}&:=(Z_k)^3\int \frac{dx}{2\pi} \dot{\rho}^{(2)}(x) G_{1,\bar{\varphi} \phi}(-x) G_{1,\bar{\varphi} \phi}(x) (g_{1,\phi\phi}^{(i)}(x)+2\pi\bar{l}^{(i)}_{1,\phi\phi}\delta(x))\,,
\end{align}
and:
\begin{align}
a_3^{(ij)}&:=i(\delta_{ij}+2\delta_{i1}\delta_{j2})(Z_k)^3\int \frac{dx}{2\pi} \rho^{(1)}(x) G_{1,\bar{\varphi} \phi}(-x) g_{1,\phi\phi}^{(i)}(x)\times g_{1,\phi\phi}^{(j)}(x)\,,\\
b_3^{(ij)}&:=i(\delta_{ij}+2\delta_{i1}\delta_{j2})(Z_k)^3\int \frac{dx}{2\pi}\dot{\rho}^{(1)}(x) G_{1,\bar{\varphi} \phi}(-x) g_{1,\phi\phi}^{(i)}(x)\times g_{1,\phi\phi}^{(j)}(x)\,,
\end{align}
\begin{align}
a_3^{\prime(ij)}&:=i(Z_k)^3 (\delta_{ij}+2\delta_{i1}\delta_{j2}) \rho^{(1)}(0) G_{1,\bar{\varphi} \phi}(0) (\bar{l}^{(i)}_{1,\phi\phi} g_{1,\phi\phi}^{(j)}(0))+g_{1,\phi\phi}^{(i)}(0)\bar{l}^{(j)}_{1,\phi\phi})\,,\\
b_3^{\prime(ij)}&:=i(Z_k)^3 (\delta_{ij}+2\delta_{i1}\delta_{j2}) \dot{\rho}^{(1)}(0) G_{1,\bar{\varphi} \phi}(0) (\bar{l}^{(i)}_{1,\phi\phi} g_{1,\phi\phi}^{(j)}(0))+g_{1,\phi\phi}^{(i)}(0)\bar{l}^{(j)}_{1,\phi\phi})\,,
\end{align}
\begin{align}
a_3^{\prime\prime(ij)}&:=i(Z_k)^3(\delta_{ij}+2\delta_{i1}\delta_{j2}) \rho^{(1)}(0) G_{1,\bar{\varphi} \phi}(0)\, \bar{l}^{(i)}_{1,\phi\phi} \bar{l}^{(j)}_{1,\phi\phi}\,,\\
b_3^{\prime\prime(ij)}&:=i(Z_k)^3(\delta_{ij}+2\delta_{i1}\delta_{j2}) \dot{\rho}^{(1)}(0) G_{1,\bar{\varphi} \phi}(0) \, \bar{l}^{(i)}_{1,\phi\phi} \bar{l}^{(j)}_{1,\phi\phi}\,.
\end{align}

Following the same strategy for the remaining diagrams, we obtain:
\begin{align}
\nonumber \vcenter{\hbox{\includegraphics[scale=0.6]{oneloop2prime.pdf} }}&+\vcenter{\hbox{\includegraphics[scale=0.6]{oneloop2prime3.pdf} }}=- \frac{72(2\pi) u_3^{(1)}u_3^{(2)} }{Z_k^2 N^2 k^3} \Big\{ \big[\delta(\Omega+\Omega^\prime)\Big[\bar{I}_3^{(1)}+\bar{I}_3^{(2)}+2(\bar{J}_3^{(11)}\\
&+\bar{J}_3^{(12)}+\bar{J}_3^{(22)})\big]+4\pi \delta(\Omega)\delta(\Omega^\prime) (\bar{K}_3^{(11)}
+\bar{K}_3^{(12)}+\bar{K}_3^{(22)})\Big\}\,, \label{contribution11}
\end{align}
\begin{align}
\nonumber\vcenter{\hbox{\includegraphics[scale=0.6]{oneloop2prime2.pdf} }}&+\vcenter{\hbox{\includegraphics[scale=0.6]{oneloop2prime4.pdf} }}= \frac{2\pi(2iu_3^{(2)})^2 }{Z_k^2 N k^3} \bigg\{ \delta(\Omega+\Omega^\prime) \bigg[\bigg(\frac{1}{N}+2\bigg)(\bar{I}_3^{(1)}+2\bar{J}_3^{(11)})+\frac{3}{N}\bar{I}_3^{(2)}\\
&+\frac{6}{N}(\bar{J}_3^{(12)}+\bar{J}_3^{(22)})\bigg]+4\pi \delta(\Omega)\delta(\Omega^\prime) \bigg[\bigg(\frac{1}{N}+2\bigg)\bar{K}_3^{(11)}+\frac{3}{N}(\bar{K}_3^{(12)}+\bar{K}_3^{(22)}) \bigg] \bigg\} \,,
\end{align}
and finally:
\begin{align}
\nonumber\vcenter{\hbox{\includegraphics[scale=0.8]{oneloop2prime5.pdf} }}&=\frac{4\pi(2iu_3^{(2)})^2 }{Z_k^2 N k^3}\bigg\{\delta(\Omega+\Omega^\prime)\big[\delta_{kl}(\bar{I}_3^{(1)}+\bar{I}_3^{(2)}+2\bar{J}_3^{(11)}+\bar{J}_3^{(12)})\\\nonumber
&+\frac{1}{N}(\bar{J}_3^{(12)}+2\bar{J}_3^{(22)})\big]+2\pi \delta(\Omega)\delta(\Omega^\prime)\big[\delta_{kl}(2\bar{K}_3^{(11)}+\bar{K}_3^{(12)})\\
&+\frac{1}{N}(\bar{K}_3^{(12)}+2\bar{K}_3^{(22)}) \big]\bigg\}\,.
\end{align}
Note that we have to be careful with the positions of gray and whites bubbles, which are for instance responsible for a factor $2$ in the first contribution to the left-hand side of equation \eqref{contribution11} concerning the second contribution. \\

On the other hand, computing the left-hand side of \eqref{floweq10} $\big(\dot{\Gamma}_k^{(2)}\big)_{\bar{\varphi}\bar{\varphi}}$, we get for vanishing fields:
\begin{align}
\nonumber\big(\dot{\Gamma}_k^{(2)}\big)_{\bar{\varphi}_k\bar{\varphi}_l}= \Big[\dot{Z}_k \delta_{kl}&+\frac{Z_k}{N} (\dot{\Delta}^{(1)}+\eta_k\Delta^{(1)})\Big] \delta(t-t^\prime)\\
&+Z_k(\dot{\Delta}^{(2)}+\eta_k\Delta^{(2)})\delta_{kl}+\frac{Z_k}{N}(\dot{\Delta}^{(3)}+\eta_k\Delta^{(3)})\,.
\end{align}
Thus, the projection into the subspace parametrized with the supersymmetric truncation \eqref{truncation4} requires:
\begin{align}
\nonumber \dot{\Delta}^{(1)}=-\eta_k \Delta^{(1)}&-\frac{36(\bar{u}_3^{(1)})^2+72\bar{u}_3^{(1)}\bar{u}_3^{(2)}+4(\bar{u}_3^{(2)})^2}{N} \big[\bar{I}_3^{(1)}+\bar{I}_3^{(2)}+2(\bar{J}_3^{(11)}+\bar{J}_3^{(12)}\\
&+\bar{J}_3^{(22)})\big]-\frac{8(\bar{u}_3^{(2)})^2}{N^2}\big[N\bar{I}_3^{(1)}+\bar{I}_3^{(2)}+2N\bar{J}_3^{(11)}+3(\bar{J}_3^{(12)} +\bar{J}_3^{(22)}) \big]\,,
\end{align}
\begin{align}
 \dot{\Delta}^{(2)}=&-(1+\eta_k) \Delta^{(2)}-2\bar{u}_4^{(2)}\Big[\Big(1+\frac{2}{N} \Big) ({I}_2+2{J}_2^{(1)})+\frac{2}{N}{J}_2^{(2)}\Big]-\frac{8(\bar{u}_3^{(2)})^2}{N} (2\bar{K}_3^{(11)}+\bar{K}_3^{(12)}) \,,
\end{align}
\begin{align}
\nonumber \dot{\Delta}^{(3)}=&-(1+\eta_k) \Delta^{(3)}-\frac{8\bar{u}^{(2)}_4}{N}\bar{J}^{(2)}-72\frac{(\bar{u}_3^{(1)})^2+2\bar{u}_3^{(1)}\bar{u}_3^{(2)}}{N}(\bar{K}_3^{(11)}+\bar{K}_3^{(12)}+\bar{K}_3^{(22)})\\
&-\frac{8(\bar{u}_3^{(2)})^2}{N}\bigg[\bigg(1+2N\bigg)\bar{K}_3^{(11)}+(4\bar{K}_3^{(12)}+5\bar{K}_3^{(22)}) \bigg] \,,
\end{align}
and imposes finally the following closed relation for the anomalous dimension:
\begin{equation}
\eta_k\equiv -\frac{8(\bar{u}_3^{(2)})^2 }{N} [\bar{I}_3^{(1)}+2\bar{J}_3^{(11)}+(\bar{I}_3^{(2)}+\bar{J}_3^{(12)})] \,,
\end{equation}
from which we can easily extract $\eta_k$:

\begin{equation}
\begin{boxed}{
\eta_k\equiv -\frac{8(\bar{u}_3^{(2)})^2 }{N} \frac{b^{(1)}+2(b^{(11)}+b^{\prime (11)})+b^{(2)}+(b^{(12)}+b^{\prime(12)})}{1+\frac{8(\bar{u}_3^{(2)})^2 }{N}(a^{(1)}+2(a^{(11)}+a^{\prime(11)})+a^{(2)}+(a^{(12)}+a^{\prime(12)}))} \,.}
\end{boxed}
\end{equation}
The flow equations for higher order interactions can be obtained in the same way. The detail being unnecessarily technical, we have placed it in the Appendix \eqref{AppA}. We introduce for this purpose a diagrammatic method, which could easily be automated by a numerical routine for further investigations (i.e. for higher truncation).

\section{First numerical investigations for $N\to \infty$}\label{sec4}

\begin{figure}
\begin{center}
\includegraphics[scale=0.4]{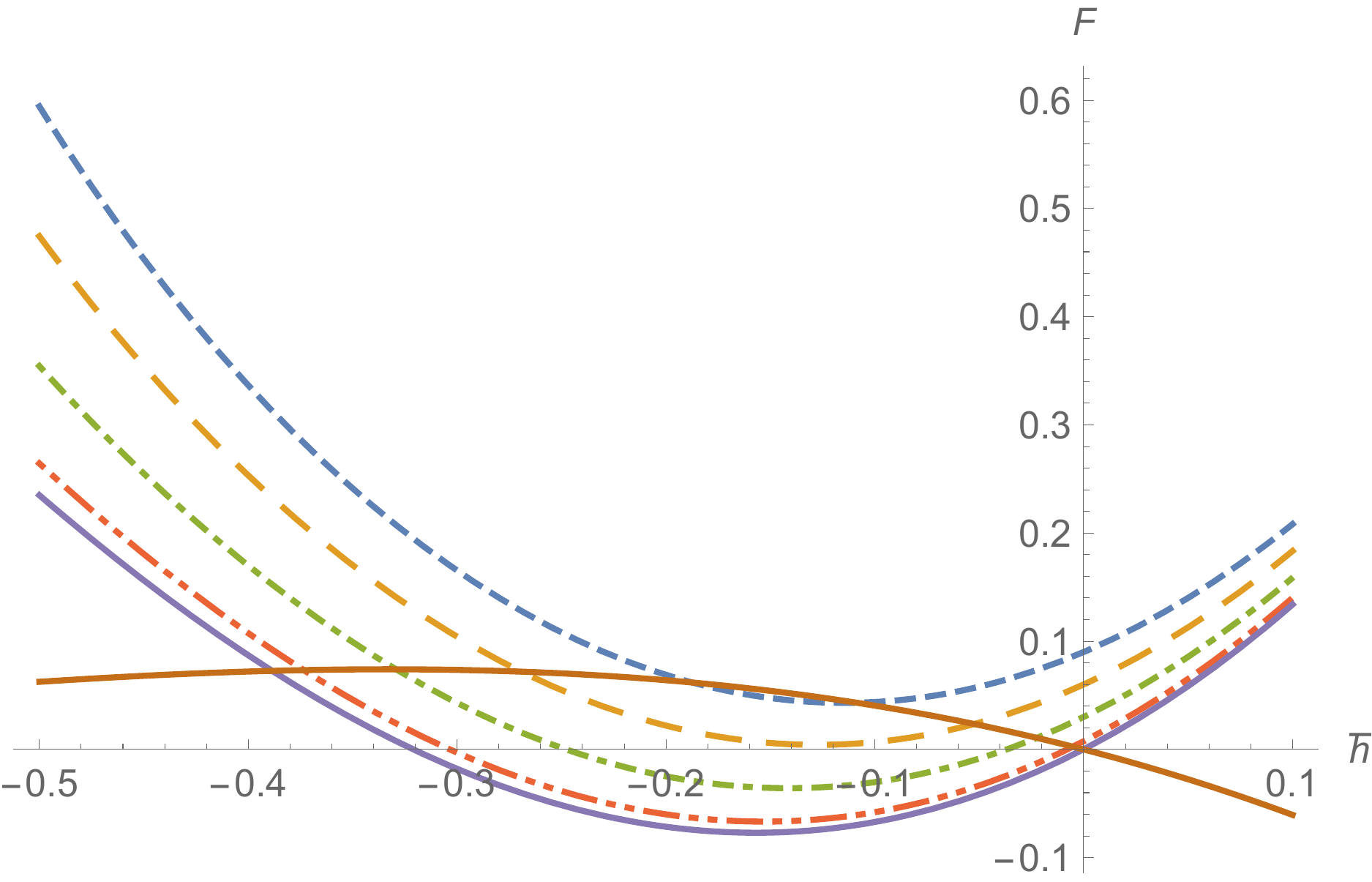}\quad \includegraphics[scale=0.4]{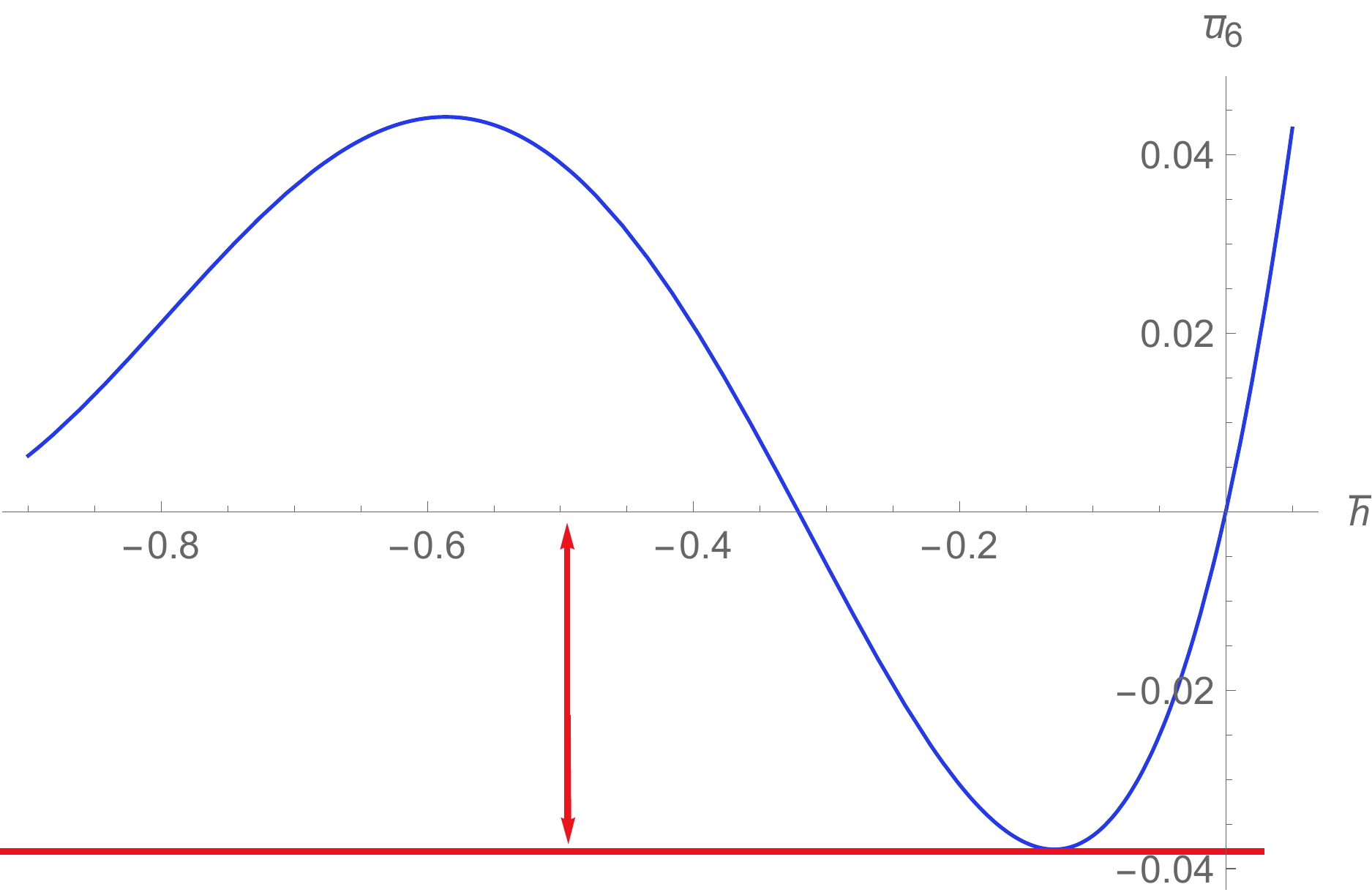}
\end{center}
\caption{On the left: $F(h,\bar{u}_6)$ for $\bar{u}_6=0$ (purple curve), $\bar{u}_6=-0.005$ (red curve), $\bar{u}_6=-0.02$ (green curve), $\bar{u}_6=-0.04$ (yellow curve) and $\bar{u}_6=-0.06$ (blue curve). The brown curve, corresponding to $\bar{u}_4^{(1)}(\bar{h}) $, the solution at the fixed point. On the right, $\bar{u}_6\equiv \bar{u}_6(\bar{h})$, given by the condition $F(h,\bar{u}_6)=0$.}\label{figTensor1}
\end{figure}

\begin{figure}
\begin{center}
\includegraphics[scale=0.4]{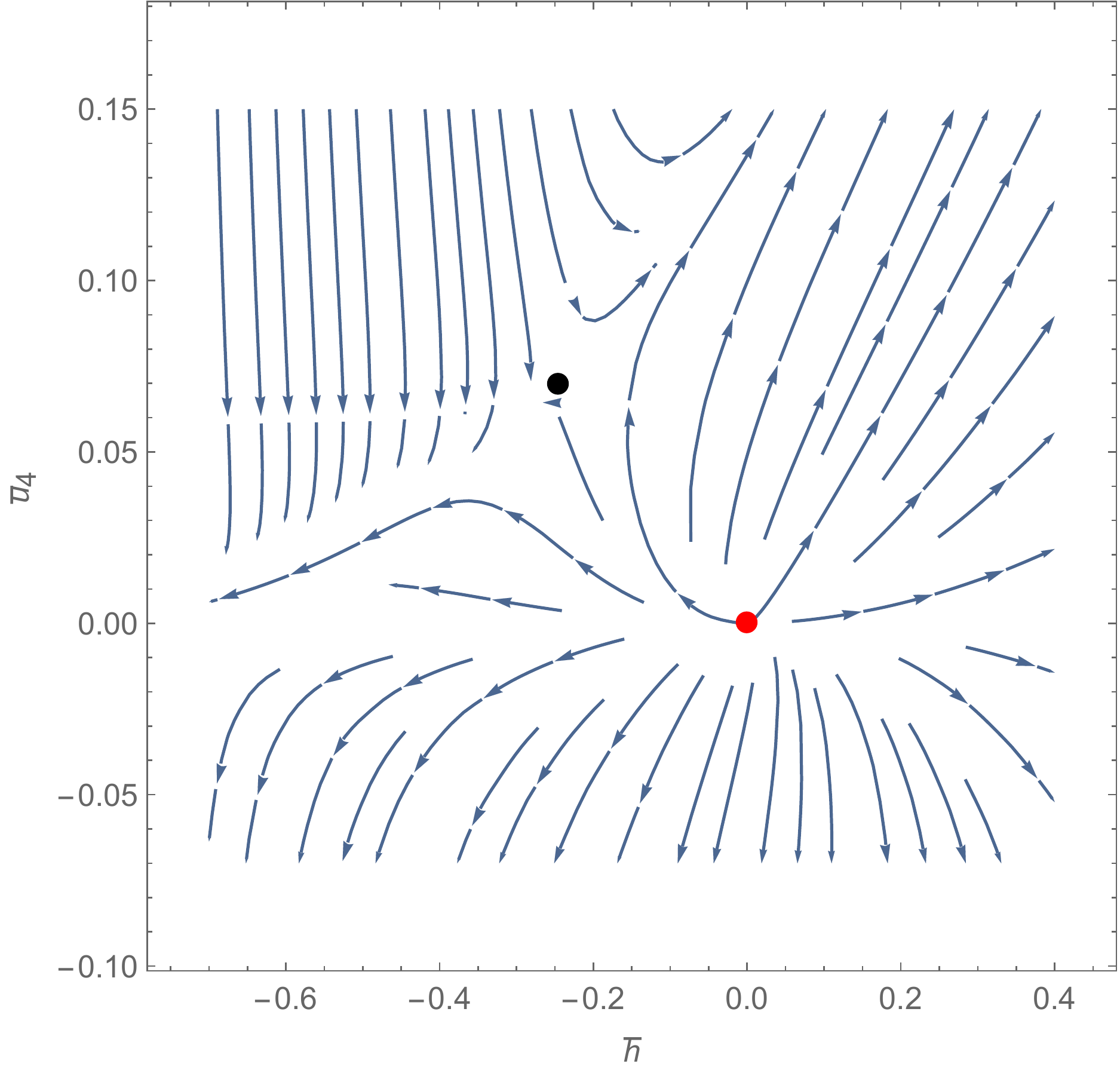} \quad \includegraphics[scale=0.4]{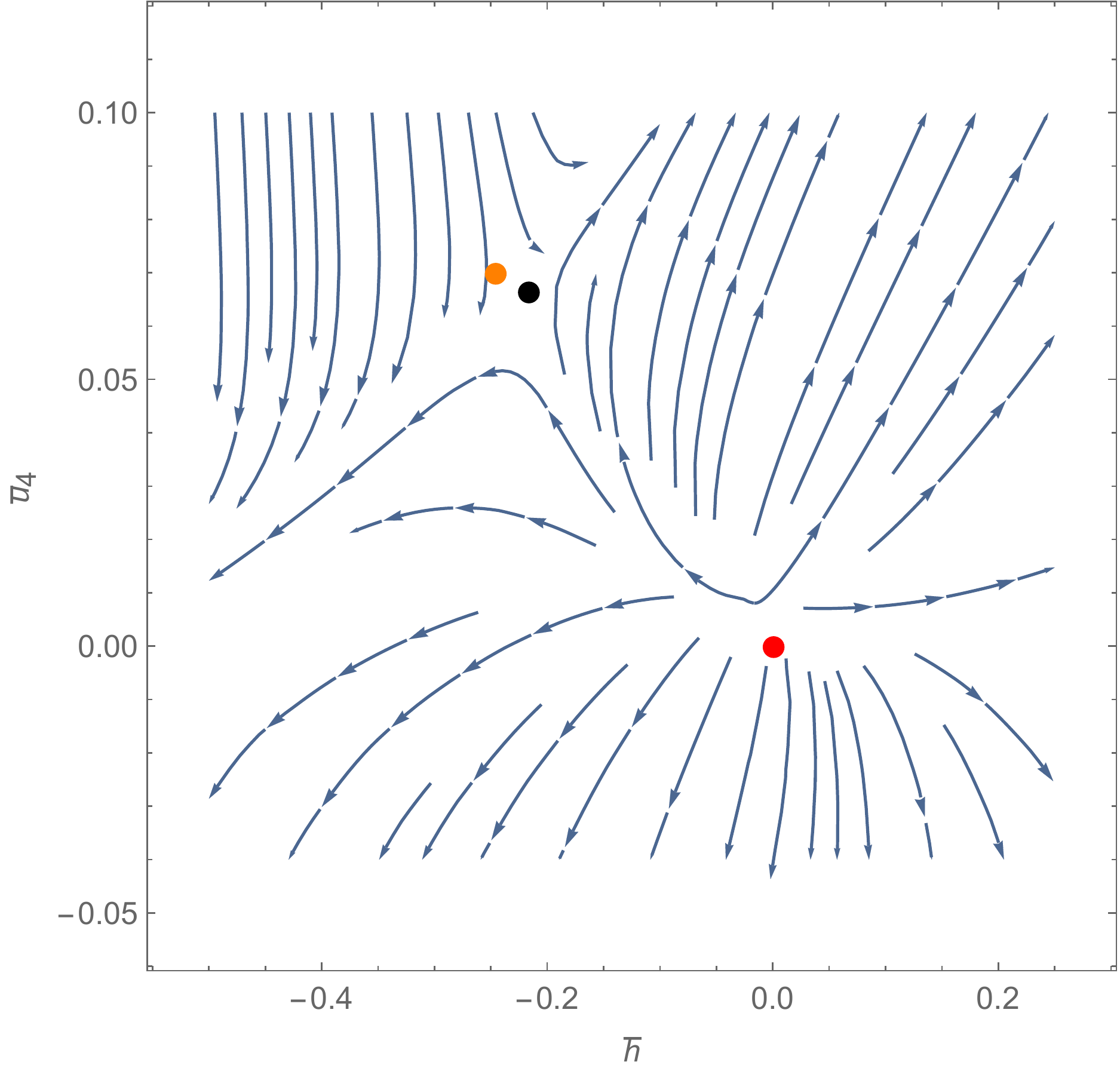}
\end{center}
\caption{On the left: Behavior of the RG flow without disorder ($u_6=0$) in the vicinity of the Wilson-Fisher fixed point (in black) and the Gaussian fixed point (in red). On the right: Projection of the behavior of the RG flow for nonzero disorder, $\vert  \bar{u}_6 \vert <\vert  \bar{u}_6^* \vert$. The position of the fixed point for $u_6=0$ is materialized by the orange point.}\label{flowTens}
\end{figure}

\begin{figure}[h!]
\begin{center}
\includegraphics[scale=0.4]{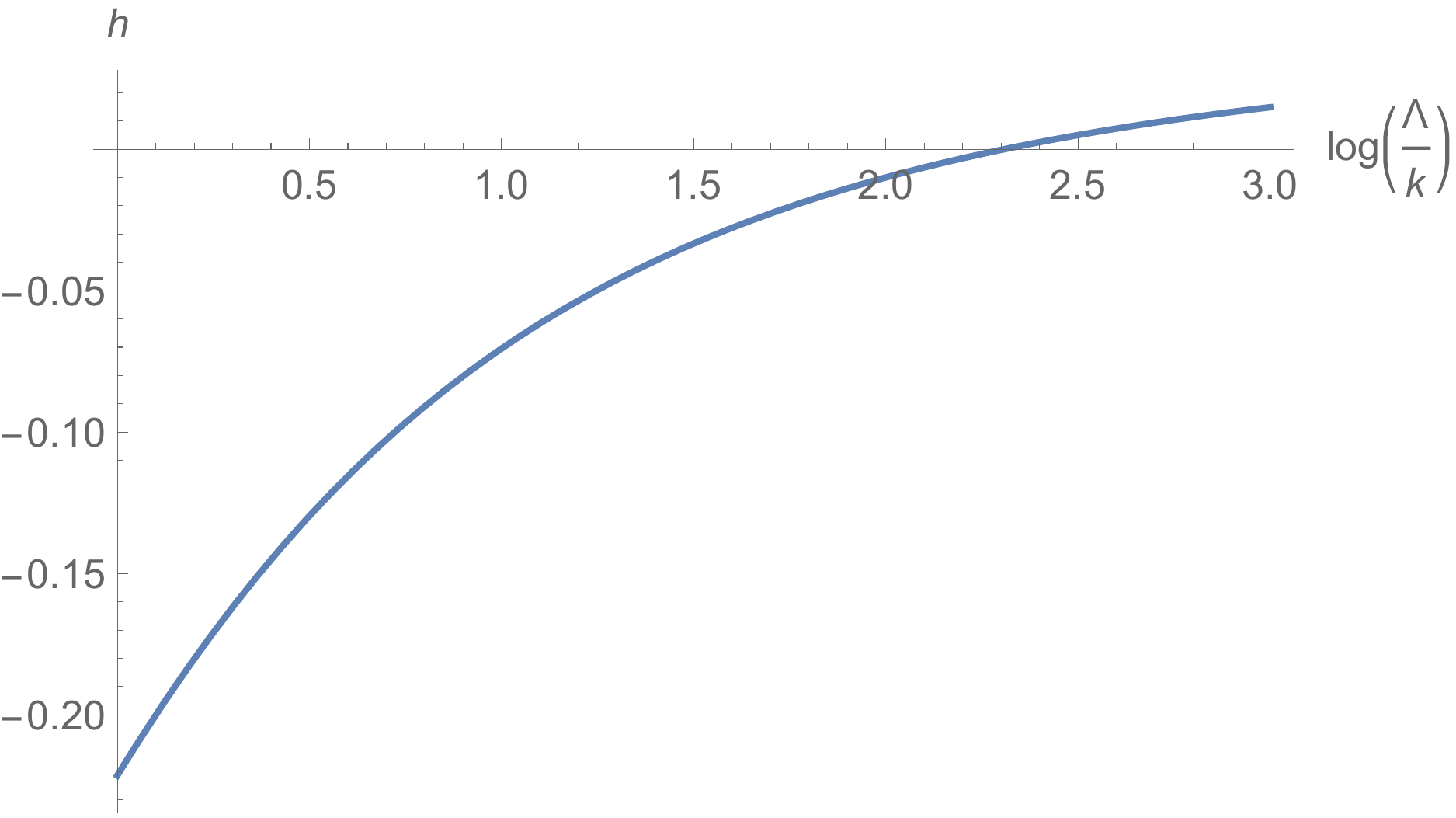}\quad \includegraphics[scale=0.4]{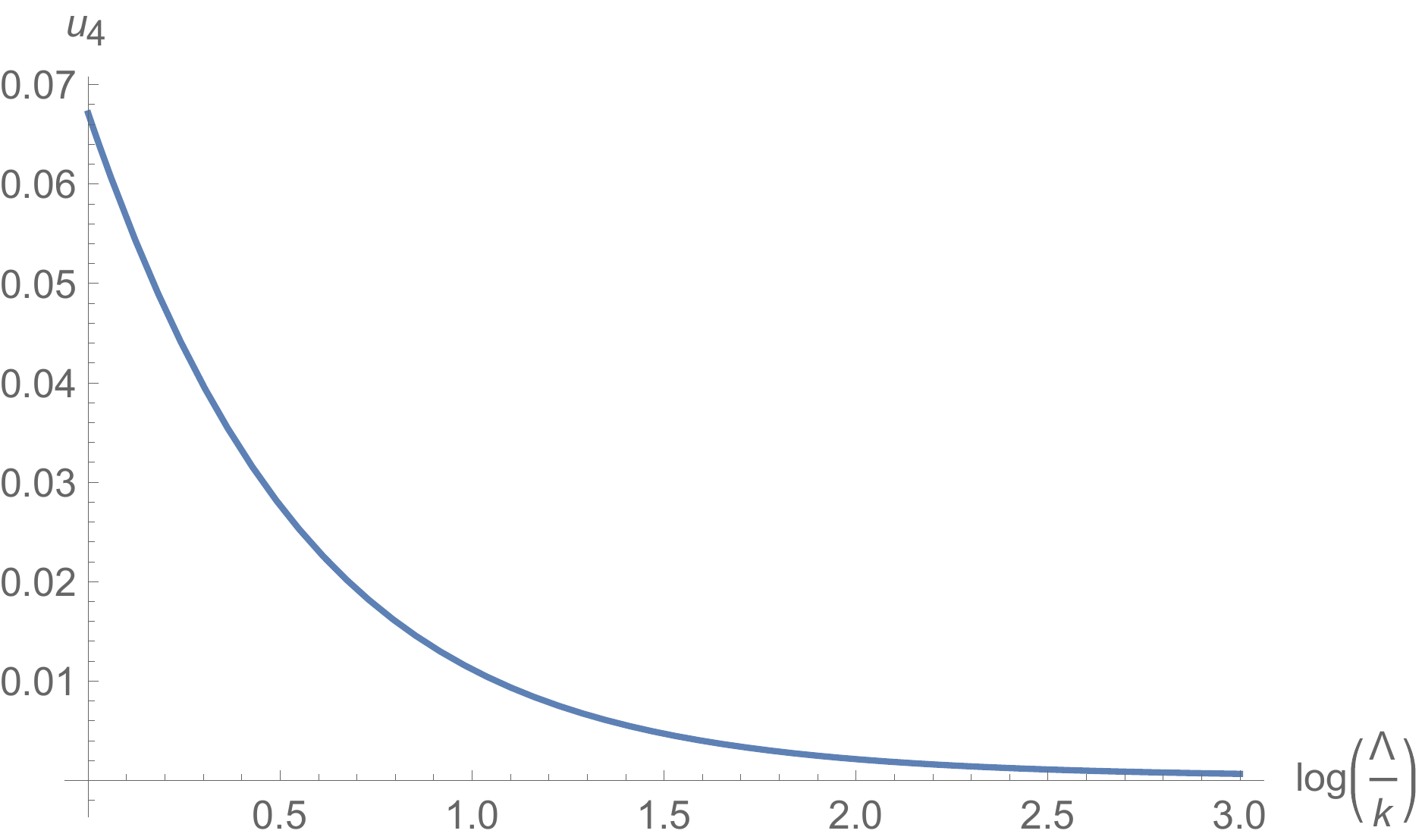}
\end{center}
\begin{center}
\includegraphics[scale=0.4]{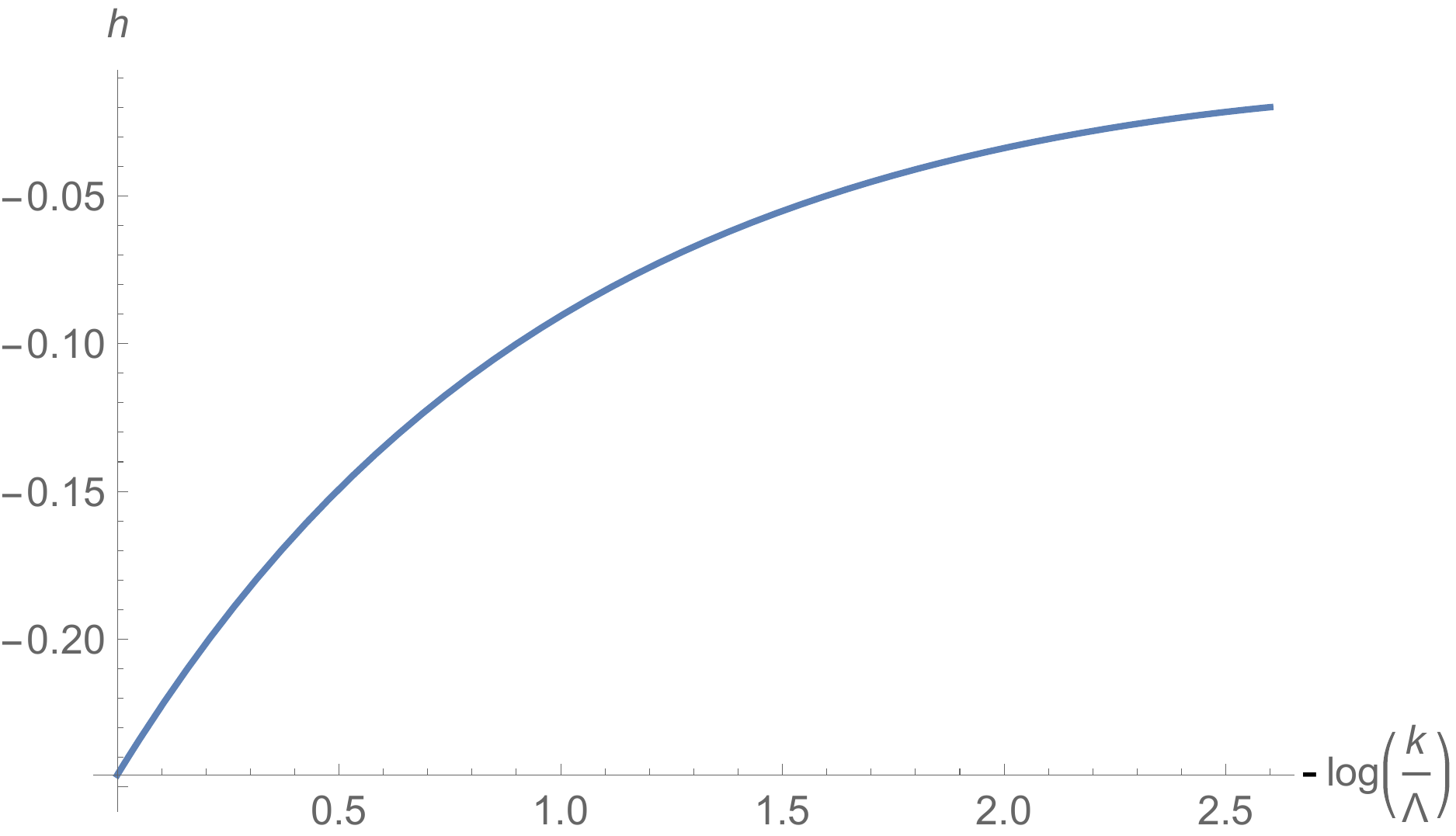}\quad \includegraphics[scale=0.4]{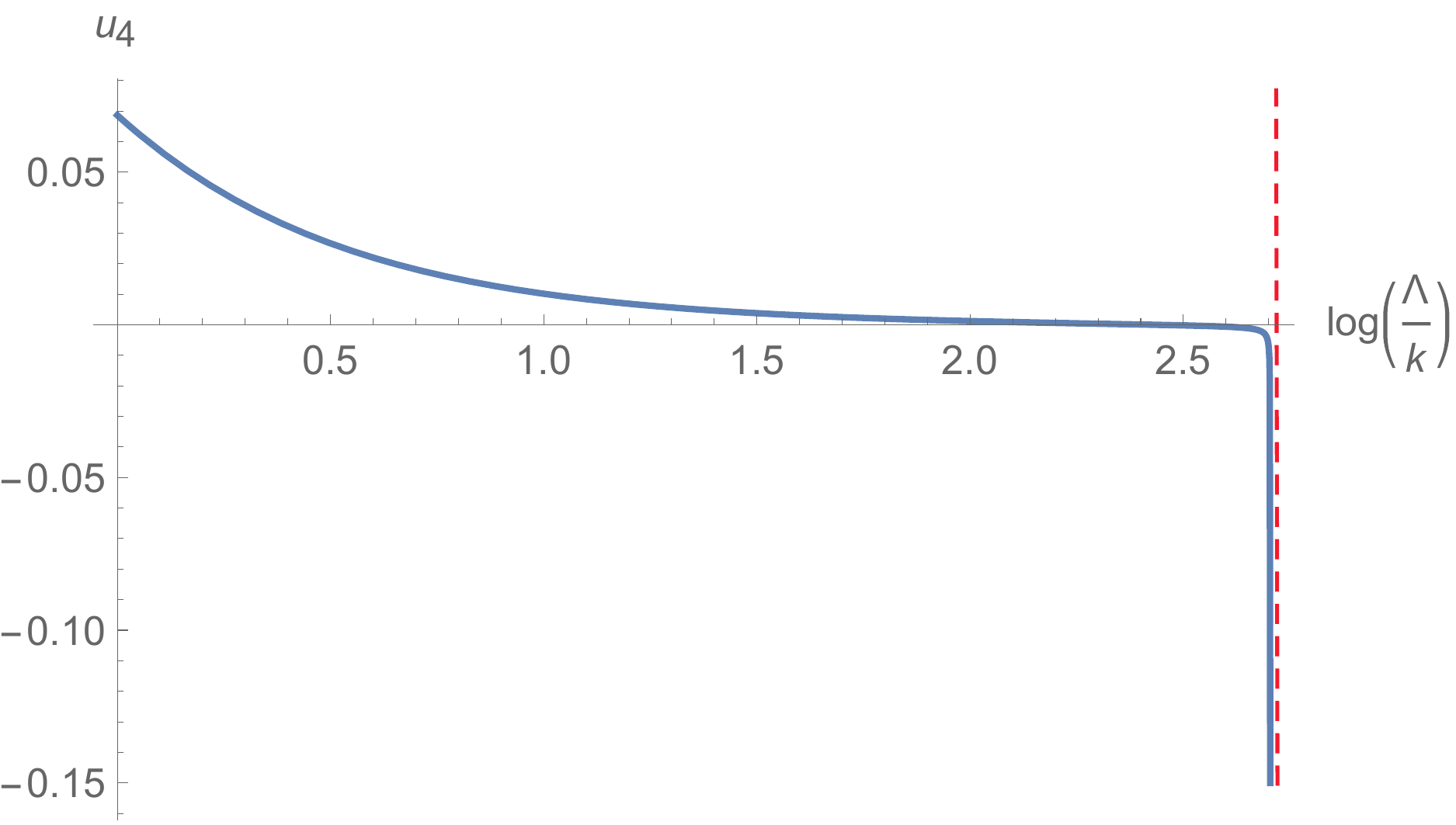}
\end{center}
\caption{On the top: Evolution of the mass ${h}$ (on left) and of quartic coupling (on right) for starting points in the vicinity of the Wilson-Fisher fixed point with vanishing disorder. On the bottom: Evolution of the mass ${h}$ (on left) and of quartic coupling (on right) for starting points in the vicinity of the Wilson-Fisher fixed point with non-vanishing disorder.}\label{trajectoriesu60}
\end{figure}

\begin{figure}[h!]
\begin{center}
\includegraphics[scale=0.4]{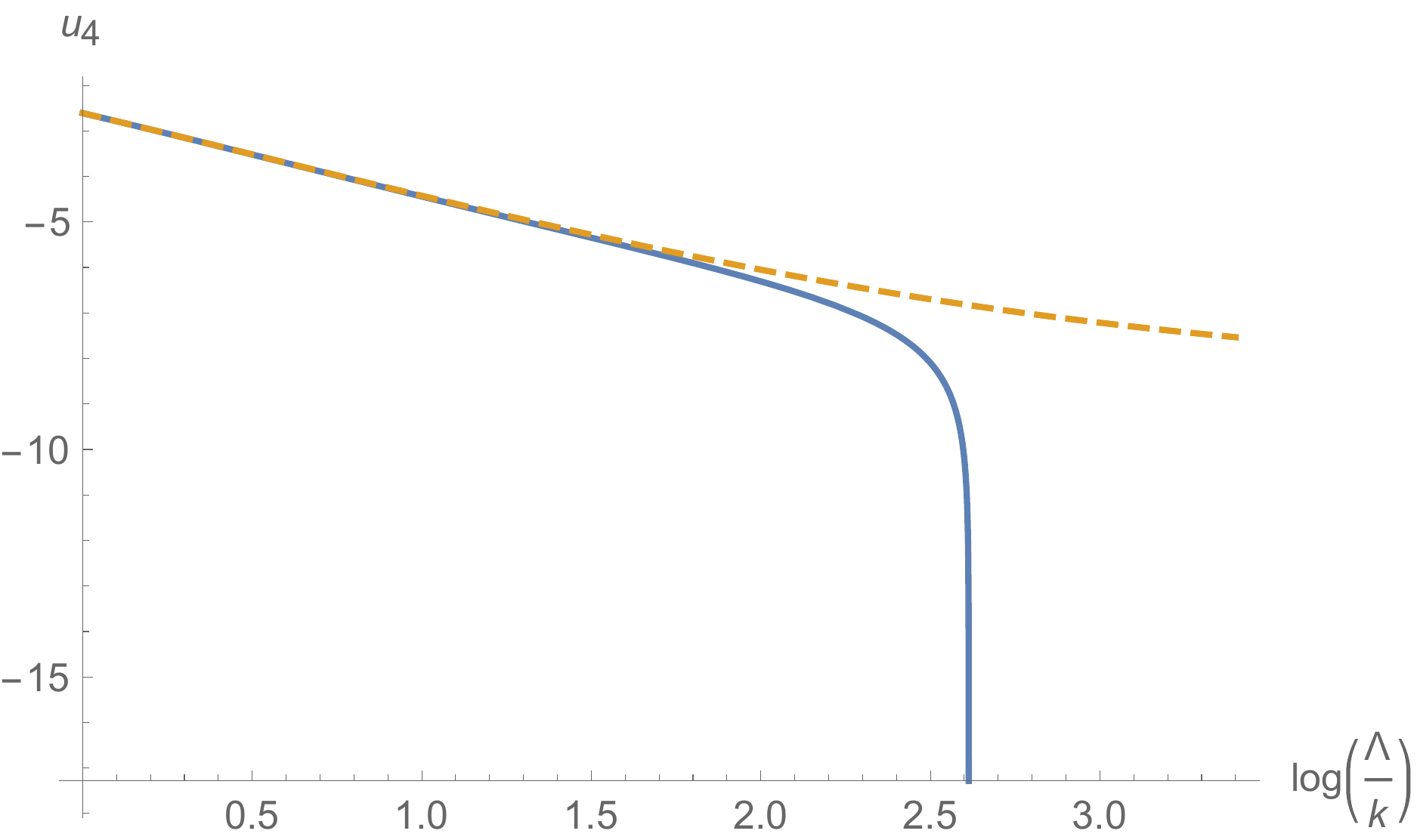}
\includegraphics[scale=0.4]{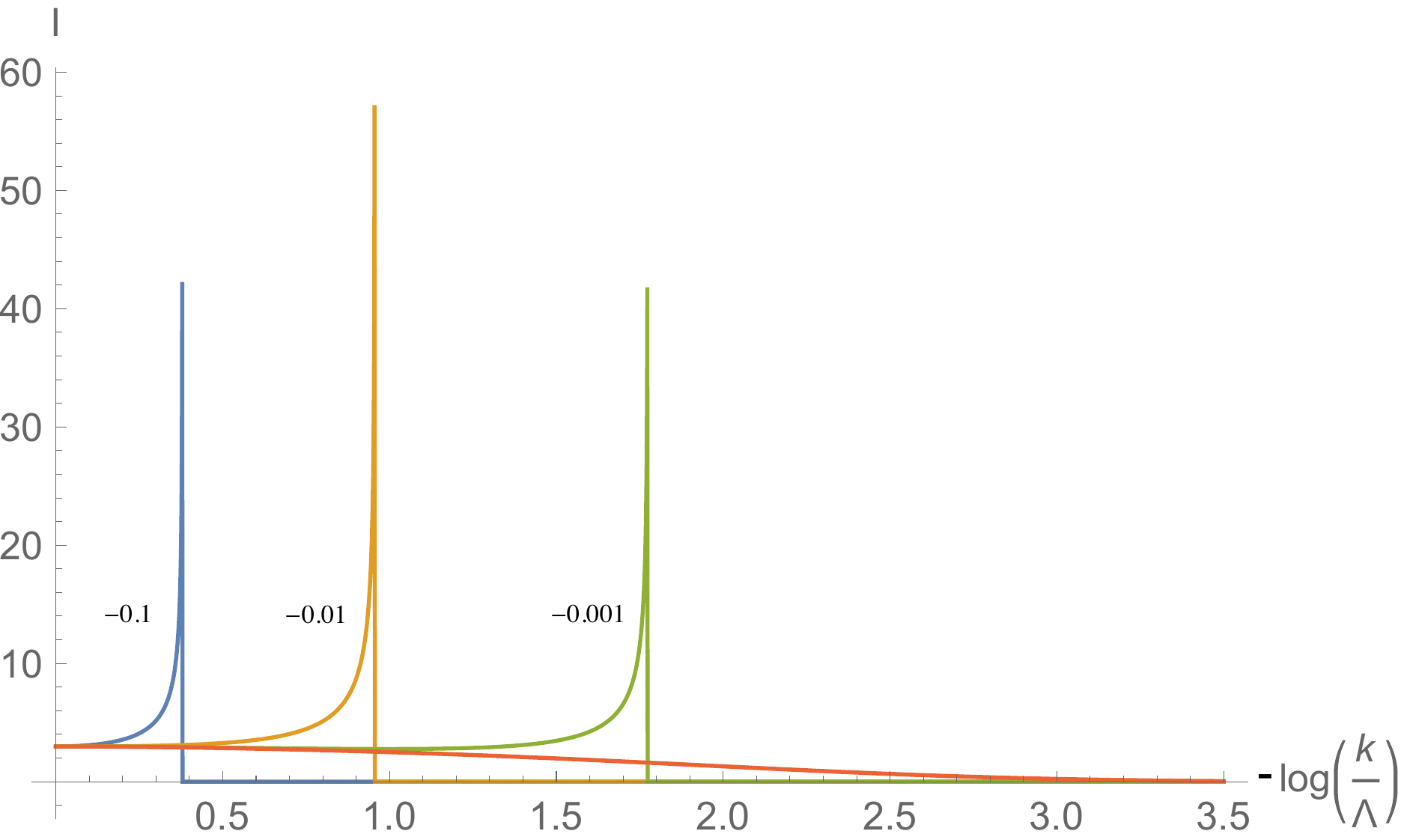}
\end{center}
\caption{On the left: Behavior of $u_4^{(1)}$ for $\bar{u}_6=-10^{-4}$ (solid curve) and $\bar{u}_6=-10^{-5}$ (dotted curve). On the right: The behavior of $\mathcal{I}$ for $\bar{u}_6(\Lambda)=-10^{-1}$ (blue curve), $\bar{u}_6(\Lambda)=-10^{-2}$ (yellow curve) and $\bar{u}_6(\Lambda)=-10^{-3}$ (green curve) and $\bar{u}_6(\Lambda)=0$ (red curve).}\label{divergencesIJ}
\end{figure}

\begin{figure}
\begin{center}
\includegraphics[scale=0.4]{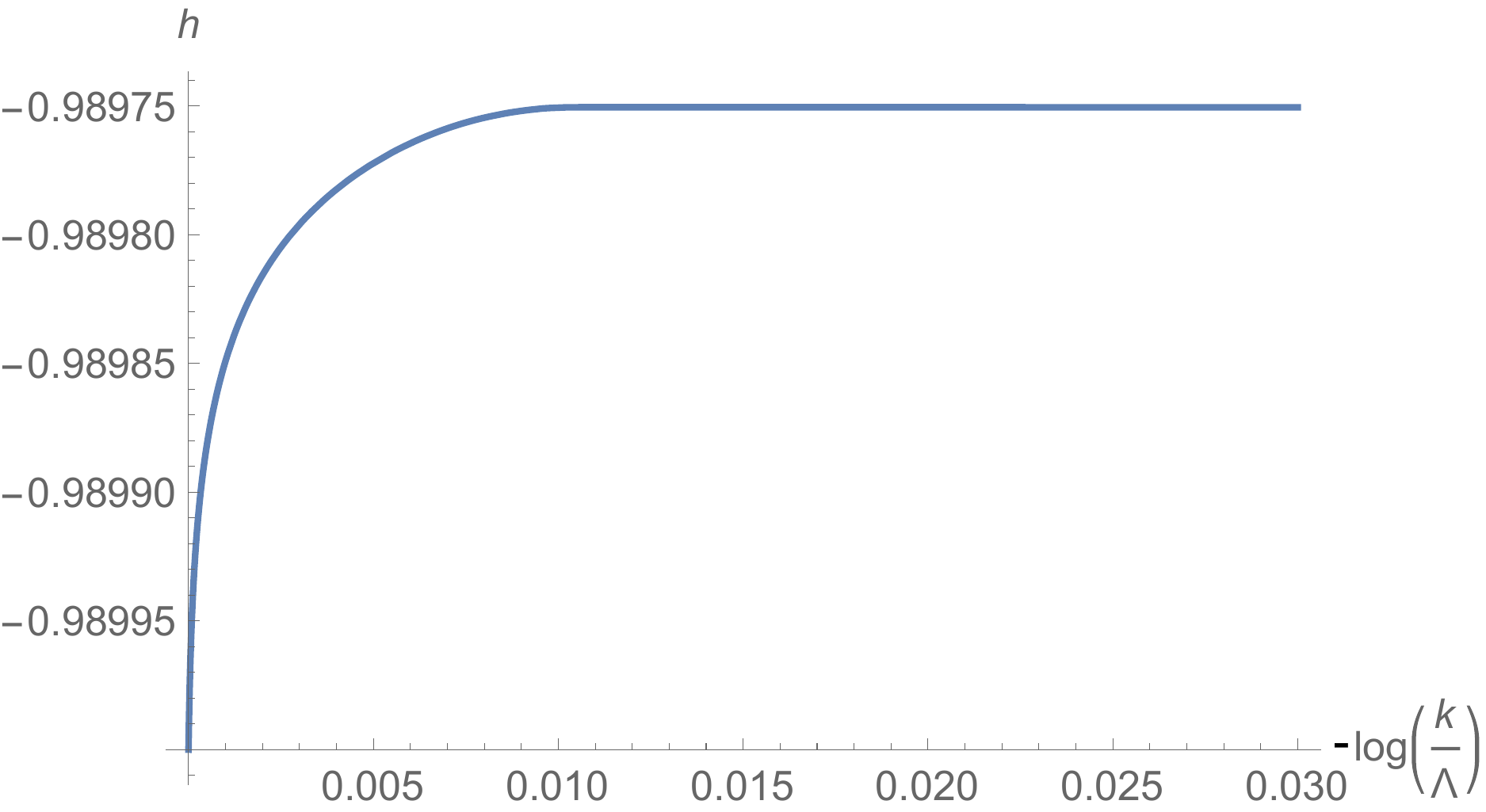}\quad \includegraphics[scale=0.4]{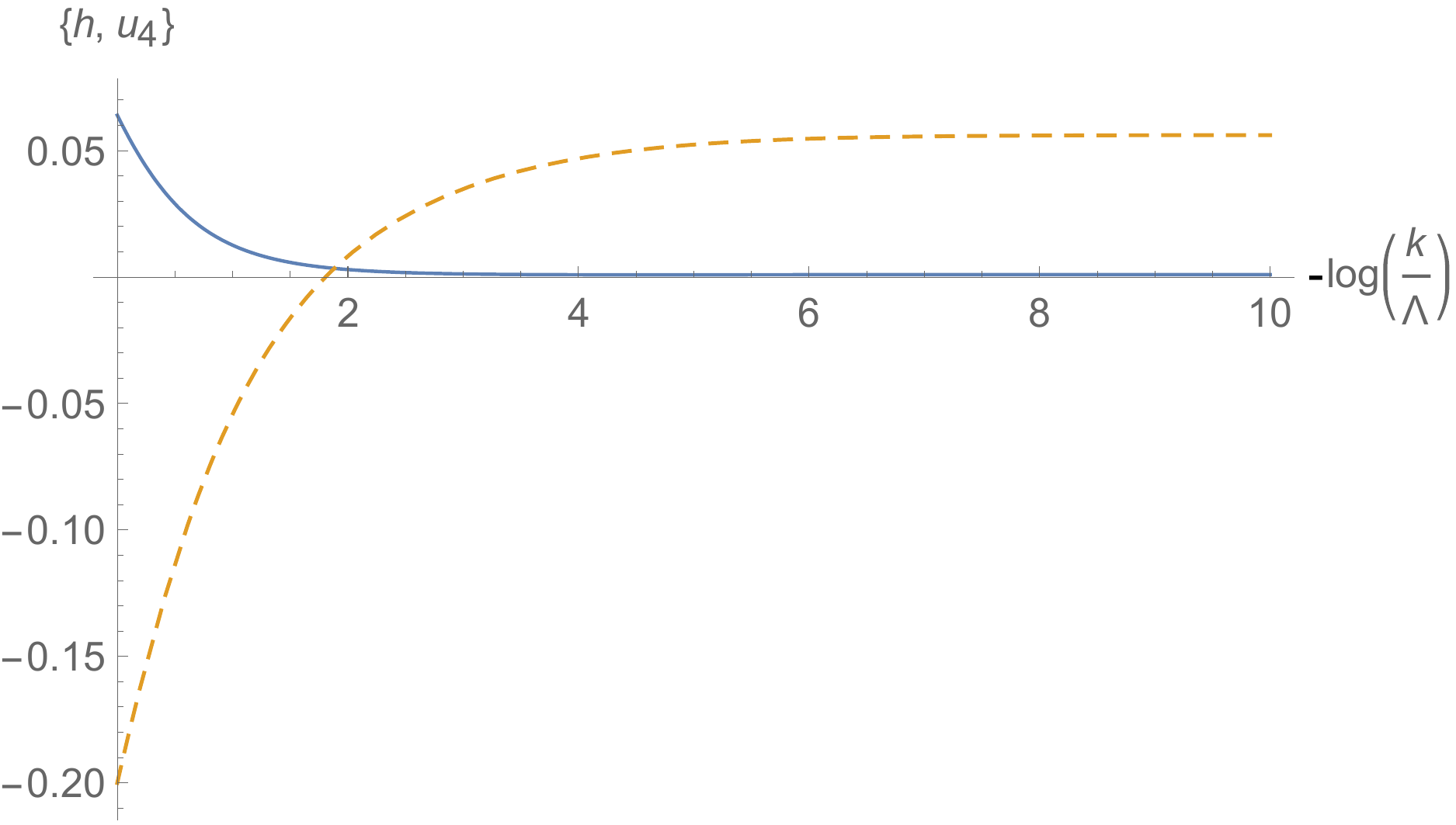}
\end{center}
\caption{On the left: Convergence toward negative mass for zero disorder. On the right: Convergence of mass (yellow curve) and quartic coupling (blue curve) with small disorder.}\label{flowbehaviorzerodiv}
\end{figure}

In this section we provide a first look at numerical investigations of the formalism that we introduced in the previous sections, and we plan to devote a companion paper for a deeper numerical analysis. For this first look we focus on the large $N$ limit, and set $J_0=0$. The last condition discard from our analysis all the odd couplings $u_3^{(1)}, u_3^{(2)}$ and so on. The remaining flow equations split in two sets. The first set, which reduces to $(\dot{h},\dot{u}_4^{(1)})$ is such that, even if initially $(h,u_4^{(1)})=(0,0)$, the corresponding flow equations does not vanish due to $u_6$. The second set on the contrary includes the remaining couplings whose flow equations remain zero if we set vanishing initial conditions. For our analysis, we discard this last class of couplings. Moreover we set $\alpha=1$, which simplifies even more the original complicated set of flow equations that we previously derived and which reduces now to:
\begin{align}
&\dot{\bar{h}}=-\bar{h}-4\bar{u}_4^{(1)}(\bar{I}_2+2\bar{J}_2)\,,\\
&\dot{\bar{u}}_4^{(1)}=-2{\bar{u}}_4^{(1)}-\frac{3}{2} \frac{\bar{u}_6}{(1+\bar{h})^2}+8(u_4^{(1)})^2(I_3^\prime+3J_3^{\prime\,(1)})\,,\\
&\dot{\bar{u}}_6=-3{\bar{u}}_6\,.
\end{align}
Note that the last equations implies that in the large $N$ limit $u_6$ remains constant. First, let us investigate the fixed point structure of the coupled system $(\dot{\bar{h}},\dot{\bar{u}}_4^{(1)})$, by considering $u_6$ as an external adjustable parameter. This system can be easily triangulated, the condition $\dot{\bar{h}}=0$ defining $\bar{u}_4^{(1)}(\bar{h})$ along the fixed point solutions, which are given by the zeros of the function $F(h,\bar{u}_6):=\dot{\bar{u}}_4^{(1)}(\bar{h},\bar{u}_4^{(1)}(\bar{h}),\bar{u}_6)$. Figure \ref{figTensor1} illustrates the behavior of the fixed point solution for several values of $\bar{u}_6$. Note that we focused on the region $h<0$, aiming to describe the spherical regime; and $\bar{u}_4^{(1)}>0$ as required by stability. On the left, the Figure \ref{figTensor1} shows the behavior of $F(h,\bar{u}_6)$ for some values of $\bar{u}_6$, which has to be negative to be interpreted as a disorder effect, see \eqref{classicalpotential}.
For $\bar{u}_6=0$, we have a single fixed point solution, reached for $\bar{h}\sim -0.24$ ($\bar{u}_4^{(1)}\sim 0.07$). It corresponds to a Wilson-Fisher fixed point, having one repulsive and one attractive direction, the corresponding critical exponents\footnote{This represents the opposite values of the stability matrix eigenvalues.} being $(-2.79, 1.14)$. Figure \ref{flowTens} (on left) illustrates the behavior of the RG flow around this point.
It is instructive to follow the behavior of some RG trajectories in the spherical sector (i.e. for $\bar{h}<0$). Figure \ref{trajectoriesu60} (on the top) show the behavior of two trajectories starting in the vicinity of the Wilson-Fisher fixed point. The mass ${h}$ reaches a plateau at a finite time whereas $u_4^{(1)}$ goes to zero. Asymptotically, and exploiting the global invariance reparametrization for fields, it is suitable to fix the asymptotic value of the effective non-zero vacuum as $\langle q^2 \rangle = N$.
Figure \ref{figTensor1} on the right shows that the Wilson-Fisher fixed point does not exist for all values of $\bar{u}_6$. Indeed, recalling that $\bar{u}_6 <0$, we show the existence of a minimum $\bar{u}_6^* \sim -0.037$. Below this value, the fixed point disappears, which is illustrated on the left of Figure \ref{figTensor1}.  Because $\vert \bar{u}_6\vert$ increases exponentially with $\ln(k/\Lambda)$, the fixed point is expected to eventually disappear in the deep IR, and the phase transitions will be first-order \cite{DeDominicisbook}. The behavior of the RG flow below this limit is illustrated in Figure \ref{flowTens} (on right), where we show that trajectories are not bounded from below. The nature of the transition can be inferred by studying the behavior of such a trajectory, for the nonzero disorder. Figure \ref{trajectoriesu60} (on the bottom) shows the typical evolution of mass and quartic coupling starting in the vicinity of the Wilson-Fisher fixed point with very small disorder $(\bar{u}_6(\Lambda) \sim -10^{-4})$. It shows that, while the behavior of the mass $h$ remains quite similar with its behavior without the disorder, the behavior of the coupling, in contrast, diverges at finite RG parameter $t=\ln(k/\Lambda)$. The origin of the divergences can be traced back to the behavior of the integrals $\bar{I}_2, \bar{J}_2, \bar{I}_3^\prime, \bar{J}_3^\prime$ involved into the flow equations.  Figure \ref{divergencesIJ} on the right shows the behavior of $\mathcal{I}(\bar{u}_2):=\bar{I}_2(\bar{u}_2)+2\bar{J}_2(\bar{u}_2)$ for different values of $\bar{u}_6(k=\Lambda)$, and on the left we superpose the behavior of $u_4^{(1)}(t)$ for $\bar{u}_6(\Lambda)=-10^{-4}$ (solid blue curve) and $\bar{u}_6(\Lambda)=-10^{-5}$ (dotted yellow curve) respectively, for $t:=\ln(k/\Lambda)$. As the disorder decreases, the singularity comes later and ends up disappearing when the disorder is canceled out. Note that this kind of divergence already occurs when the disorder is canceled out, along the attractive direction (in green in the Figure \ref{flowTens}) pointing towards the negative masses. The flow diverges in that case because it reaches the singularity of the integrals. As the disorder  $ \vert \bar{u}_6 \vert $ remains smaller enough some trajectories have time to escape the divergence (Figure \ref{flowbehaviorzerodiv} (on right)). However, as soon as it is large enough, and in particular for $\vert \bar{u}_6 \vert>\vert \bar{u}_6^* \vert$; trajectories fail to converge. Note that this happens at finite scale $t_*=\ln \left(\frac{\bar{u}_6(\Lambda)}{\bar{u}_6^*}\right)$. Moreover, it diverges earlier and earlier as the disorder increases, and the supersymmetric RG flow fails to provide a suitable description (i.e. Ward identities \eqref{WT2} cannot be integrated along with the flow beyond a finite time interval, which collapses to zero). Because of the relation between time reversal symmetry and supersymmetry, we expect that such a divergent behavior signals the end of the equilibrium regime\footnote{Note that such a singular behavior at finite scale is generally encountered for disorder systems \cite{Gredat_2014, Tissier_2011,Tissier_2012}.}, and a dynamical breaking of ergodicity, which is not controlled by a fixed point and thus have to be of the first order. Note that because the flow for coupling is unbounded from below, the ground state fails to be normalizable (at least for the quartic truncation). Finally, note also that trajectories in the vicinity of the Wilson-Fisher point always converge for the zero disorder case, as Figure \ref{flowbehaviorzerodiv} (on the left) shows. Moreover, note that, for $u_6=0$, we cannot reach the region $u_4^{(1)}<0$ from the region $u_4^{(1)}>0$.
\medskip

\begin{figure}
\begin{center}
\includegraphics[scale=0.4]{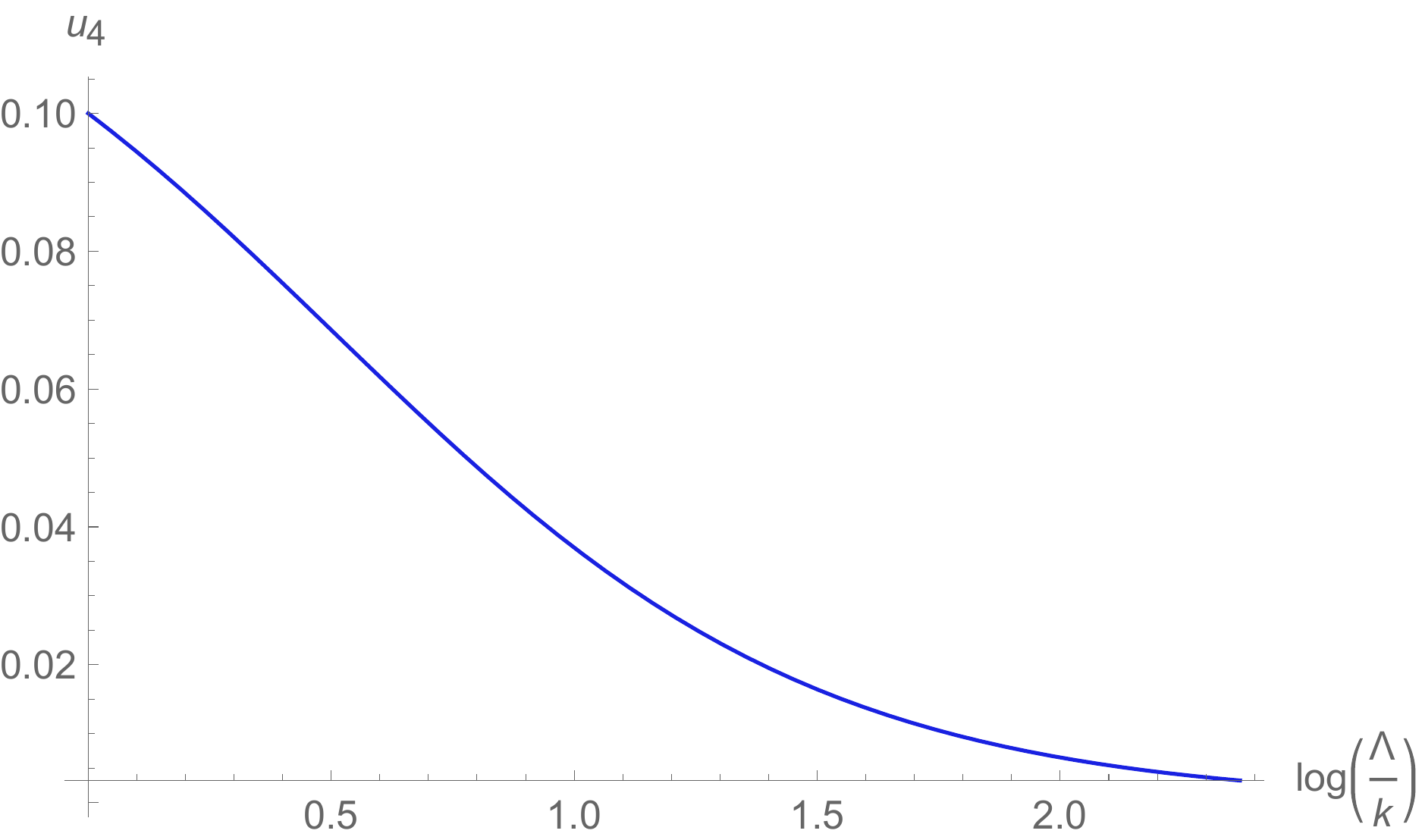}
\includegraphics[scale=0.4]{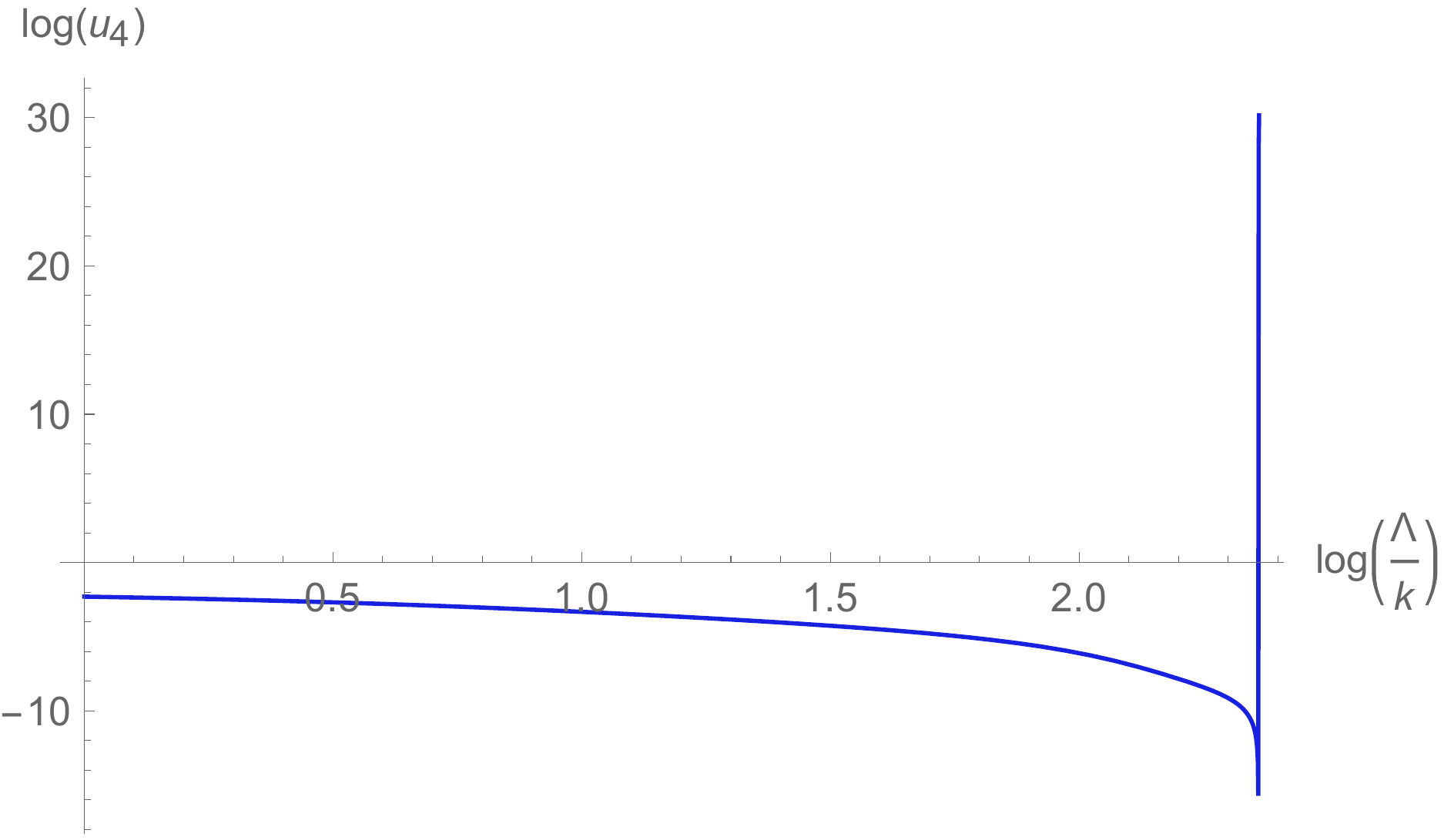}
\end{center}
\caption{On the right: Finite time scale singularities for some initial condition in the spherical region, using non-zero classical field expansion. On the left: Convergence for zero disorder.}\label{Finitescale}
\end{figure}

Such a conclusion obviously has to be confirmed with more sophisticated approaches. For instance, our field expansion around zero classical field does not seem truly suitable to describe the spherical model. Indeed, one could envisage a development around the vacuum of the local part of the effective potential. This, essentially, corresponds to $V[\Phi]$ in equation \eqref{classicalpotential}. In terms of the field $\phi$, it corresponds to a linear interaction for the response field $\sim \tilde{\varphi}V^\prime[\phi]+(\bar{\psi} \psi) V^{\prime\prime}[\phi]$. Let us focus on the quartic potential \eqref{classicalpotential}. It has two minima, for $\phi^2=0$ and $\phi^2=-2Dh_1N/h_1$, both making the linear coupling vanish in the response field. The solution $\phi=0$ moreover extremizes the second derivative $V^{\prime\prime}[\phi]$, $V^{\prime\prime\prime}[\phi=0]=0$. Because we cannot use $O(N)$ symmetry to choose the non-zero vacuum along a single axis, we impose $\phi^2=r^2 N$ and $\sum_{i=1}^N \phi_i\sim0$. With this respect, for $h<0$, the vanishing solution becomes a minimum of $V^{\prime\prime}[\phi]$, because $V^{\prime\prime}[\phi\neq 0]=0$ while $V^{\prime\prime}[\phi=0]=h$.  Thus, the zero vacuum is expected to be well convergent with this respect (see \cite{Synatschke_2009,wilkins2021functional,wilkins2021functional2}). Let us briefly consider the non-zero vacuum $\phi^2=r^2 N$ for the following truncation:
\begin{equation}
\Gamma_k[\mathcal{M}]:=\Gamma_{k,\text{kin}}[\mathcal{M}]+\Gamma_{k,\text{int}}[\mathcal{M}]\,,
\end{equation}
where $\Gamma_{k,\text{kin}}$ is again given by \eqref{truncationkin} and
\begin{equation}
\Gamma_{k,\text{int}}[\mathcal{M}]=i\int dz V_1[\mathcal{M}(z)]-\frac{u_6(k)}{N^2} \int dz dz^\prime (\mathcal{M}(z) \cdot \mathcal{M}(z^\prime))^3\,,
\end{equation}
with:
\begin{equation}
V_1(\phi)=\frac{h(k)}{2}\phi\cdot \phi+\frac{u_4(k)}{4N}(\phi\cdot \phi)^2\,.
\end{equation}
Taking the second derivative with respect to classical (super-)field of the Wetterich equation \eqref{Wetterich}, we get schematically:
\begin{equation}
\dot{\Gamma}_k^{(2)}=-\frac{1}{2}\STr R_k G \Gamma^{(4)} G+\STr R_k G \Gamma^{(3)}G \Gamma^{(3)}G\,.
\end{equation}
From the truncation that we consider, odd functions can be discarded in the large $N$ limit. The computation of $G^{-1}_{k,\bar{\varphi}\phi}$ follows the strategy explained in the previous section, except that we have to keep into account the contribution of the quartic interaction evaluated for $\phi^2=r^2(k) N$. In replacement of \eqref{A} and \eqref{B}, we get for the diagonal parts of $A$ and $B$:
\begin{equation}
A_{k,\bar{\varphi}_i\phi_j}^{\text{diag}}(\omega)=Z_k(1+\rho_k^{(2)}(\omega)-2\pi u_6 (r^2)^2 \delta(\omega))\delta_{ij}\,,
\end{equation}
and
\begin{equation}
B_{k,\bar{\varphi}_i\phi_j}^{\text{diag}}(\omega)=Z_k(\omega+ih+ir^2u_4+i\rho_k^{(1)}(\omega))\delta_{ij}\,,
\end{equation}
which, evaluated for $r^2=-h/u_4$ reduce to:
\begin{equation}
A_{k,\bar{\varphi}_i\phi_j}^{\text{diag}}(\omega)=Z_k(1+\rho_k^{(2)}(\omega)-2\pi u_6 (h/u_4)^2 \delta(\omega))\delta_{ij}\,.
\end{equation}
and
\begin{equation}
B_{k,\bar{\varphi}_i\phi_j}^{\text{diag}}(\omega)=Z_k(\omega+i\rho_k^{(1)}(\omega))\delta_{ij}\,.
\end{equation}
The off-diagonal terms behaving like $\sim \frac{\phi_i\phi_j}{N}$, with $\phi_i\sim \mathcal{O}(1)$. 
The resulting flow equations can be derived following the diagrammatic expressions given in \ref{AppA} and numerically investigated. Once again, finite time scale divergences appear, as Figure \ref{Finitescale} shows explicitly for some initial conditions in the spherical region. Our methods remain highly descriptive, as we mentioned before, and should be improved to make useful physical predictions beyond this finite scale divergence effect, which is however a general feature for disordered systems \cite{Gredat_2014, Tissier_2011,Tissier_2012}.

\section{RG for the $p=2$ models} \label{sec5}

In this section we briefly consider the case $p=2$ for which equation \eqref{eq1} reduces to:
\begin{equation}
\frac{dq_i}{dt} =- \sum_j J_{ij} q_j(t)-V^\prime(Q)q_i(t)+\eta_i(t) \,, \label{eq1bis}
\end{equation}
for $\overline{J_{ij}J_{kl}}=\frac{\lambda^2}{2N}(\delta_{ik}\delta_{jl}+\delta_{jk}\delta_{il})$, $\overline{J_{ij}}=0$ and delta-correlated noise (see \eqref{deltanoise}). We especially focus on the quartic case,
\begin{equation}
V^\prime(Q)=\kappa_1 +\frac{\kappa_2}{N} Q^2\,,
\end{equation}
which goes toward the spherical model in the large $N$ limit for $\kappa_1< 0$ and $\kappa_2 >0$, $\kappa_1/\kappa_2=-\mathcal{O}(1)$.  Note that this model can be solved exactly \cite{DeDominicisbook,Cugliandolo2,Cugliandolo3}, and is know to exhibit a weak ergodicity breaking, as long has the laps time in the two point correlation remains smaller than the age of the system. For long time, the system behave likely as a disguised ferromagnet rather than a truly glassy system. Second order phase transitions has been pointed out for this system, investigated for instance through the Ruelle formalism in \cite{vanDuijvendijk2010}. We recover it here within our renormalization group formalism. In particular, we show that the divergences exist only for $N\to \infty$, and is governed by a true interacting fixed point.
\medskip

Integrating out the disorder, we get the following effective potential, in replacement of \eqref{classicalpotential}:
\begin{align}
\overline{W_0}[\mathcal{M}]:=&-i\frac{\lambda}{8N} \int dz^\prime \big(\Phi(z)\cdot \Phi(z^\prime)\big)^2+\frac{\kappa_1}{2} \Phi^2(z)+\frac{1}{2D}\frac{\kappa_2}{4N}(\Phi^2)^2(z)\,.\label{classicalpotentialmatrix}
\end{align}
The truncation that we consider is as soon as given by formula \eqref{truncation3}, but with the effective potential:
\begin{align}
\nonumber U_k[\mathcal{M}(z)]=&\frac{u_2}{2} \big(\mathcal{M}(z)\cdot \mathcal{M}(z)\big)+ \frac{u_{41}}{N} (\mathcal{M}^2)^2(z)
+ i\frac{u_{42}}{N} \int dz^\prime \big(\mathcal{M}(z)\cdot \mathcal{M}(z^\prime)\big)^2\,.
\end{align}
The flow equations can be easily derived following the method detailed in the previous section (section \ref{sec3}). Note that because the distribution of the disorder is centered, all the odd functions vanish and $\eta_k=0$. For the other functions, we get from the strategy detailed in Appendix \ref{AppA}:
\begin{equation}
\dot{\bar{u}}_2=-\bar{u}_2-\frac{4\bar{u}_{41}}{N} \big[(2+N)\big[\bar{I}_2+2\bar{J}_2^{(1)}\big]+2\bar{J}_2^{(2)}\, \big] -\frac{\alpha\bar{u}_{42}}{(\bar{u}_2+\alpha)^2}+\frac{\bar{u}_{42}}{N}(2\bar{K}_2+\bar{K}_2^{\prime})\,,
\end{equation}
\begin{equation}
\dot{\bar{u}}_{41}=-2\bar{u}_{41}+\frac{8 \bar{u}_{41}^2}{N} (N+4)\big[\bar{I}_3^\prime+3\bar{J}_3^{\prime (1)}+3\bar{J}_3^{\prime (2)}\, \big]+\frac{24 \alpha \bar{u}_{42}\bar{u}_{41}}{N}\frac{N+2}{(\bar{u}_2+\alpha)^3}\,,
\end{equation}
\begin{equation}
\dot{\bar{u}}_{42}=-2\bar{u}_{42}+\frac{6\alpha \bar{u}_{42}^2}{N}\frac{1+4N}{(\bar{u}_2+\alpha)^3}+\frac{48 \bar{u}_{41}\bar{u}_{42}}{N} \big[\bar{I}_3^\prime+3\bar{J}_3^{\prime (1)}\big]\,,\label{equ42}
\end{equation}
where $J_2^{(1)}$, $J_2^{(2)}$, $I_3^\prime$, $J_3^{\prime (1)}$ and $J_3^{\prime (2)}$ have been defined in section \ref{sec3}, equations \eqref{J21}, \eqref{J22}, \eqref{Iprime3}, \eqref{Jprime31} and \eqref{Jprime32}, up to the replacements $h\to u_2$ and $\Delta^{(1)}=\Delta^{(2)}=\Delta^{(3)}=0$. Moreover:

\begin{equation}
\bar{K}_2:= Z_k^2 \int \frac{dx}{2\pi}(\eta_k \rho^{(1)}(x)+ \dot{\rho}^{(1)}(x) ) G_{1,\bar{\varphi} \phi}(x)G_{1,\bar{\varphi} \phi}(-x)\,,
\end{equation}
and
\begin{equation}
\bar{K}_2^{\prime}:= -iZ_k^2 \int \frac{dx}{2\pi}(\eta_k \rho^{(2)}(x)+ \dot{\rho}^{(2)}(x) ) G_{1,\bar{\varphi} \phi}(-x) (g_{1,\phi\phi}^{(1)}(x)+2\pi\bar{l}^{(1)}_{1,\phi\phi}\delta(x))\,.
\end{equation}

\begin{remark}
It is instructive to compare the equations \eqref{equ6} and \eqref{equ42}, corresponding to tensor and matrix models respectively. In the first case, in the large $N$ limit, $\dot{u}_6\to 0$, and $u_6$ can be viewed as a constant parameter along the flow. By contrast, $\dot{u}_{42} = \mathcal{O}(u_{42}^2)$ in the large $N$ limit, and the behavior of ${u}_{42}$ remains non-trivial. A moment of reflection helps seeing that this difference can be traced back to the fact that, for $n\geq 5$ the flow equation for $\Gamma_k^{(n)}$ becomes linear in $\Gamma_k^{(n)}$.
\end{remark}
\bigskip

\begin{figure}
\begin{center}
\includegraphics[scale=0.6]{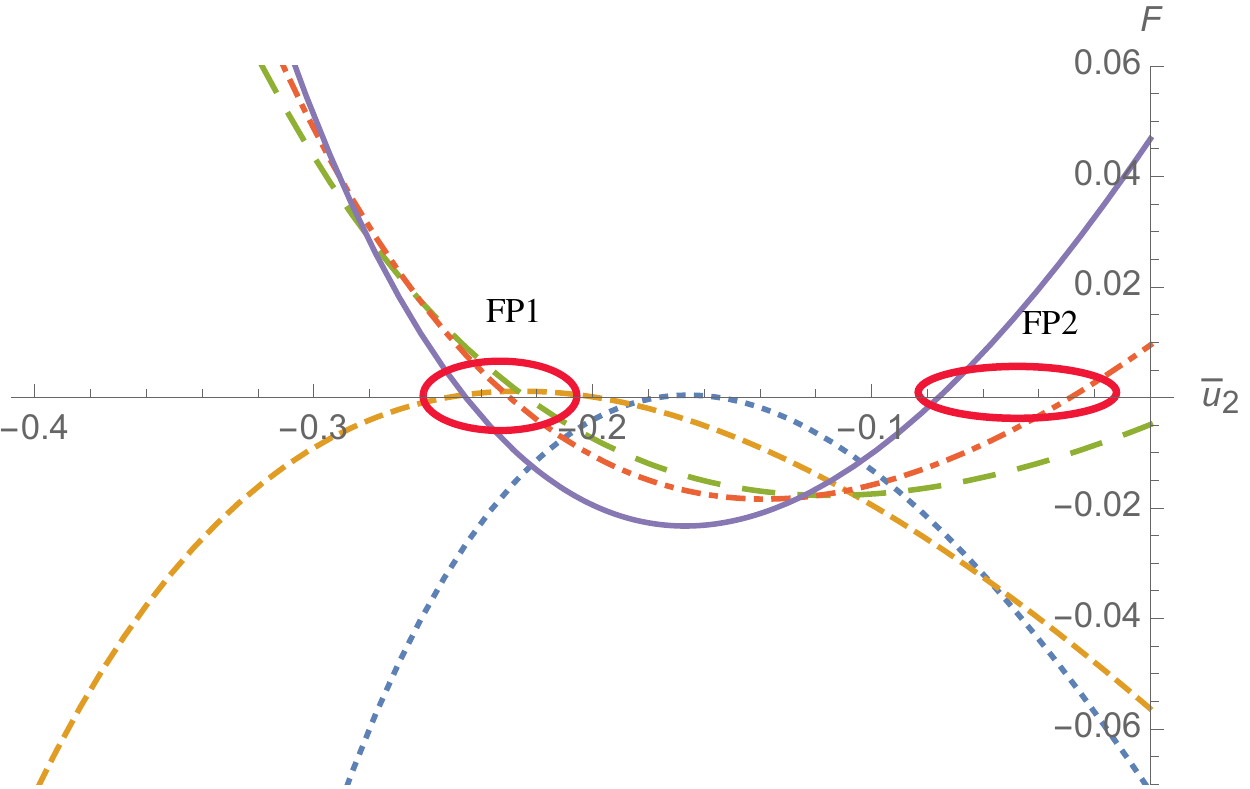} \quad \includegraphics[scale=0.6]{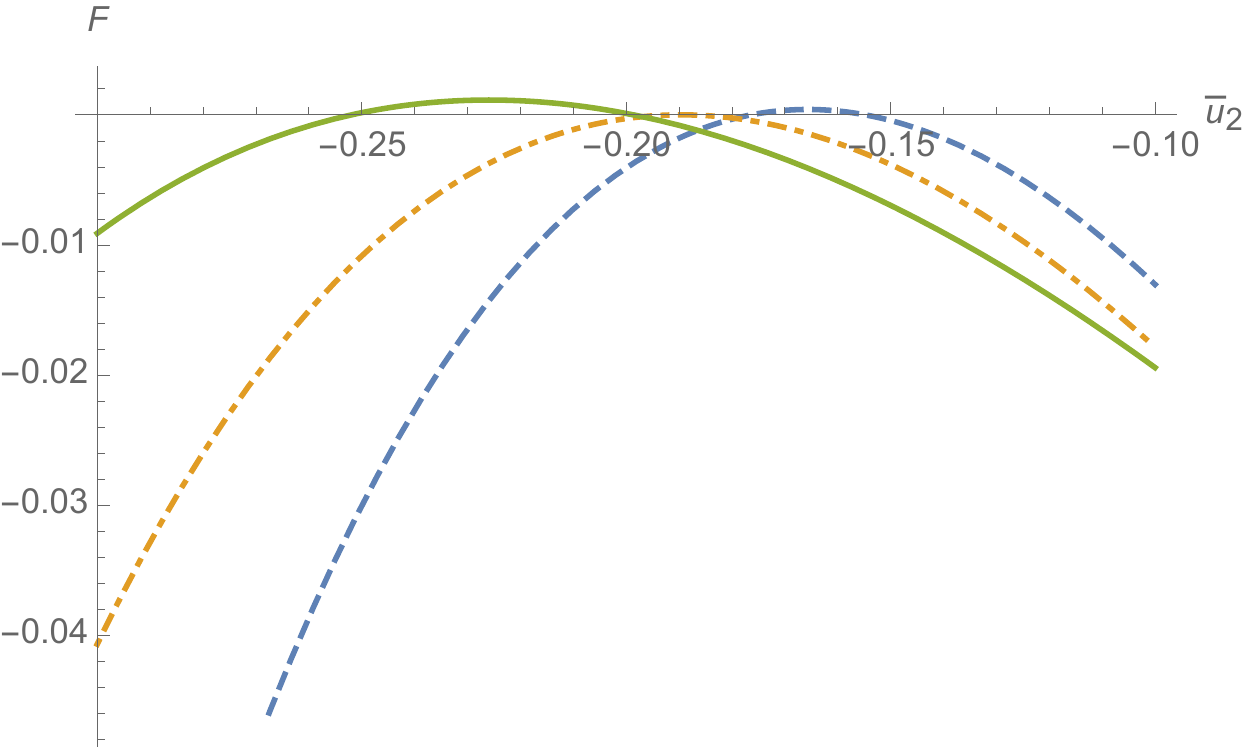}
\end{center}
\caption{On the left: Fixed point solutions in the region $u_2<0$ for $N=1$ (blue curve), $N=3$ (yellow curve), $N=10$ (green curve), $N=15$ (red curve) and $N=\infty$ (purple curve). For $N$ large enough, two fixed points appears, labeled as $FP1$ and $FP2$. On the right, fixed point solutions for $N=1,2,3$, respectively in blue, yellow and green. The case $N=11$ being special from the fact that it corresponds to the limit case where only one fixed point (FP1) appears.}\label{figFPMAT}
\end{figure}

\begin{figure}
\begin{center}
\includegraphics[scale=0.4]{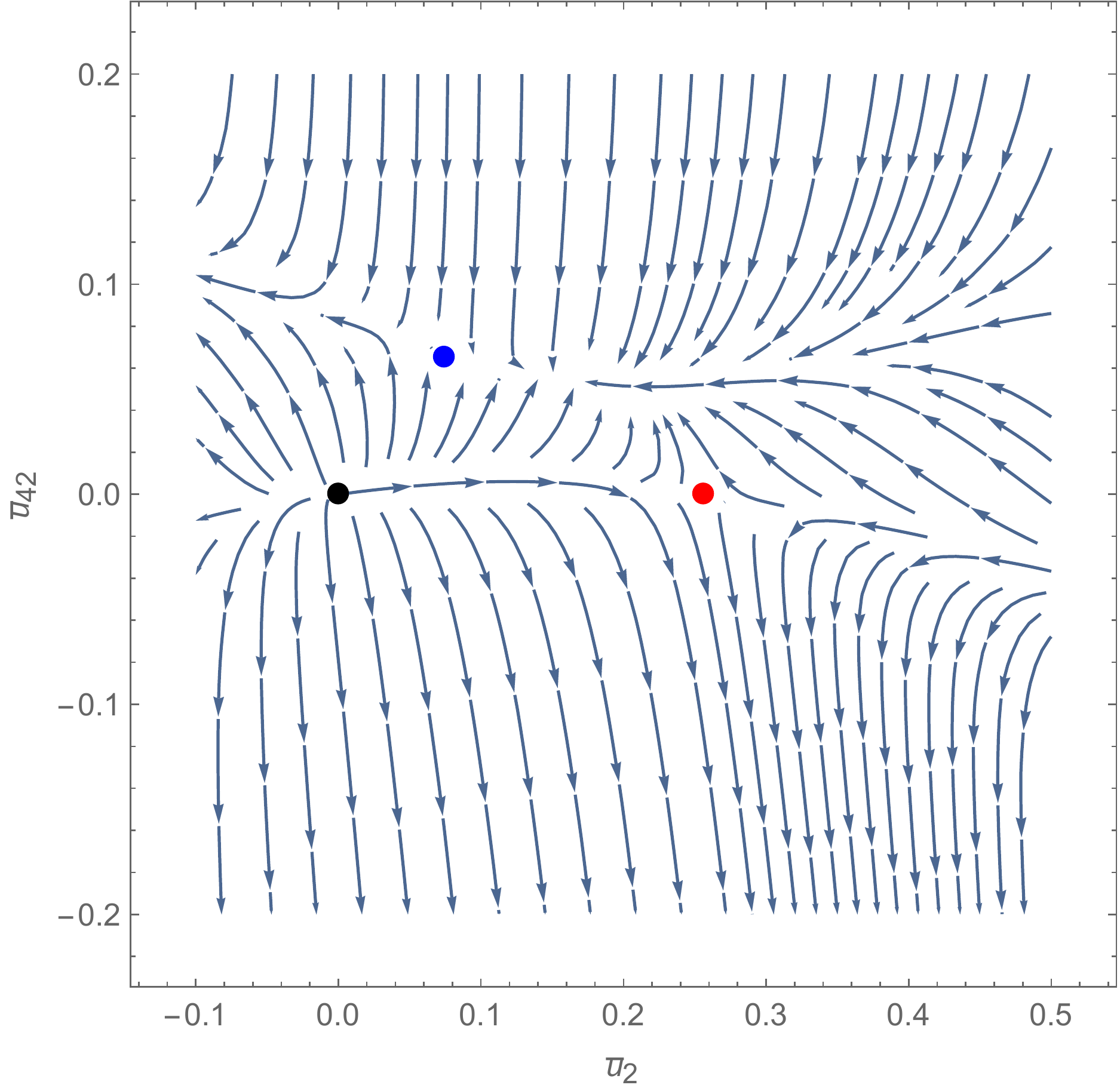} \quad\includegraphics[scale=0.4]{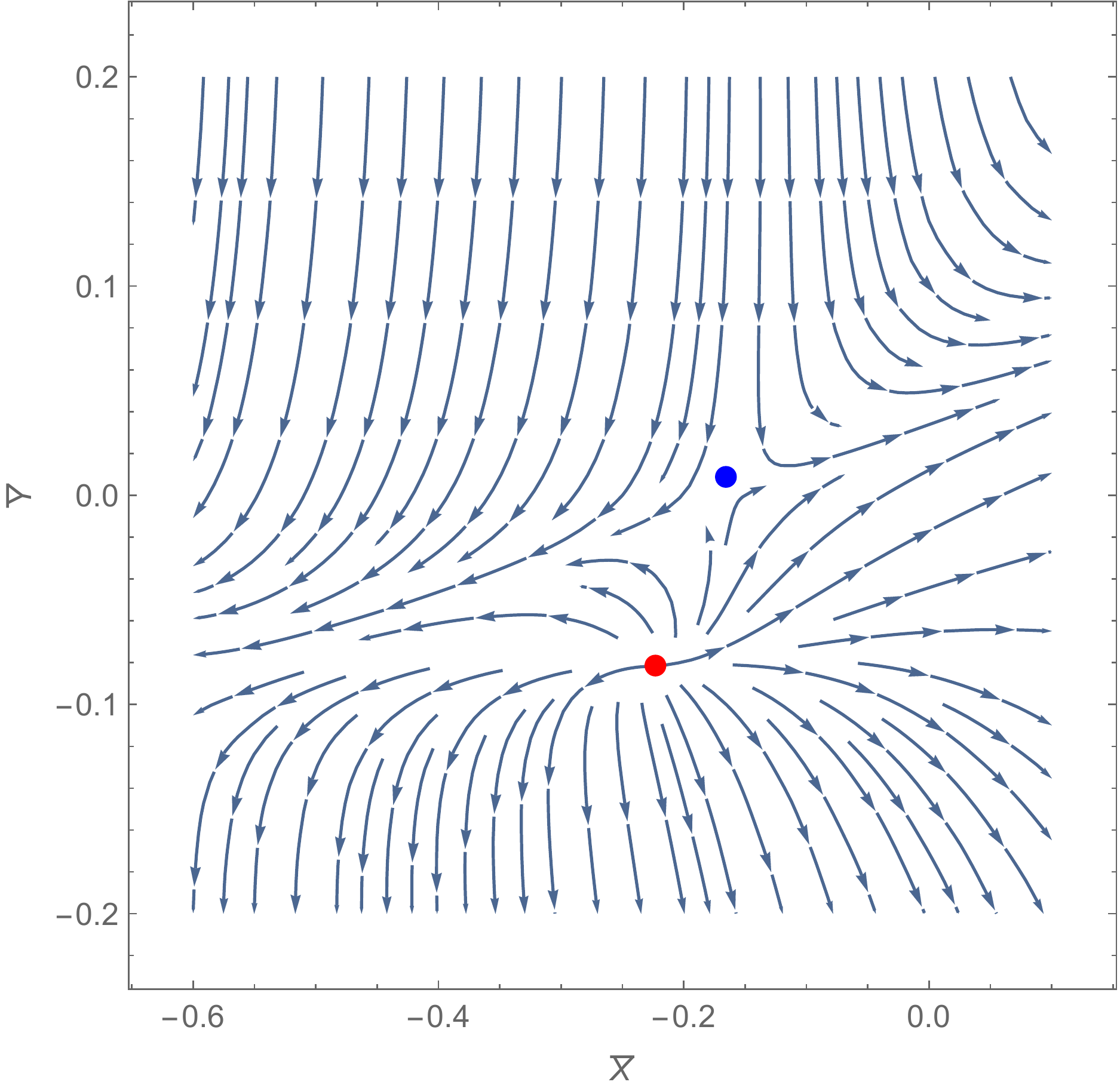}
\end{center}
\caption{Behavior of the large $N$ RG flow. On the left: Projection of the flow into the plane passing through FP1 (red dot), FP2 (blue dot) and the Gaussian fixed point (black dot). On the right: Dynamical projection of the flow into the plane spanned by the two relevant directions of FP1 (red dot).}\label{FlowMatLargeN}
\end{figure}

\begin{figure}
\begin{center}
\includegraphics[scale=0.4]{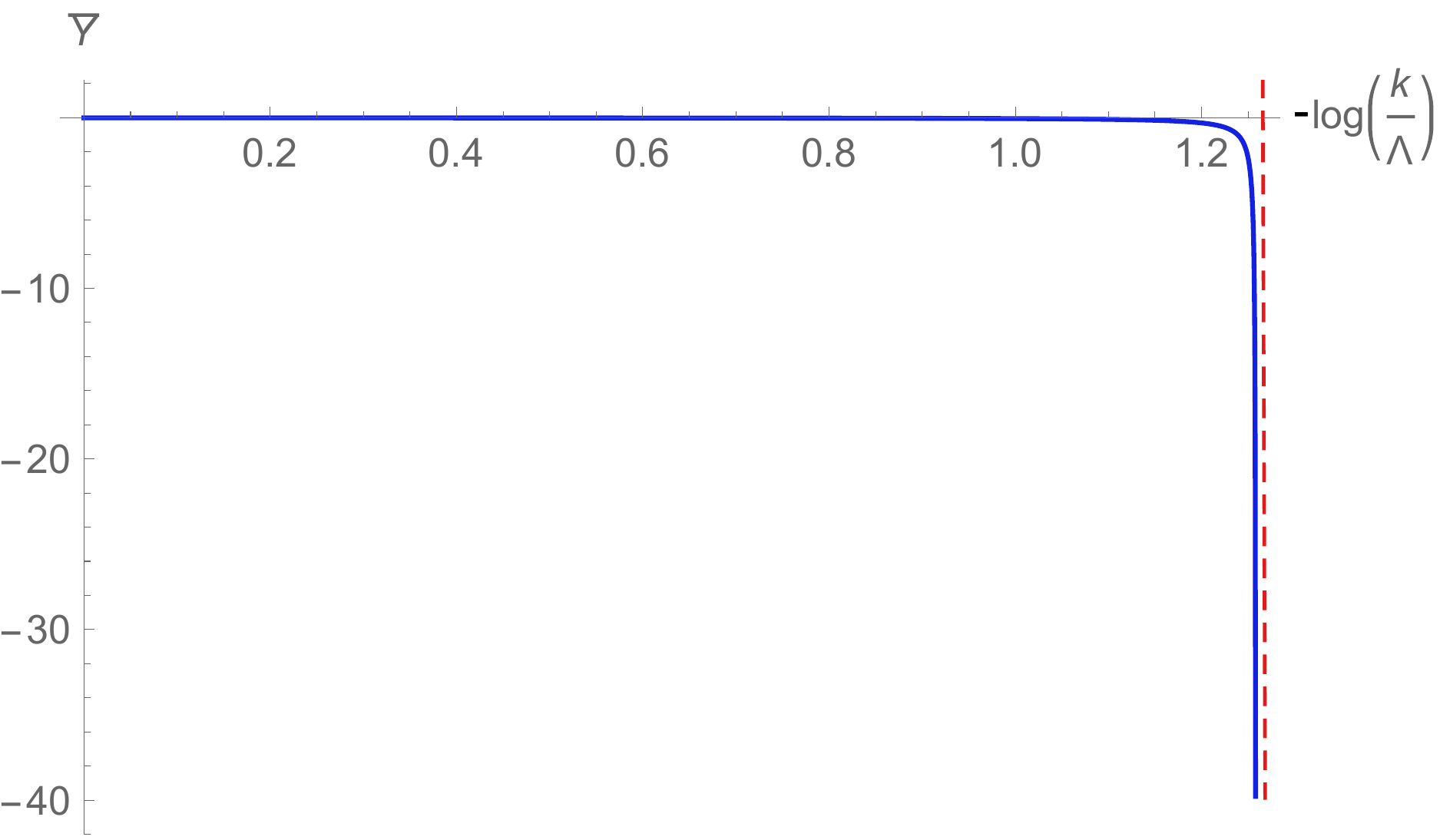} \quad\includegraphics[scale=0.4]{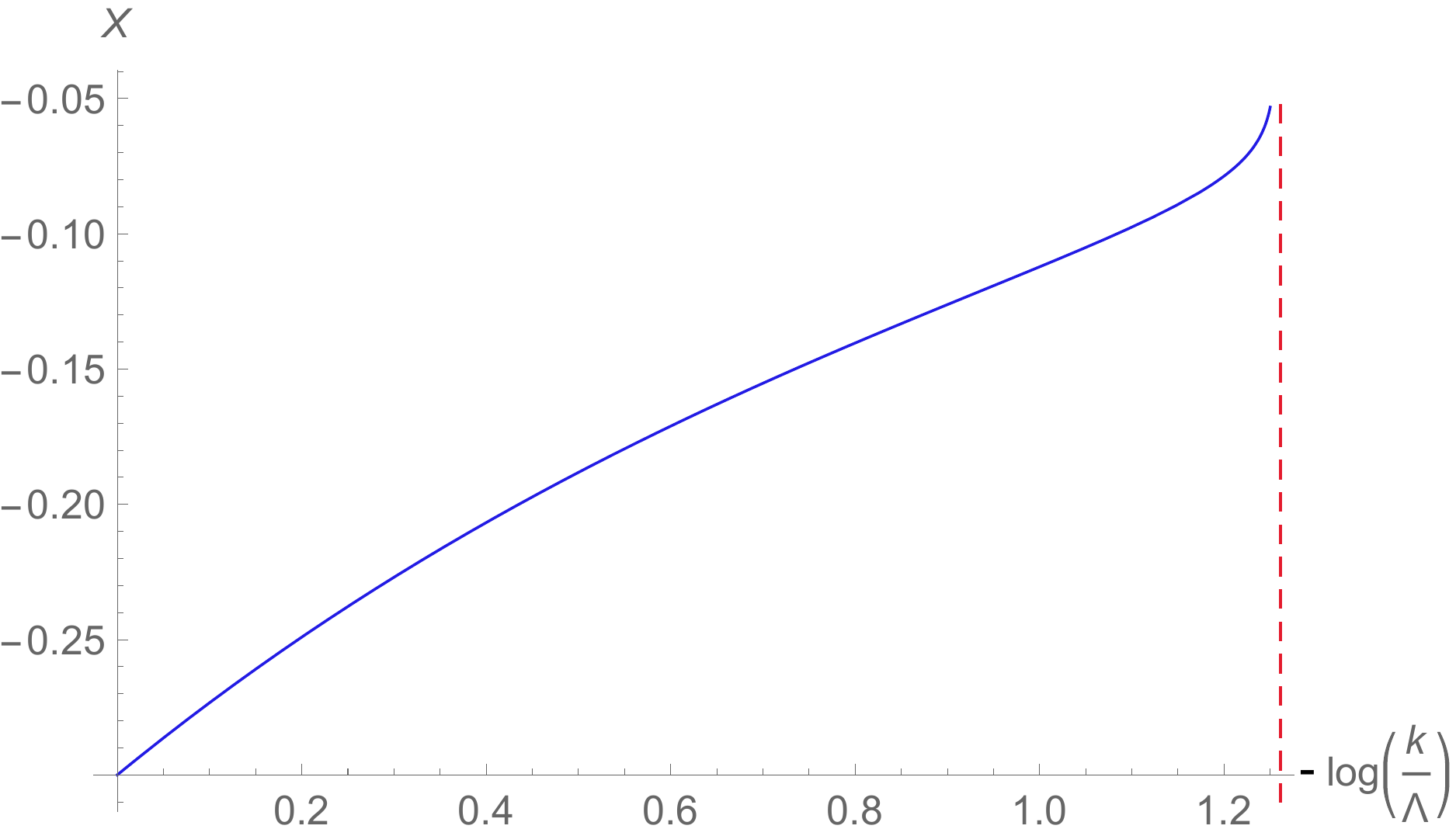}
\end{center}
\caption{On left: Finite scale singularity at large N of the eigencoupling $\bar{Y}$ along a trajectories starting in the vicinity of the fixed point FP1 (the red trajectories on Figure \ref{FlowMatLargeN}). On right, the behavaior of the eigencoupling $X$ along the same trajectory. }\label{FiniteTimeDiv}
\end{figure}

\begin{figure}
\begin{center}
\includegraphics[scale=0.52]{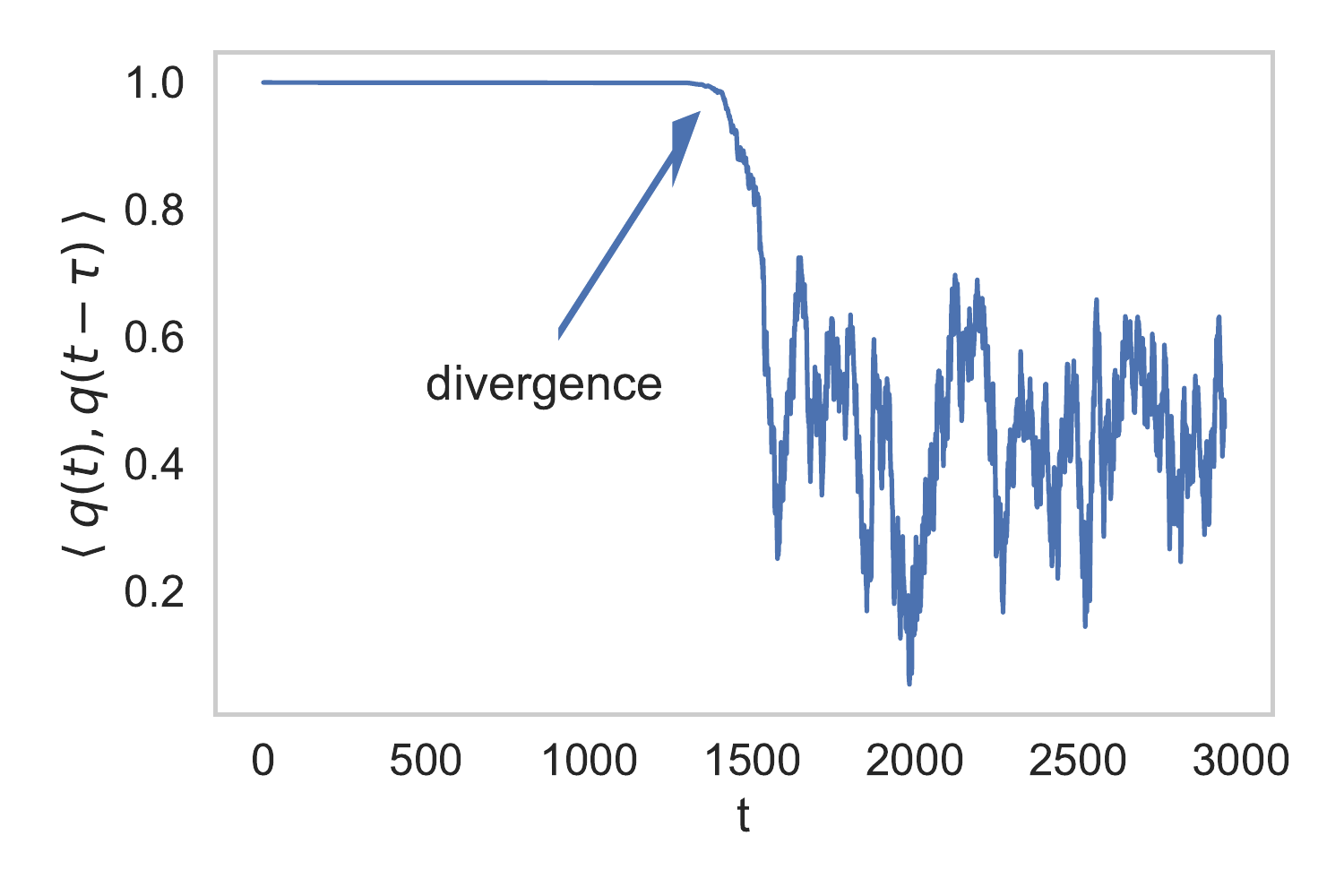}  \includegraphics[scale=0.52]{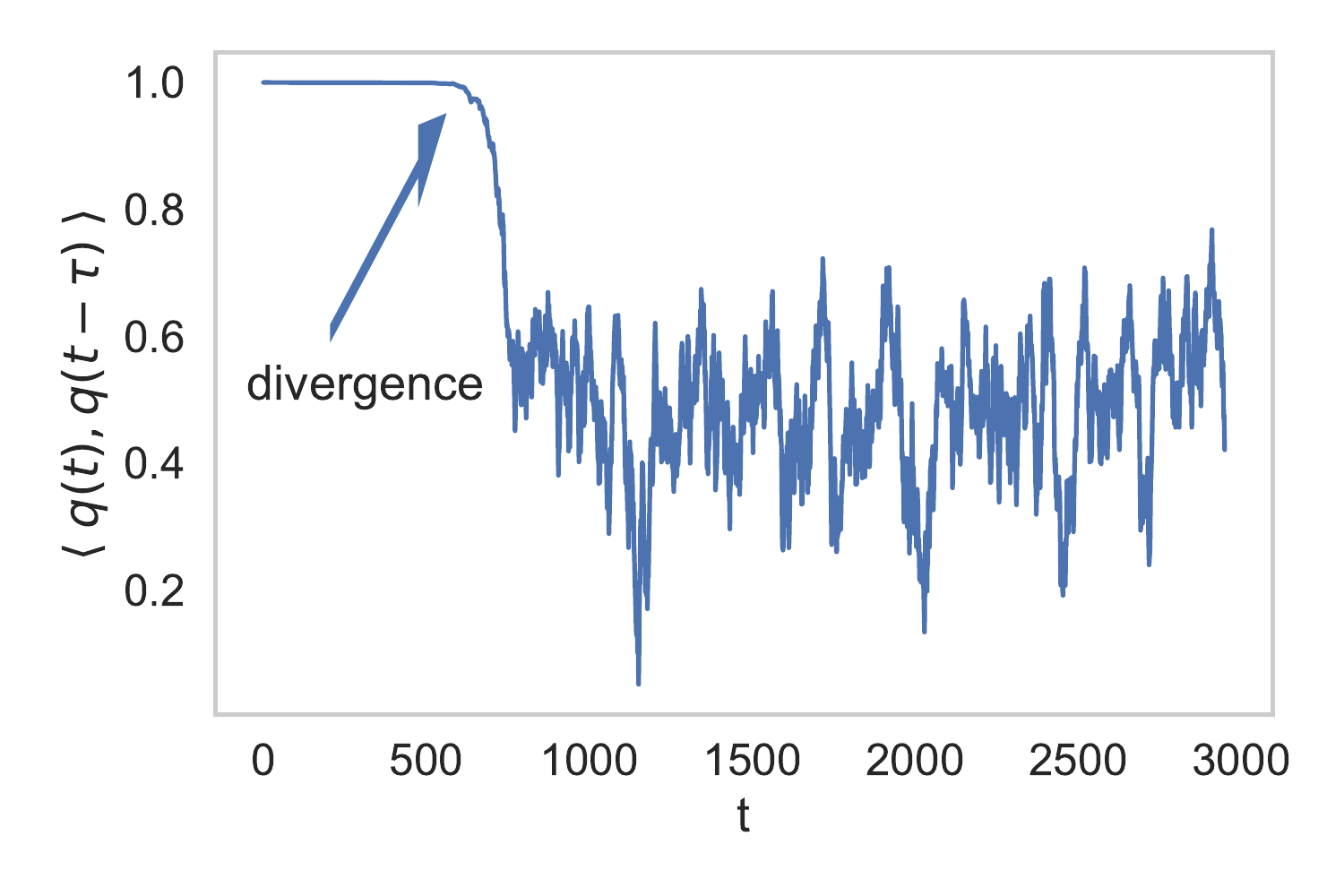}
\end{center}
\caption{Simulation of the two-point correlation function $\langle q(t) , q(t-\tau) \rangle$ in function of the time for $N=200$ and $\tau=50$. On the left: $D=0.01$ and on the right $D=0.1$. We observe a divergence of the two point correlation of $q$ which seems to be related to the singularity identified theoretically.}\label{Divergenceq}
\end{figure}

\noindent
For numerical investigations, we use the same strategy as for tensors, but in this case we will consider also solutions for finite $N$. The fixed point solution can be investigated by triangulation; using the first equation, and assuming $u_{41}\neq 0$, we may extract $\bar{u}_{42}\equiv \bar{u}_{42}(\bar{u}_2,\bar{u}_{41})$, solving the second solution we get finally: $\bar{u}_{41}\equiv \bar{u}_{41}(\bar{u}_2)$, $\bar{u}_{42}\equiv \bar{u}_{42}(\bar{u}_2)$; and the fixed point solutions can be found by investigating the zero of the function $F(\bar{u}_{2})=\dot{\bar{u}}_{42}(\bar{u}_2,\bar{u}_{41}(\bar{u}_{2}),\bar{u}_{42}(\bar{u}_2))$. Figure \ref{figFPMAT} shows zeros of $F$ for different $N$. For $N$ large enough, we get two zeros, labeled respectively as FP1 and FP2 on the Figure. The value reached for $\bar{u}_{42}$ at $FP2$ is positive, has the wrong sign, and can be interpreted as a disorder effect, whereas $\bar{u}_{41}$ is almost zero. Moreover, the flow equations show that $\dot{\bar{u}}_{42}\propto {\bar{u}}_{42}$; and the flow cannot reach the region $u_{42}<0$ from the region $u_{42}>0$. Critical exponents at FP1 are  $\Theta_1:=(-2.8, 2.0, 1.2)$, and $\bar{u}_{42}\approx 0^-$, $\bar{u}_{41}\approx 0.07$. It has thus one irrelevant and two relevant directions; eigenvectors $(v_1,v_2,v_3)$ being essentially parallel to the original axis $(u_2,u_{41},u_{42})$ around Gaussian fixed point:
\begin{align}
v_1&\sim (-0.17, 0.73, 0.65)\\
v_2&\sim (0,0,1.0)\\
v_3&\sim (0.29, 0.32, 0.90)\,.
\end{align}
Interestingly, $u_{42}$ is one of the unstable direction. Figure \ref{FlowMatLargeN} (on left) shows the behavior of the RG flow for large $N$ in the plan  crossing through FP1, FP2 and the Gaussian fixed point (on left). The figure on the right shows the behavior of the RG flow in the plan spanned by the two relevant directions of FP1, the \textit{eigencouplings} $(X,Y)$ being defined by projection along the directions $v_2$ and $v_3$ (suitably orthogonalized by standard Schmidt procedure):
\begin{align}
X&\approx 0.97 u_2+0.23u_{41}\,,\\
Y&\approx 0.15u_2-0.64 u_{41}+0.75u_{42}\,,
\end{align}
the remaining coordinate $Z$ along the irrelevant direction $v_1$ being set to its value at the fixed point. 
To understand the role played by the fixed point FP1, it is instructive to investigate the behavior of the couplings along a typical trajectory, as we did for $p=3$. Figure \ref{FiniteTimeDiv} shows the behavior of the eigencouplings $Y$ and $X$ on the plan of Figure \ref{FlowMatLargeN} along the red trajectory. As for $p=3$, we show that the trajectory exhibit a singularity at a finite time scale.  This behavior, typical of disordered systems \cite{Gredat_2014, Tissier_2011,Tissier_2012} is, as for the tensor cases interpreted as the signal that supersymmetry and equilibrium conditions cannot survive for arbitrary large time scale, the lack of analyticity ensuring that Ward identities fail to be integrated along RG trajectories. In order to connect this singularity to the concrete behavior of a vector obeying the Langevin dynamics in Equation \ref{eq1}, we plotted in Figure \ref{Divergenceq} the two-point correlation in function of the time for  $N=1000$ (considered as large N) and $\tau=50$. We observe the existence of two phases, a phase of convergence where the $2$-points correlation is $\mathcal{O}(1)$, that is followed by a phase where the trajectory becomes divergent. We see in the Figure \ref{Divergenceq} that the divergent phase appears later for $D=0.01$ (left) than for $D=0.1$ (right), which is in accordance with the previous insight that a smaller $D$ leads to a later divergence. Such a behavior moreover seems to be a consequence of the large $N$ limit, numerical integration of the flow for small $N$ showing, on the contrary, good convergence properties, as we can see on Figure \ref{FiniteTimeDiv} (on right).

\begin{figure}
\begin{center}
\includegraphics[scale=0.4]{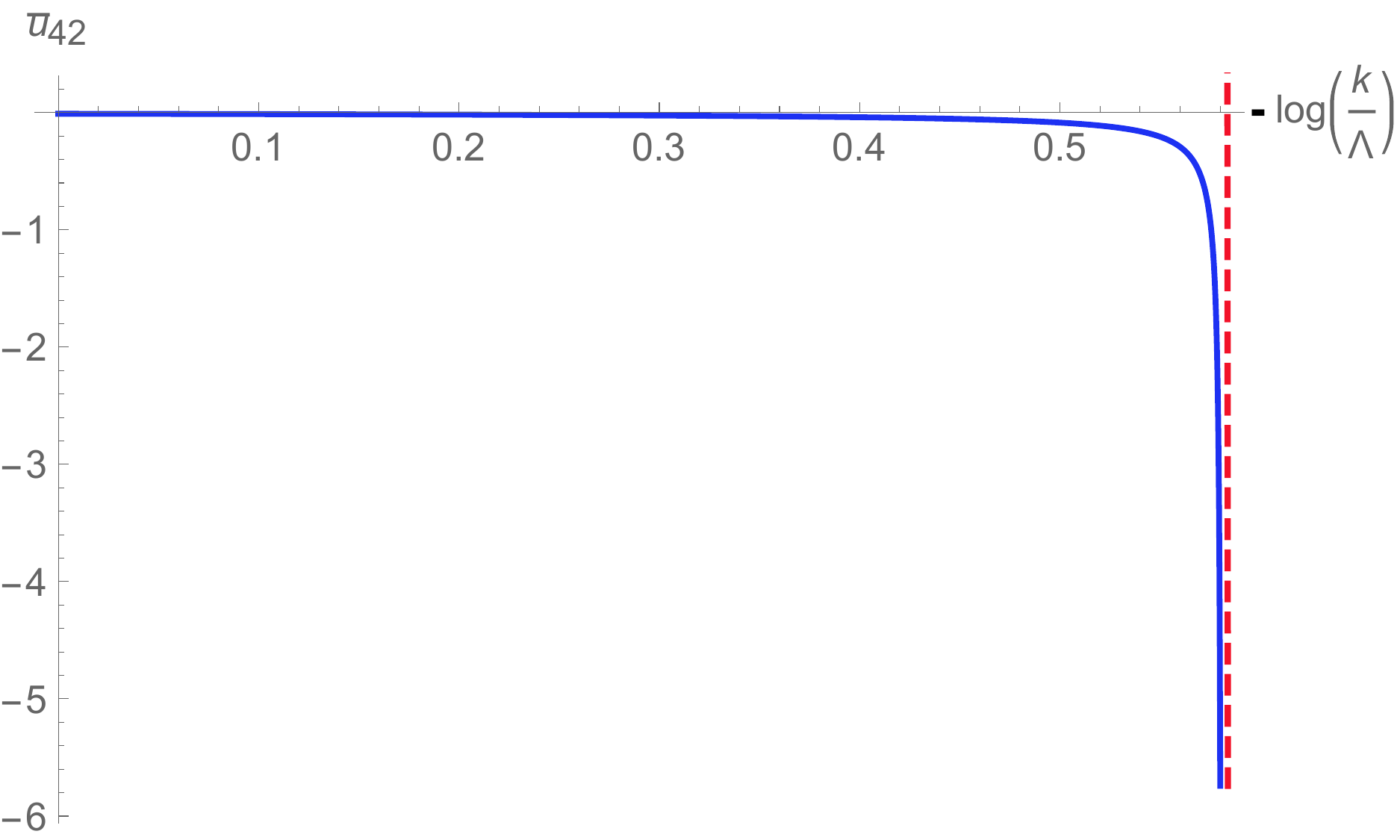} \quad \includegraphics[scale=0.4]{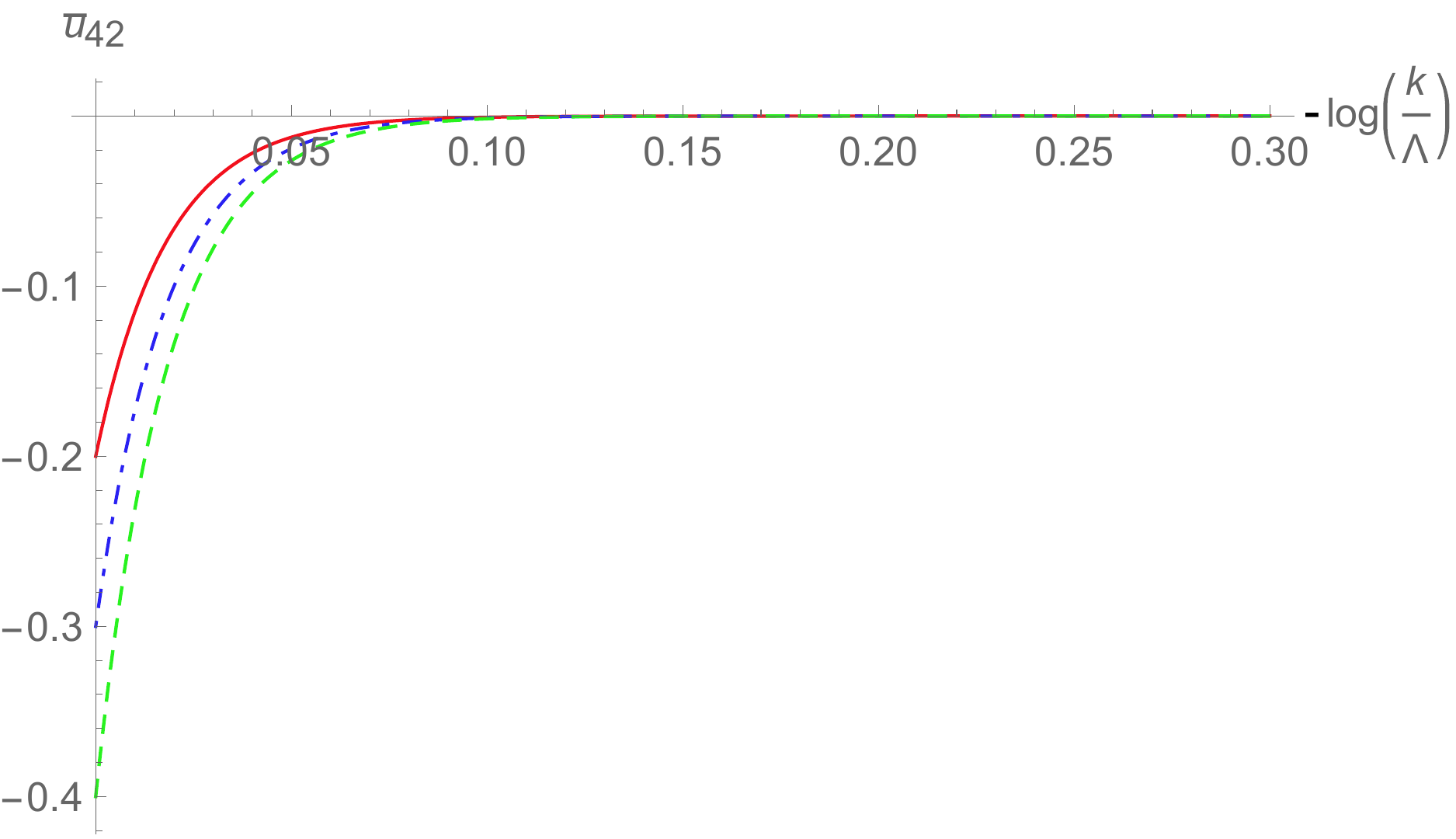}
\end{center}
\caption{On the left: the behavior of the disorder coupling $\bar{u}_{42}$ in the limit $N\to \infty$ which becomes singular at finite time scale. On the right: the behavior of the disorder coupling for $N=100$ for three different initial conditions for $u_{42}(\Lambda)$ (the remaining couplings being initialized in the vicinity of the non Gaussian fixed point FP1).}\label{FiniteNcurves}
\end{figure}

\section{Conclusions and outlooks}\label{sec6}

In this work, we provided a nonperturbative functional renormalization group framework based on a coarse-graining in time for a special kind of dynamical systems described by disordered Langevin equations.  These results aim to bring new insights and expand our knowledge on the dynamics of gradient descent (in particular Langevin dynamics) within high dimensional non convex landscapes similar to glassy systems. The understanding of these dynamics in these landscapes is central and essential for multiple physical and concrete applications, such as biological \cite{Biolgarel1988mean} and neuroscience systems \cite{Neurobick2020understanding}, material physical \cite{materkotliar2004strongly}, optimization in data science \cite{optimorland1985mean}, machine learning algorithms \cite{machmei2019mean}, etc. We focused on the equilibrium regime and constructed explicit supersymmetric approximate solutions within a self-consistent framework based on a multi-local expansion. Keeping interactions up to order two in that expansion, we were able to construct flow equations describing the evolution of effective couplings. Planning to provide a deeper numerical investigation of the resulting equation in a companion work, we mainly studied the case of a centered disorder distribution and for a minimal truncation in the large $N$ limit. We moreover essentially studied the symmetric phase expansion on the $p=3$ and $p=2$ models, where the disorder is respectively incarnated as a rank 3 random tensor and as a random matrix. In both cases, we showed that effective potential becomes nonanalytic at a finite time scale. This singular behavior implies that Ward-Takahashi identities fail to be continued beyond this point, and supersymmetry breaks down. Because of the close relation between supersymmetry and time reversal symmetry, we interpret this singular behavior as the signal of breaking of ergodicity, where the equilibrium description fails to be suitable for the system. Besides these similarities, the $p=2$ and $p=3$ distinguish from the nature of the transition. In the $p=2$ case, we showed that the transition is controlled by a non-Gaussian fixed point whose disorder coupling is one direction of instability. In contrast, this fixed point disappears at a finite time scale for $p=3$, and the transition must be of the first order. These observations moreover seem to be in accordance with literature \cite{Billoire_2005,Cavagna_2009,Yeo_2020} that used other methods. However, our conclusions at this stage remain very dependent on the approximations that we have used, and we have already planned several follow-ups to go further. For example, computations that we considered could be quite easily automatized, i.e. performed by a computer program; allowing to investigate larger multi-locals truncations. Another interesting avenue for the case $p=2$ would be to work in the space of the eigenvalues of the disorder, the distribution of which at this limit is described by the distribution of Wigner \cite{Potters1, Wigner1}. In action, the disorder term will then behave like an effective moment term, for a $1+1$ dimensional (super-)field theory with finite volume. One could thus consider a division into scale at the same time in frequencies and eigenvalues. All these questions and more will be addressed in our next work. Note that the case of non-equilibrium process and taking into account the case where $N$ is finite will be considered in forthcoming investigation.  \\

\noindent
\textbf{Acknowledgments:} V.L thanks Laetitia Mercey for its advices on the form of the manuscript.

\newpage
\begin{appendices}

\section{Higher order flow equations: diagrammatic method}\label{AppA}

In this section we present the diagrammatic method discussed in the paper to obtain higher order flow equations. 

\subsection{Naive power counting}
Let us consider an interaction with coupling $g$, made as a product of $n$ fields $\phi$, $m$ fields $\bar{\varphi}$ and $l$ time integrals. To be dimensionless, the dimension of the coupling $g$ must be equals to:
\begin{equation}
[g]=\frac{1}{2}n-\frac{1}{2}m+l\,.
\end{equation}
Note that if each time integral is the remaining part of an integral over the supercoordinate $z$, $l$ must be equal to $m$,
\begin{equation}
[g]=\frac{1}{2} (n+m) > 0\,.
\end{equation}
In the language of field theory, this means that all the interactions are relevant.

\subsection{Graphical representation for higher flow equations}
\begin{figure}
\includegraphics[scale=1]{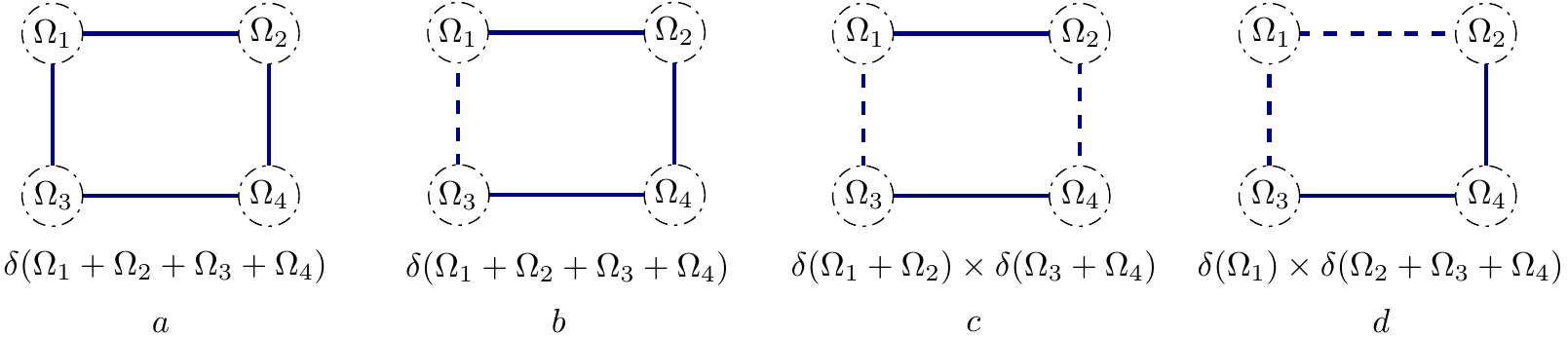}
\caption{Local structure of a typical effective bubble graph build of four vertices and four effective edges, the local ones being materialized with a solid blue edge whereas the non-local ones are labeled with a dashed blue edge. Each vertex is labeled with the total external momenta $\Omega_i$. For effective bubble graphs $a$ and $b$, all the external momenta satifies a global conservation: $\Omega_1+\Omega_2+\Omega_3+\Omega_4=0$. For effective bubble graphs $c$ and $d$ however, external momenta partition in two disjoint conserved sets. Then, $\Omega_1=0$ and $\Omega_2+\Omega_3+\Omega_4=0$ for graph $c$, $\Omega_1+\Omega_2=0$ and $\Omega_3+\Omega_4=0$ for graph $d$.}\label{figlocal}
\end{figure}

For the derivation of the higher flow equations, it is suitable to use of a graphical representation, like the one used for equations \eqref{graph1} and \eqref{graph2}. To this end, we introduce the following rules:

\begin{itemize}
\item To each field $\psi$, $\bar{\psi}$, $\phi$ and $\bar{\varphi}$ we associate respectively the symbols $\vartriangle$, $\blacktriangle$, {\large$\bullet$} and {\tiny{$\blacksquare$}}.

\item To each sum as $\sum_{i} \mathcal{M}_{\alpha\,i}:= \mathcal{M}_{\alpha\,\bullet}$ we associate an isolated vertex of type corresponding to the index $\alpha$.

\item To each scalar product of type $\mathcal{M}_\alpha \cdot \mathcal{M}_\beta$ we associate a solid edge bounded by vertices corresponding to the indices $\alpha$ and $\beta$.

\item The vertex corresponding to fields sharing the same superpath coordinate $z$ are surrounded with a closed dotted path materializing the integral $\int dz$. We call bubbles these closed paths.
\end{itemize}
\begin{figure}
\begin{align*}
\includegraphics[scale=1]{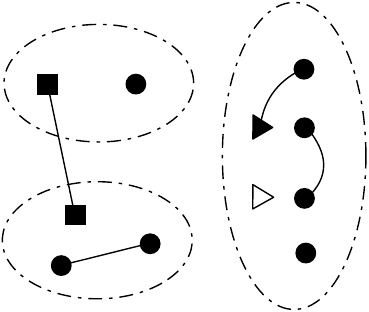}
\end{align*}
\caption{A typical example of bubble graph, corresponding to an interaction involving 11 fields. Note that a bubble cannot contain more than one black square or a pair of triangles.}\label{figgraphs1}
\end{figure}
We call \textit{local bubble graph} such a construction and Figure \ref{figgraphs1} provides an example of such a bubble graph. As an illustration, let us consider the coupling $J= \int dz dz^\prime \mathcal{M}_{\bullet}^2(z) \big(\mathcal{M}(z)\cdot \mathcal{M}(z^\prime)\big) \mathcal{M}_{\bullet}(z^\prime)$ involved in the truncation \eqref{truncation4}. We have five vertices, two of them linked with a solid edge and the others being isolated. These vertices moreover leave in the interior of two distinct bubbles, corresponding to the integrals $\int dt$ and $\int dt^\prime$. Using this graphical representation, $J$ decomposes as:
\begin{align}
\nonumber J&=\vcenter{\hbox{\includegraphics[scale=0.7]{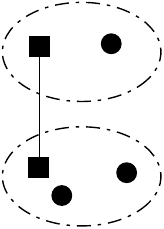} }}+\vcenter{\hbox{\includegraphics[scale=0.7]{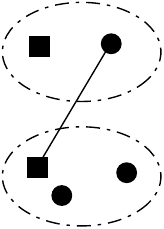} }}+2\vcenter{\hbox{\includegraphics[scale=0.7]{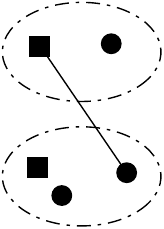}}}+2\vcenter{\hbox{\includegraphics[scale=0.7]{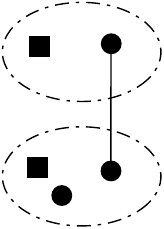} }}\\\nonumber
&+2\vcenter{\hbox{\includegraphics[scale=0.7]{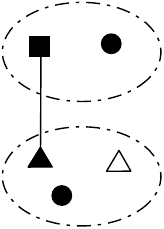} }}+2\vcenter{\hbox{\includegraphics[scale=0.7]{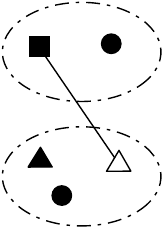} }}
+2\vcenter{\hbox{\includegraphics[scale=0.7]{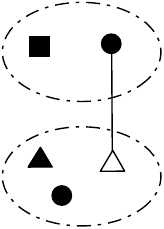} }}+\vcenter{\hbox{\includegraphics[scale=0.7]{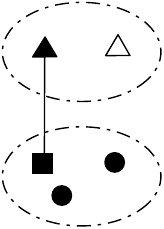} }}
\\\nonumber
&+\vcenter{\hbox{\includegraphics[scale=0.7]{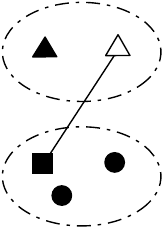} }}+2\vcenter{\hbox{\includegraphics[scale=0.7]{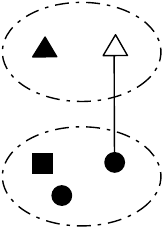} }}+2\vcenter{\hbox{\includegraphics[scale=0.7]{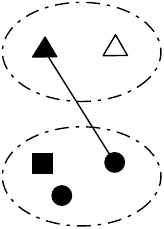} }}+2\vcenter{\hbox{\includegraphics[scale=0.7]{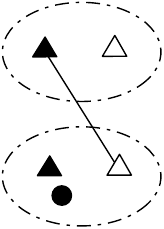} }}\\
&+2\vcenter{\hbox{\includegraphics[scale=0.7]{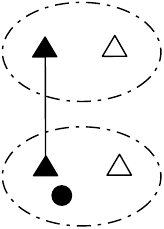} }}+2\vcenter{\hbox{\includegraphics[scale=0.7]{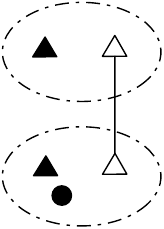} }}+2\vcenter{\hbox{\includegraphics[scale=0.7]{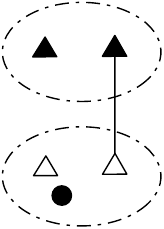} }}+2\vcenter{\hbox{\includegraphics[scale=0.7]{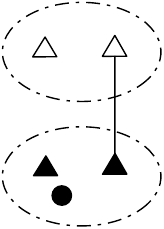} }}\,.
\end{align}
where the numerical coefficient in front of each diagram's counts the number of equivalent configurations. In the rest of this paper, we use this graphical notation to represent as well as the effective interactions and the corresponding effective vertices $\Gamma^{(n)}_k$, the context avoiding ambiguities. Using this graphical notation, flow equations will take the form of an equality between sums of graphs. On the left-hand side the bubble graphs involved in the truncation, weighted with $\beta$-functions, and on the right-hand side a series of graphs built as  bubble graphs contracted with effective propagators. We will denote as \textit{effective edges} the effective propagators, and to simplify the notations we materialize them as a solid thick edge. Following these rules, the first term on the right-hand side of equation \eqref{graph2} must we read as:
\begin{equation}
\vcenter{\hbox{\includegraphics[scale=1]{oneloop2.pdf} }}\equiv \vcenter{\hbox{\includegraphics[scale=1]{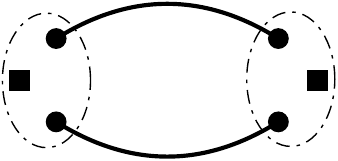} }}\,. \label{effectiveloop}
\end{equation}
By contrast, we call such a diagram an \textit{effective bubble graph}. Note that we have to keep in mind that into a loop, one of the solid thick edges corresponds to the propagator $G_k\dot{\textbf{R}}_k G_k$ when the others correspond only to $G_k$. The next step is to define a rule allowing to identify on both sides the contributions corresponding to the same $\beta$-function. To this end, we notice that the components of the effective propagators are of two kinds (see \eqref{decomplocal}), either \textit{local}, proportional to $\delta(\omega+\omega^\prime)$ or \textit{non-local}, proportional to $\delta(\omega)\delta(\omega^\prime)$. Now, let us consider an effective bubble graph involving a single loop of length $n$: $\mathcal{L}_n=\{\ell_1,\cdots, \ell_n\}$. Each of these edges $\ell_i$ can be local or non-local. If all of them are local, the resulting effective bubble graph is local as well (i.e. all external momenta are constrained by a single global Dirac delta). From connectivity, if one of these edges is non-local, the chain of local edges build a spanning tree, and the resulting diagram remains local. If at least two effective edges become non-local however, the resulting effective bubble graph becomes a product of local effective bubble graphs (i.e. the external momenta splits into two disjoint sets, both of them constrained by a Dirac delta). See Figure \ref{figlocal}. We denote as \textit{local components} each sub-local bubble graph. Finally, each of these local components can be expanded in powers of external momenta, the first term of this expansion corresponding to the \textit{ultra-local} approximation.
\medskip

As an example, let us consider the effective loop \eqref{effectiveloop}. In this equation, the loop depends on the external momenta $\Omega$ (see equation \eqref{gamma33}), shared by the black squares, and do not correspond with a local bubble graph. From the previous discussion, we must have, formally:
\begin{equation}
\vcenter{\hbox{\includegraphics[scale=0.9]{oneloop22.pdf} }}\equiv K_1(\Omega)\delta(\Omega+\Omega^\prime) +K_2 \delta(\Omega)\delta(\Omega^\prime)\,,
\end{equation}
where $K_1$ denotes the local component built with local edges. Expanding the effective loop around vanishing external momenta $\Omega$ to obtain the ultra-local approximation, we get:
\begin{equation}
K_1(\Omega)= K_1(0)+ \Omega K_1^\prime(0)+ \mathcal{O}(\Omega^2)\,,
\end{equation}
and the ultra-local approximation of the effective bubble graph follows:
\begin{equation}
\vcenter{\hbox{\includegraphics[scale=0.9]{oneloop22.pdf} }}=K_2\times \left\{ \vcenter{\hbox{\includegraphics[scale=1]{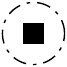} }} \vcenter{\hbox{\includegraphics[scale=1]{vertexsingle.pdf}}}\right\} +K_1(0) \,\times\, \vcenter{\hbox{\includegraphics[scale=1]{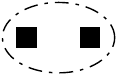} }}+ \mathcal{O}(\Omega)\,.
\end{equation}
Note that as soon as we projected into the superfield components, the bubbles materialize time rather than supercoordinate integrations. In general, we introduce the definition:
\begin{definition}
Let us consider a graph $\mathcal{G}=\{\mathcal{G}_\alpha,\ell_a\}$ built as a set of bubble graphs $\mathcal{G}_\alpha$ and effective edges $\ell_a$. The boundary graph $\partial \mathcal{G}$ of $\mathcal{G}$ is the bubble graph obtained from $\mathcal{G}$ by:
\begin{enumerate}
\item Deleting all the effective edges $\{\ell_a\}$ as well as their boundary vertices.
\item We merge all the bubbles linked by at least one effective line.
\end{enumerate}
\end{definition}
Within this definition, it is clear that the boundary bubble graph of $\mathcal{G}$ is nothing but the local approximation of $\mathcal{G}$. We thus define the following projection rule: From a given flow equation involving a few numbers of effective bubble graphs on the right-hand side, we must have to:
\begin{enumerate}
\item Identify on both sides coefficients in front of graphs having the same boundary bubble graph.
\item Expands the loops in the power of the external momenta and keeping only the zero-order term of the expansion.
\end{enumerate}
As a first application of the graphical method that we propose, we consider the flow equation for the coupling $h$, which behaves like a mass term in the field theory vocabulary. From equation \eqref{floweq2pts1}, considering the pair $\bar{\varphi}\phi$ for external fields; we get:
\begin{equation}
\big(\dot{\Gamma}_k^{(2)}\big)_{\bar{\varphi}_i\phi_j}=i \dot{h} \delta_{ij} \delta(\omega+\omega^\prime) + \mathcal{O}(\omega,\mathcal{M}),
\end{equation}
Thus, keeping only diagrams having the same boundary graphs on both sides, we arrive to the equation:
\begin{equation}
\dot{h}\, \vcenter{\hbox{\includegraphics[scale=0.9]{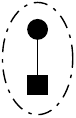} }} = -\frac{u_4^{(1)}}{2N} \Bigg\{\vcenter{\hbox{\includegraphics[scale=0.8]{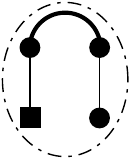}}} +\vcenter{\hbox{\includegraphics[scale=0.8]{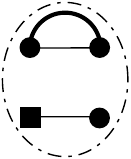}}} \Bigg\}-i \frac{u_4^{(2)}}{2N} \Bigg\{\vcenter{\hbox{\includegraphics[scale=0.8]{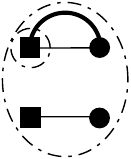}}} +\vcenter{\hbox{\includegraphics[scale=0.8]{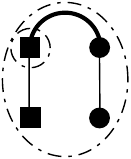}}} \Bigg\}+ \frac{2i(u_3^{(2)})^2}{N^2}\, \vcenter{\hbox{\includegraphics[scale=0.8]{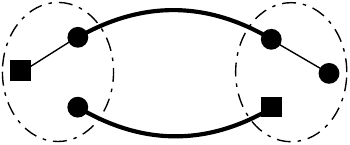}}}\,,
\end{equation}
where on the left-hand side we assume to keep only diagrams having the same boundaries as the bubble graph on the right-hand side. Note moreover that we discarded the combinatorial coefficients in front of each bubble graphs for convenience. The first two diagrams have been computed previously, we get:
\begin{equation}
\vcenter{\hbox{\includegraphics[scale=0.7]{vertex411.pdf}}} = \frac{16}{k} (I_2 +2J_2^{(1)})\,,\quad \vcenter{\hbox{\includegraphics[scale=0.7]{vertex412.pdf}}} = \frac{8N}{k}(\bar{I}_2+2\bar{J}_2^{(1)})+\frac{16}{k}\bar{J}_2^{(2)}\,,
\end{equation}
\begin{equation}
\vcenter{\hbox{\includegraphics[scale=0.7]{vertex413.pdf}}}=- \frac{2i\alpha N}{Z_k k} \frac{1+\eta_k}{(\bar{h}+\alpha)^2}\,,\quad \vcenter{\hbox{\includegraphics[scale=0.7]{vertex414.pdf}}}=- \frac{2i \alpha}{Z_k k} \frac{1+\eta_k}{(\bar{h}+\alpha)^2}\,,
\end{equation}
and finally:
\begin{equation}
\vcenter{\hbox{\includegraphics[scale=0.7]{vertex331.pdf}}}=-\frac{4iN}{Z_k^2k^2} (\bar{I}_3^\prime+ 2 \bar{J}_3^{\prime (1)})\,,
\end{equation}
where:
\begin{equation}
\bar{I}_3^\prime=iZ_k^3\int \frac{dx}{2\pi}(\eta_k \rho^{(2)}(x)+\dot{\rho}^{(2)}(x)) (G_{1,\bar{\varphi}\phi}(x))^2 G_{1,\bar{\varphi}\phi} (-x)\,, \label{Iprime3}
\end{equation}
and:
\begin{equation}
\bar{J}_3^{\prime (1)} =Z_k^3\int \frac{dx}{2\pi}(\eta_k \rho^{(1)}(x)+\dot{\rho}^{(1)}(x))  \vert G_{1,\bar{\varphi}\phi}(x))\vert^2(g_{1,\phi\phi}^{(1)}(x)+2\pi \bar{l}_{1,\phi\phi}^{(1)} \delta(x))\,.\label{Jprime31}
\end{equation}
For the dimensionless coupling $\bar{h}:= h/k$, we thus obtain:
\begin{align}
\nonumber\dot{\bar{h}}=-(1+\eta_k)\bar{h}- \frac{4\bar{u}_4^{(1)}}{N}& \left[ (2+N)(\bar{I}_2 +2\bar{J}_2^{(1)})+2 \bar{J}_2^{(2)} \right]\\
&-\frac{\alpha \bar{u}_4^{(2)}}{N} (N+1) \frac{1+\eta_k}{(\bar{h}+\alpha)^2}+\frac{8(\bar{u}_3^{(2)})^2}{N}(\bar{I}_3^\prime+ 2 \bar{J}_3^{\prime (1)})\,.
\end{align}
As another explicit example of the use of this diagrammatic technique, let us consider the flow equations for $3$-point functions. From the exact flow equation, we easily deduce:
\begin{align}
\dot{\Gamma}_k^{(3)}= \STr \dot{\tilde{\textbf{R}}}_k G_k \bigg(&-\frac{1}{2} \Gamma^{(5)}_k +3\Gamma^{(4)}_kG_k\Gamma_k^{(3)}-3 \Gamma^{(3)}_kG_k \Gamma_k^{(3)}G_k\Gamma_k^{(3)} \bigg)G_k\,.
\end{align}
Using truncation \eqref{truncation4} and focusing on the component $\Gamma_{k, \phi\phi \bar{\varphi}}^{(3)}$, we get essentially two kinds of contributions, whose boundary graphs correspond respectively to couplings $u_3^{(1)}$ and $u_3^{(2)}$. Note that bosonic and fermionic loops cancels must cancel exactly, like $L_B$ and $L_F$ in equations \eqref{LB} and \eqref{LF}. For this reason we must have the diagrammatic equality:
\begin{equation}
\vcenter{\hbox{\includegraphics[scale=0.8]{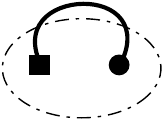} }}+\vcenter{\hbox{\includegraphics[scale=0.8]{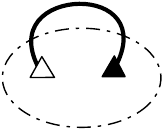} }}\equiv 0\,.
\end{equation}
For the remaining couplings, we get the diagrammatic expansions:
\begin{align}\label{x23}
\nonumber i \frac{\dot{u}_3^{(2)}}{N}\, \vcenter{\hbox{\includegraphics[scale=0.7]{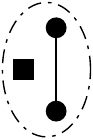} }} &=\frac{1}{2} \frac{u_5^{(1)}}{N^3} \, \vcenter{\hbox{\includegraphics[scale=0.6]{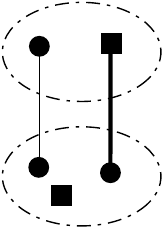} }} - \frac{iu_5^{(3)}}{2N^3}\, \vcenter{\hbox{\includegraphics[scale=0.7]{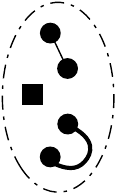} }}- \frac{iu_5^{(2)}}{2N^3} \Bigg\{ \vcenter{\hbox{\includegraphics[scale=0.7]{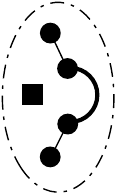} }} +\vcenter{\hbox{\includegraphics[scale=0.7]{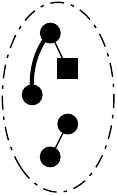} }} \Bigg\}\\\nonumber
&+\frac{9i(u_3^{(2)})^3}{N^3}\Bigg\{ \vcenter{\hbox{\includegraphics[scale=0.6]{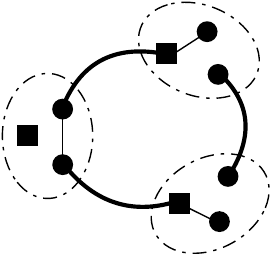} }}+\vcenter{\hbox{\includegraphics[scale=0.6]{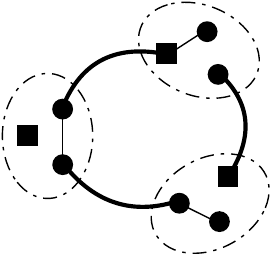} }} \Bigg\}-\frac{3 u_4^{(1)}u_3^{(1)}}{N^3} \, \vcenter{\hbox{\includegraphics[scale=0.6]{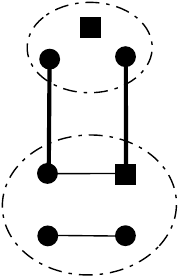} }} \\\nonumber
&-\frac{3 u_4^{(1)}u_3^{(2)}}{N^2} \Bigg\{ \vcenter{\hbox{\includegraphics[scale=0.6]{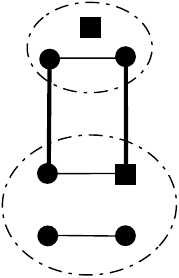} }}+\vcenter{\hbox{\includegraphics[scale=0.6]{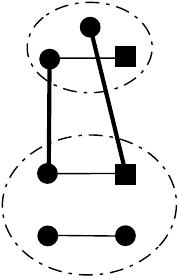} }}+\vcenter{\hbox{\includegraphics[scale=0.6]{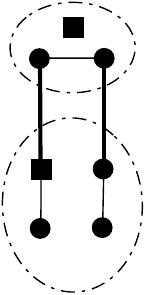} }}\Bigg\}-\frac{3i u_4^{(2)}u_3^{(1)}}{N^3} \, \vcenter{\hbox{\includegraphics[scale=0.6]{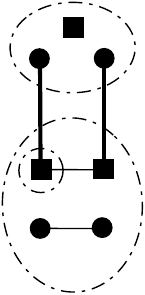} }}\\
& -\frac{3i u_4^{(2)}u_3^{(2)}}{N^3} \Bigg\{ \vcenter{\hbox{\includegraphics[scale=0.6]{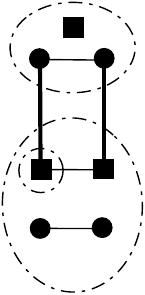} }}+\vcenter{\hbox{\includegraphics[scale=0.6]{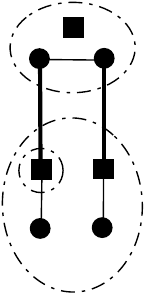} }} +\vcenter{\hbox{\includegraphics[scale=0.6]{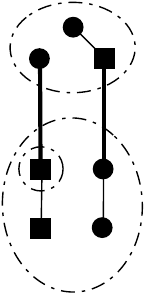} }} \Bigg\}\,.
\end{align}
Translating this  equation in formula, we get the flow equation of the coupling $\bar{u}_3^{(2)}$:
\begin{align}
\nonumber\dot{\bar{u}}_3^{(2)}=&-\frac{3}{2}(1+\eta_k){\bar{u}}_3^{(2)}-\frac{\bar{u}_5^{(1)}}{N} \frac{1+\eta_k}{(\bar{h}+\alpha)^2}-\frac{3\bar{u}_5^{(3)}}{N} [ \bar{I}_2+2\bar{J}_2^{(1)}+2J_2^{(2)}] \\\nonumber
&-\frac{4\bar{u}_5^{(2)}}{N}[\bar{I}_2+2\bar{J}_2^{(1)}+\bar{J}_2^{(2)}]+\frac{36 (\bar{u}_3^{(2)})^3}{N} [\bar{I}_4+3\bar{J}_4^{(1)}+\bar{J}_4^{(2)}]
\\\nonumber
&+\frac{72 \bar{u}_4^{(1)}\bar{u}_3^{(1)}}{N}[\bar{I}_3^\prime+2\bar{J}_3^{\prime (1)}+3\bar{J}_3^{\prime (2)}]+\frac{12 (N+3)\alpha \bar{u}_4^{(2)}\bar{u}_3^{(2)} }{N} \frac{1+\eta_k}{(\bar{h}+\alpha)^3}\\
&+\frac{24 \bar{u}_4^{(1)}\bar{u}_3^{(2)}}{N} \left[(2+N)(\bar{I}_3^\prime+3\bar{J}_3^{\prime (1)})+2\bar{J}_3^{\prime (2)} \right]+\frac{36\alpha \bar{u}_4^{(2)}\bar{u}_3^{(1)}}{N}\frac{1+\eta_k}{(\bar{h}+\alpha)^3}\,,
\end{align}
where:
\begin{equation}
J_3^{\prime (2)} =-{Z_k}^3 \int \frac{dx}{2\pi}(\eta_k \rho^{(1)}(x)+\dot{\rho}^{(1)}(x)) \vert G_{1,\bar{\varphi}\phi}(x) \vert^2(g_{1,\phi\phi}^{(2)}(x)+2\pi \bar{l}_{1,\phi\phi}^{(2)} \delta(x))\,,\label{Jprime32}
\end{equation}
\begin{equation}
\bar{I}_4=(Z_k)^4\int \frac{dx}{2\pi} \vert G_{1,\bar{\varphi}\phi}(x) \vert^4\Big(\eta_k \rho^{(2)}(x)+\dot{\rho}^{(2)}(x)\Big)\,,
\end{equation}
and:
\begin{align}
\bar{J}_4^{(1)}=&- (Z_k)^4\int \frac{dx}{2\pi} \Big(\eta_k \rho^{(1)}(x)+\dot{\rho}^{(1)}(x)\Big) \vert G_{1,\bar{\varphi}\phi}(x) \vert^2 \times G_{1,\bar{\varphi}\phi}(x) \Big( g_{1,\phi\phi}^{(1)}(x)+2\pi \bar{l}_{1,\phi\phi}^{(1)} \delta(x) \Big)\,, \label{J4}
\end{align}
\begin{align}
\bar{J}_4^{(2)}=&- (Z_k)^4\int \frac{dx}{2\pi} \Big(\eta_k \rho^{(1)}(x)+\dot{\rho}^{(1)}(x)\Big) \vert G_{1,\bar{\varphi}\phi}(x) \vert^2 \times G_{1,\bar{\varphi}\phi}(x) \Big( g_{1,\phi\phi}^{(2}(x)+2\pi \bar{l}_{1,\phi\phi}^{(2)} \delta(x) \Big)\,.
\end{align}

\begin{remark}\label{reality}
One can suspect that imaginary contributions occur due to the presence of factors $i$. However, it is not hard to check that purely imaginary contributions cancel for each loop. This comes from the observation that denominators can be always rewritten as even functions of $\omega$. Because each $\omega$ share a factor $i$, imaginary contributions are even in $\omega$ in the numerator and cancels by symmetry.
\end{remark}

In the same manner the diagrammatic representation of the flow from $u_3^{(1)}$ to $u_6$ are given by
\begin{align}
\nonumber &i \frac{\dot{u}_3^{(1)}}{N^2}\, \vcenter{\hbox{\includegraphics[scale=0.6]{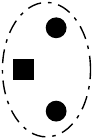} }} =\frac{1}{2} \frac{u_5^{(1)}}{N^3} \Bigg\{\, \vcenter{\hbox{\includegraphics[scale=0.5]{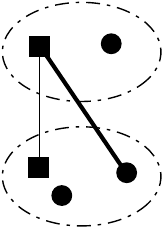} }}+\vcenter{\hbox{\includegraphics[scale=0.5]{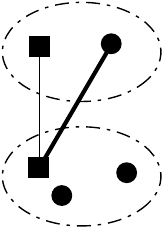} }}+\vcenter{\hbox{\includegraphics[scale=0.5]{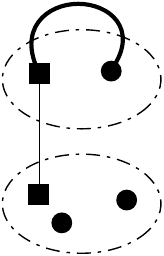} }}+\vcenter{\hbox{\includegraphics[scale=0.5]{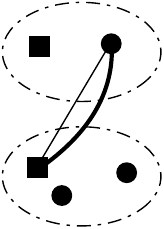} }}+\vcenter{\hbox{\includegraphics[scale=0.5]{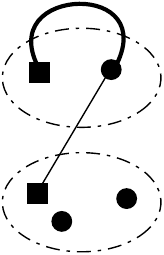} }}+\vcenter{\hbox{\includegraphics[scale=0.5]{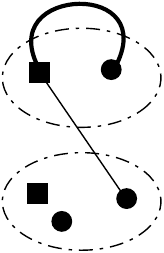} }}+\vcenter{\hbox{\includegraphics[scale=0.5]{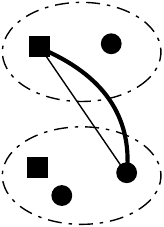} }} \\\nonumber
&+\vcenter{\hbox{\includegraphics[scale=0.5]{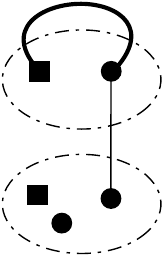} }}+\vcenter{\hbox{\includegraphics[scale=0.5]{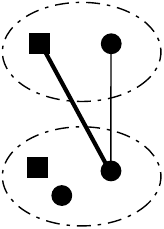} }}+\vcenter{\hbox{\includegraphics[scale=0.5]{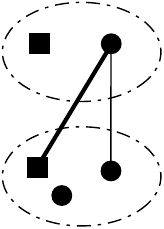} }}+\vcenter{\hbox{\includegraphics[scale=0.5]{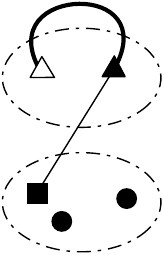} }}+\vcenter{\hbox{\includegraphics[scale=0.5]{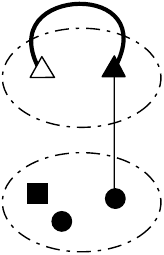} }}+\vcenter{\hbox{\includegraphics[scale=0.5]{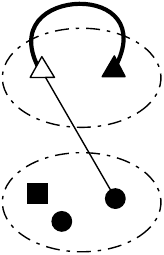} }}+\vcenter{\hbox{\includegraphics[scale=0.5]{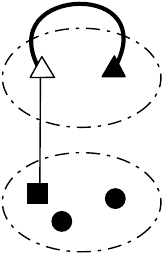} }} \Bigg\} - \frac{iu_5^{(2)}}{2N^2}\,\vcenter{\hbox{\includegraphics[scale=0.5]{figureMeta3pts52.pdf} }} \\\nonumber
&- \frac{iu_5^{(3)}}{2N^3}\, \Bigg\{ \vcenter{\hbox{\includegraphics[scale=0.5]{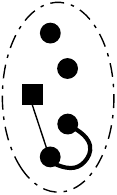} }}+\vcenter{\hbox{\includegraphics[scale=0.5]{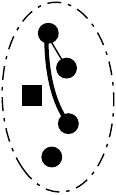} }} +\vcenter{\hbox{\includegraphics[scale=0.5]{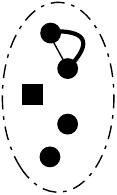} }} \Bigg\} +\frac{3i(u_3^{(1)})^3}{N^6} \, \Bigg\{ \vcenter{\hbox{\includegraphics[scale=0.5]{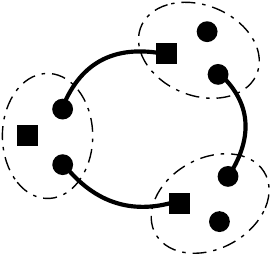} }} \Bigg\}+\frac{9i(u_3^{(1)})^2u_3^{(2)}}{N^5}\Bigg\{\vcenter{\hbox{\includegraphics[scale=0.5]{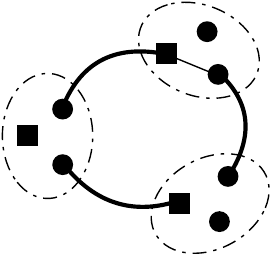} }} \\\nonumber
&+\vcenter{\hbox{\includegraphics[scale=0.5]{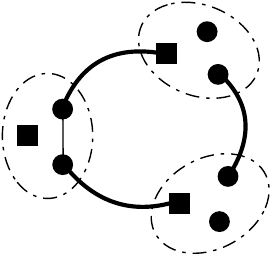} }}+\vcenter{\hbox{\includegraphics[scale=0.5]{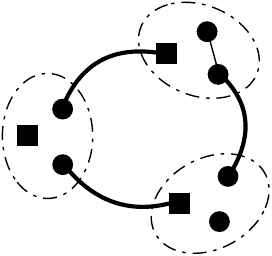} }}+\vcenter{\hbox{\includegraphics[scale=0.5]{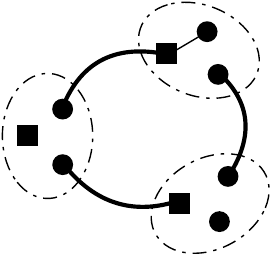} }}+\vcenter{\hbox{\includegraphics[scale=0.5]{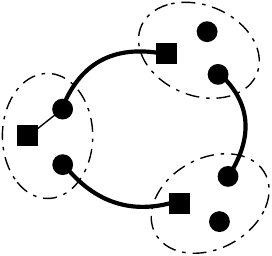} }} \Bigg\} +\frac{9i(u_3^{(2)})^2u_3^{(1)}}{N^4}\Bigg\{ \vcenter{\hbox{\includegraphics[scale=0.5]{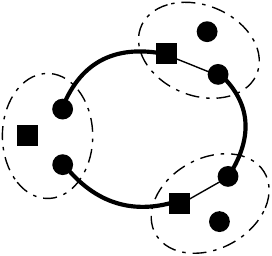} }}\\\nonumber
&+\vcenter{\hbox{\includegraphics[scale=0.5]{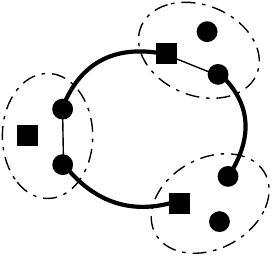} }}+\vcenter{\hbox{\includegraphics[scale=0.5]{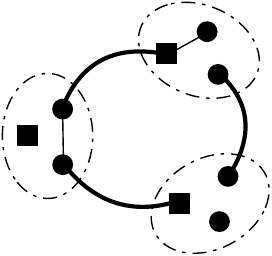} }} +\vcenter{\hbox{\includegraphics[scale=0.5]{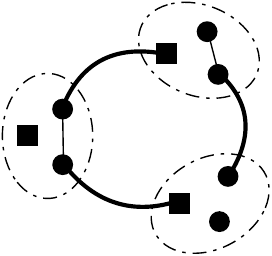} }} +\vcenter{\hbox{\includegraphics[scale=0.5]{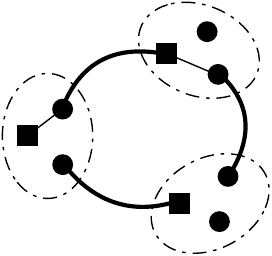} }}+\vcenter{\hbox{\includegraphics[scale=0.5]{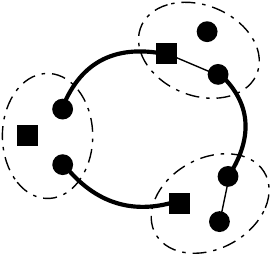} }} +\vcenter{\hbox{\includegraphics[scale=0.5]{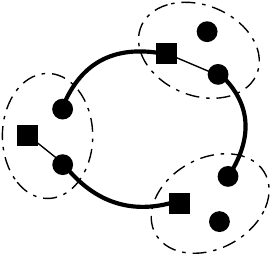} }}\\\nonumber
&+\vcenter{\hbox{\includegraphics[scale=0.5]{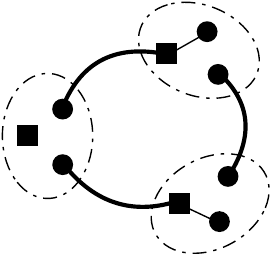} }} +\vcenter{\hbox{\includegraphics[scale=0.5]{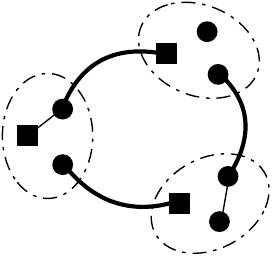} }} \Bigg\} +\frac{3i(u_3^{(2)})^3}{N^3} \Bigg\{\vcenter{\hbox{\includegraphics[scale=0.5]{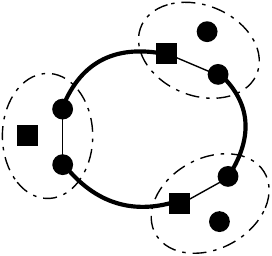} }}+\vcenter{\hbox{\includegraphics[scale=0.5]{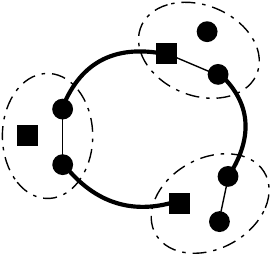} }}+\vcenter{\hbox{\includegraphics[scale=0.5]{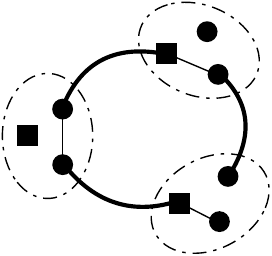} }}\\\nonumber
&+\vcenter{\hbox{\includegraphics[scale=0.5]{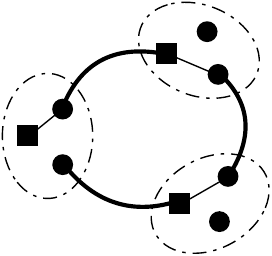} }}+\vcenter{\hbox{\includegraphics[scale=0.5]{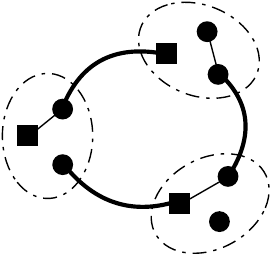} }} +\vcenter{\hbox{\includegraphics[scale=0.5]{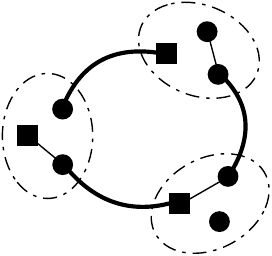} }}+\vcenter{\hbox{\includegraphics[scale=0.5]{figureMeta3pts13bis173.pdf} }}+\vcenter{\hbox{\includegraphics[scale=0.5]{figureMeta3pts13bis174.pdf} }}\Bigg\} \\
&-\frac{3 u_4^{(1)}u_3^{(1)}}{N^3} \Bigg\{ \vcenter{\hbox{\includegraphics[scale=0.5]{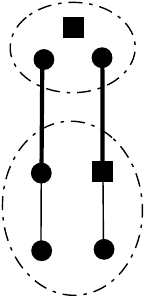} }} +\vcenter{\hbox{\includegraphics[scale=0.5]{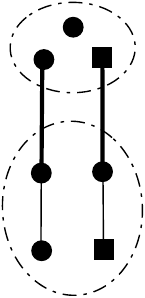} }} \Bigg\}-\frac{3i u_4^{(2)}u_3^{(1)}}{N^3} \Bigg\{ \vcenter{\hbox{\includegraphics[scale=0.5]{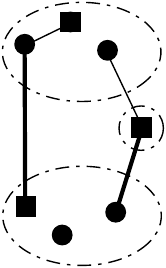} }}+\vcenter{\hbox{\includegraphics[scale=0.5]{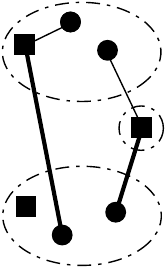} }} +\vcenter{\hbox{\includegraphics[scale=0.5]{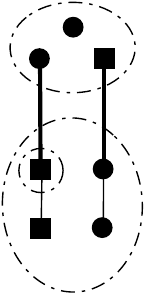} }}\Bigg\}\,. \label{eqU31}
\end{align}
\begin{align}
\nonumber i\frac{\dot{u}_4^{(1)}}{N} \vcenter{\hbox{\includegraphics[scale=0.8]{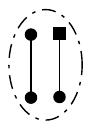} }}&= \frac{u_6}{2N^2}\Bigg\{\vcenter{\hbox{\includegraphics[scale=0.5]{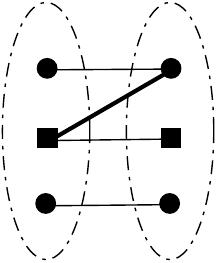} }}+\vcenter{\hbox{\includegraphics[scale=0.5]{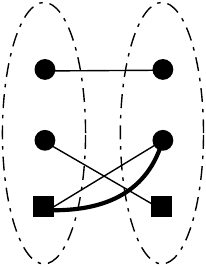} }}+\vcenter{\hbox{\includegraphics[scale=0.5]{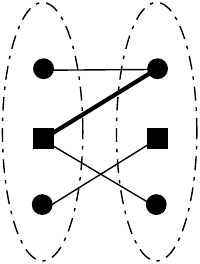} }} \Bigg\}-\frac{3(u_4^{(1)})^2}{N^2} \Bigg\{ \vcenter{\hbox{\includegraphics[scale=0.5]{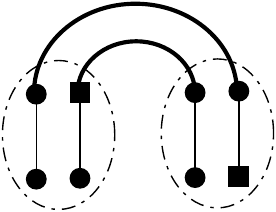} }}\\\nonumber
&+\vcenter{\hbox{\includegraphics[scale=0.5]{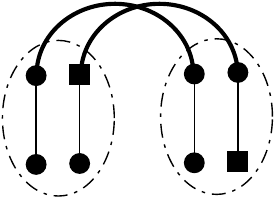} }}+\vcenter{\hbox{\includegraphics[scale=0.5]{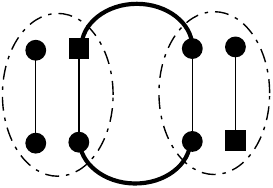} }} \Bigg\} - \frac{6iu_4^{(1)}u_4^{(2)}}{N^2}\Bigg \{ \vcenter{\hbox{\includegraphics[scale=0.5]{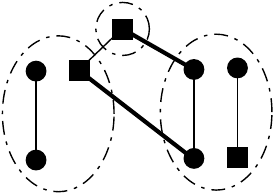} }}+\vcenter{\hbox{\includegraphics[scale=0.5]{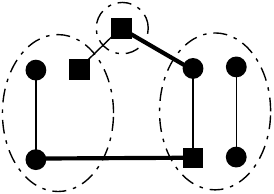} }} \\\nonumber
&+\vcenter{\hbox{\includegraphics[scale=0.5]{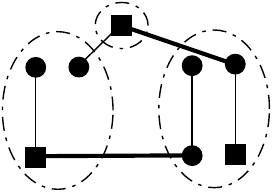} }} +\vcenter{\hbox{\includegraphics[scale=0.5]{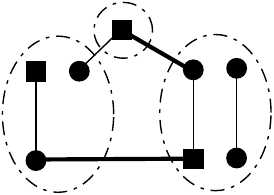} }} \Bigg\} +\frac{18i (u_3^{(2)})^2 u_4^{(1)}}{N^3}\, \Bigg\{ \vcenter{\hbox{\includegraphics[scale=0.5]{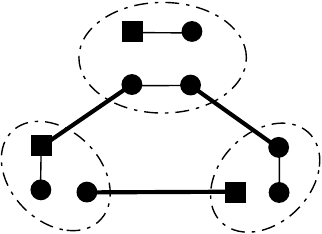} }} +\vcenter{\hbox{\includegraphics[scale=0.5]{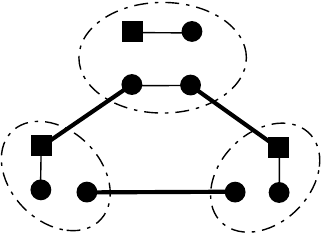} }} \\\nonumber
&+\vcenter{\hbox{\includegraphics[scale=0.5]{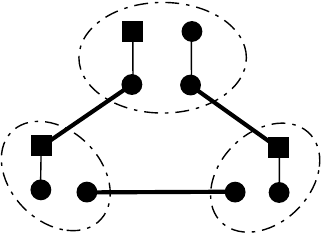} }}+\vcenter{\hbox{\includegraphics[scale=0.5]{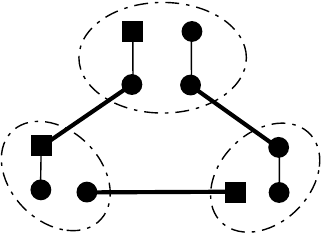} }}+\vcenter{\hbox{\includegraphics[scale=0.5]{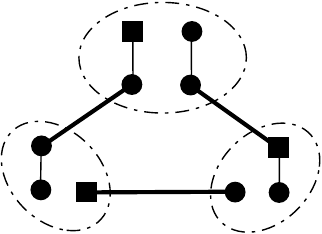} }}+\vcenter{\hbox{\includegraphics[scale=0.5]{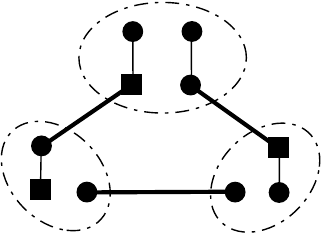} }}+\vcenter{\hbox{\includegraphics[scale=0.5]{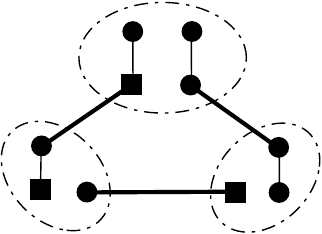} }}\\\nonumber
&+\vcenter{\hbox{\includegraphics[scale=0.5]{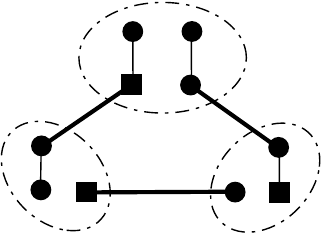} }}+\vcenter{\hbox{\includegraphics[scale=0.5]{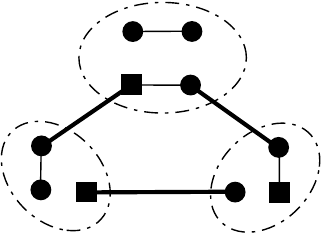} }}+\vcenter{\hbox{\includegraphics[scale=0.5]{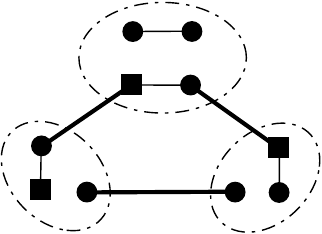} }}+\vcenter{\hbox{\includegraphics[scale=0.5]{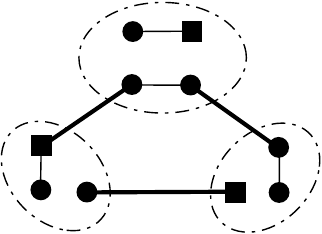} }}+\vcenter{\hbox{\includegraphics[scale=0.5]{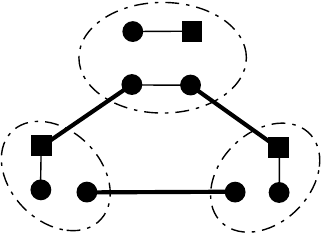} }}\Bigg\} \\
&+\frac{12 (u_3^{(2)})^4}{N^4}\, \Bigg\{ \vcenter{\hbox{\includegraphics[scale=0.5]{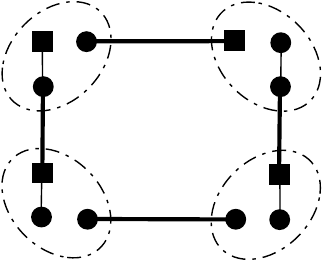} }}+\vcenter{\hbox{\includegraphics[scale=0.5]{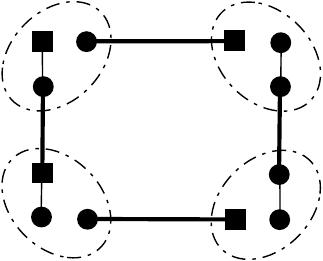} }}+\vcenter{\hbox{\includegraphics[scale=0.5]{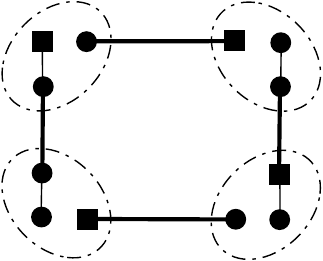} }} \Bigg\} \,,
\end{align}
\begin{align}
\nonumber -\frac{\dot{u}_4^{(2)}}{N} \vcenter{\hbox{\includegraphics[scale=0.8]{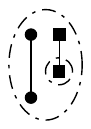} }}&= \frac{u_6}{2N^2}\Bigg\{ \vcenter{\hbox{\includegraphics[scale=0.5]{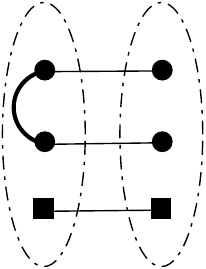}}}+\vcenter{\hbox{\includegraphics[scale=0.5]{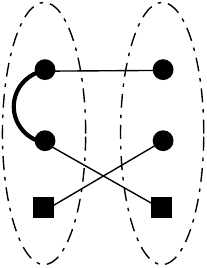} }} \Bigg\}+ \frac{6(u_4^{(2)})^2}{N^2} \Bigg \{ \vcenter{\hbox{\includegraphics[scale=0.5]{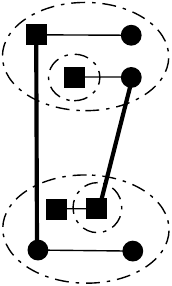} }} +\vcenter{\hbox{\includegraphics[scale=0.5]{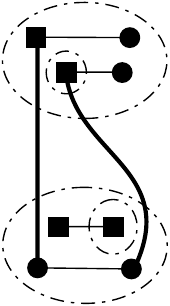} }} +\vcenter{\hbox{\includegraphics[scale=0.5]{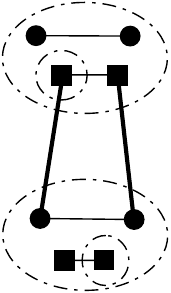} }} \Bigg\}\\\nonumber
&- \frac{6iu_4^{(1)}u_4^{(2)}}{N^2} \Bigg\{\vcenter{\hbox{\includegraphics[scale=0.5]{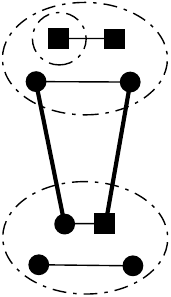} }}+\vcenter{\hbox{\includegraphics[scale=0.5]{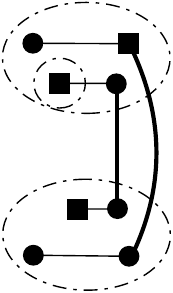} }} \Bigg\}+\frac{18 (u_3^{(2)})^2u_4^{(2)}}{N^3} \Bigg\{ \vcenter{\hbox{\includegraphics[scale=0.5]{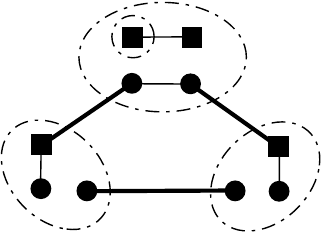} }}+\vcenter{\hbox{\includegraphics[scale=0.5]{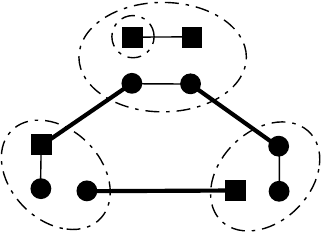} }} \\
&+\vcenter{\hbox{\includegraphics[scale=0.5]{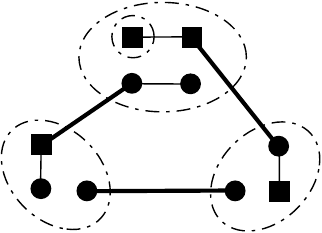} }}+\vcenter{\hbox{\includegraphics[scale=0.5]{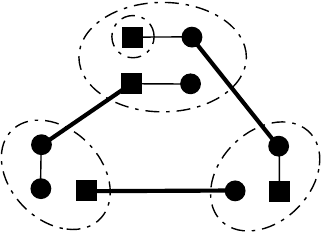} }} \Bigg\}\,,
\end{align}
\begin{align}
\nonumber-\frac{\dot{u}_5^{(1)}}{N^3} \vcenter{\hbox{\includegraphics[scale=0.6]{figureMeta11.pdf} }}&= -\frac{5i u_6 u_3^{(1)}}{N^4} \,\vcenter{\hbox{\includegraphics[scale=0.5]{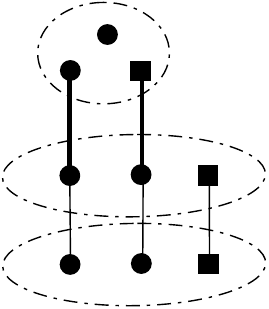} }} -\frac{10i u_5^{(1)} u_4^{(1)}}{N^4} \Bigg\{
\vcenter{\hbox{\includegraphics[scale=0.5]{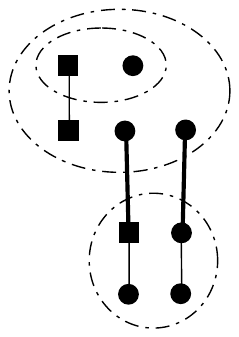} }} + \vcenter{\hbox{\includegraphics[scale=0.5]{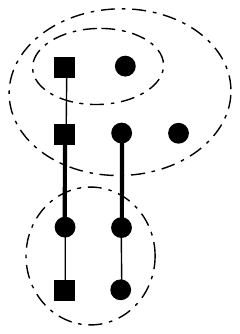} }} + \vcenter{\hbox{\includegraphics[scale=0.5]{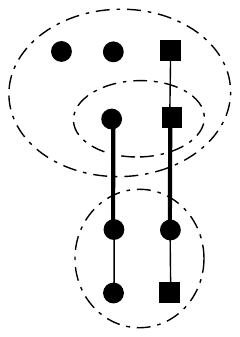} }}+ \vcenter{\hbox{\includegraphics[scale=0.5]{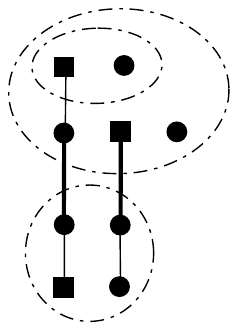} }}\Bigg\}\\\nonumber
&-\frac{30 u_5^{(1)} (u_3^{(1)})^2 }{N^7} \Bigg\{ \vcenter{\hbox{\includegraphics[scale=0.5]{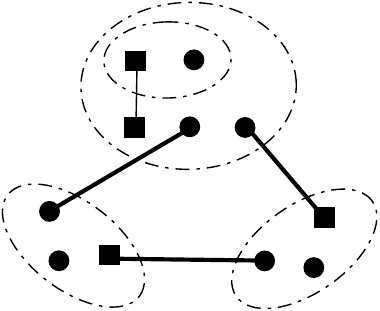} }} + \vcenter{\hbox{\includegraphics[scale=0.5]{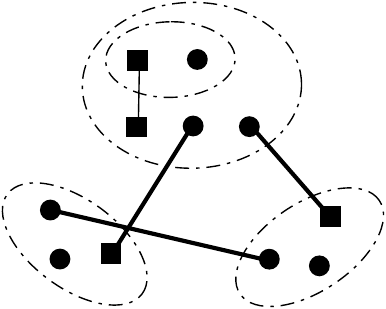} }} \Bigg\}\\\nonumber
& -\frac{60 u_5^{(1)} u_3^{(1)} u_3^{(2)}}{N^6} \Bigg\{ \vcenter{\hbox{\includegraphics[scale=0.5]{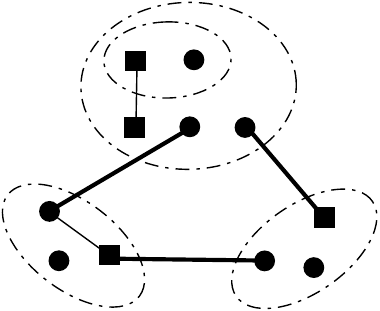} }}+\vcenter{\hbox{\includegraphics[scale=0.5]{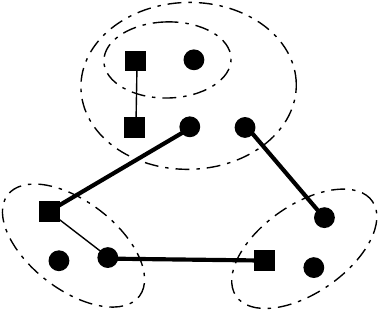} }}+\vcenter{\hbox{\includegraphics[scale=0.5]{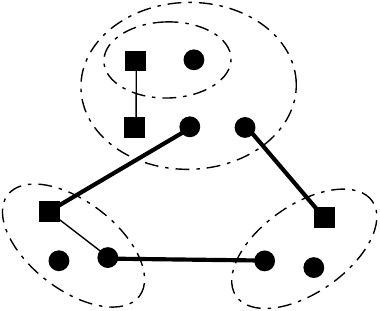} }}+\vcenter{\hbox{\includegraphics[scale=0.5]{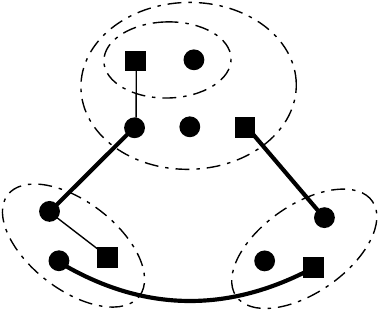} }}\\\nonumber
&+\vcenter{\hbox{\includegraphics[scale=0.5]{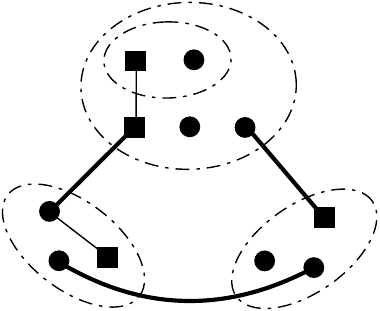} }}+\vcenter{\hbox{\includegraphics[scale=0.5]{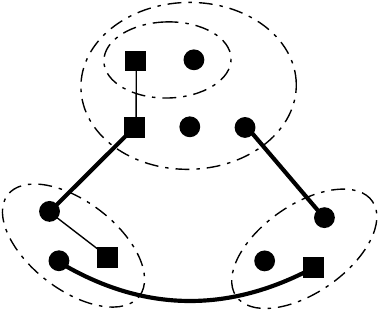} }} +\vcenter{\hbox{\includegraphics[scale=0.5]{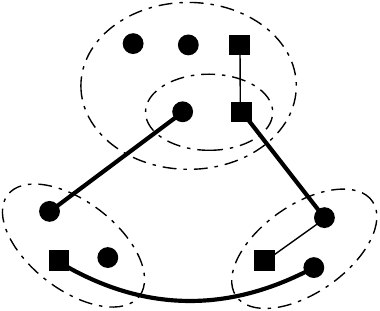} }}+\vcenter{\hbox{\includegraphics[scale=0.5]{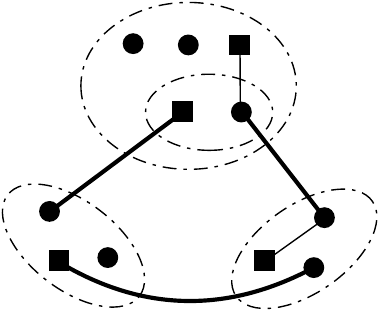} }}+\vcenter{\hbox{\includegraphics[scale=0.5]{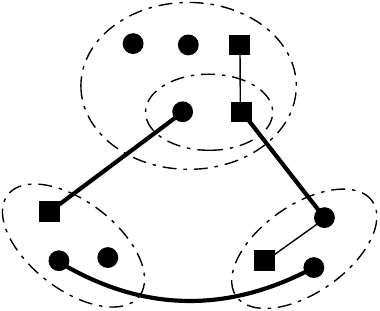} }}\Bigg\}\\\nonumber
&-\frac{30 u_5^{(1)} (u_3^{(2)})^2}{N^5} \, \Bigg\{ \vcenter{\hbox{\includegraphics[scale=0.5]{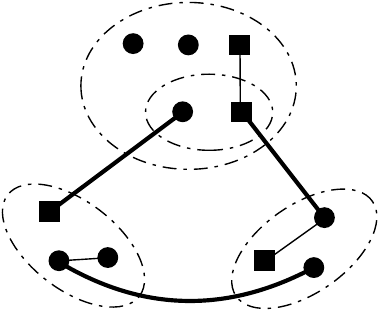} }}+\vcenter{\hbox{\includegraphics[scale=0.5]{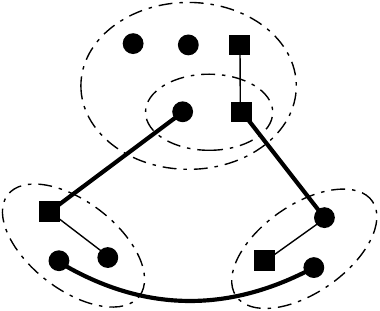} }}+\vcenter{\hbox{\includegraphics[scale=0.5]{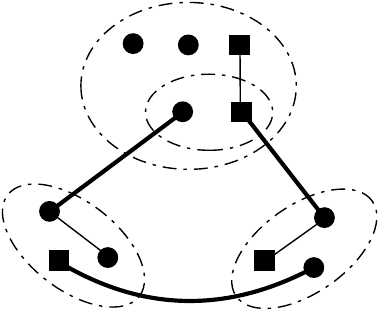} }}+\vcenter{\hbox{\includegraphics[scale=0.5]{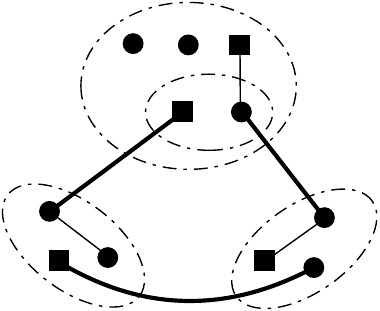} }} \\
&+\vcenter{\hbox{\includegraphics[scale=0.5]{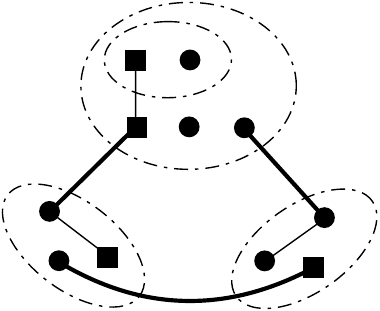} }}+\vcenter{\hbox{\includegraphics[scale=0.5]{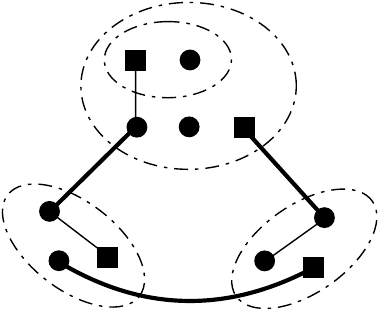} }}+\vcenter{\hbox{\includegraphics[scale=0.5]{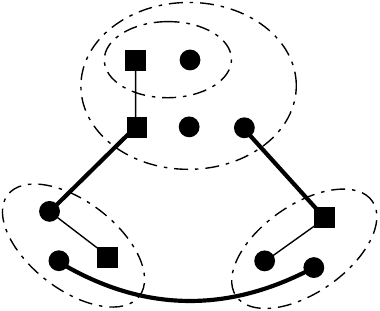} }}\Bigg\}\,.
\end{align}
\bigskip

\begin{align}
\nonumber \frac{i \dot{u}_5^{(2)}}{N^2} \vcenter{\hbox{\includegraphics[scale=0.7]{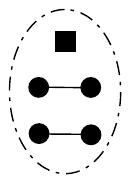} }}&= -\frac{5i u_6 u_3^{(1)}}{N^4} \, \vcenter{\hbox{\includegraphics[scale=0.5]{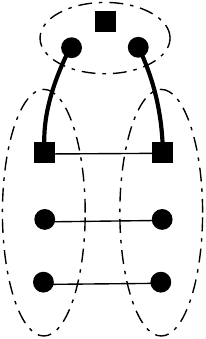} }} -\frac{5i u_6 u_3^{(2)}}{N^3}\,\Bigg\{\vcenter{\hbox{\includegraphics[scale=0.5]{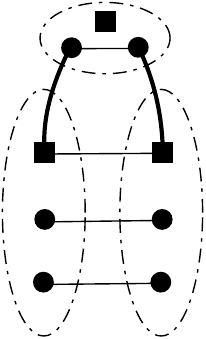} }}+\vcenter{\hbox{\includegraphics[scale=0.5]{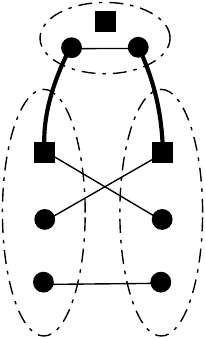} }}+\vcenter{\hbox{\includegraphics[scale=0.5]{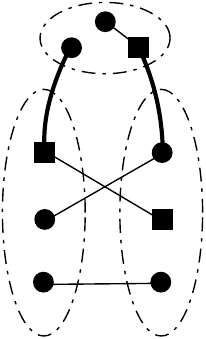} }} \Bigg\} \\\nonumber
&-\frac{10 u_5^{(2)} u_4^{(1)}}{N^3} \Bigg\{ \vcenter{\hbox{\includegraphics[scale=0.5]{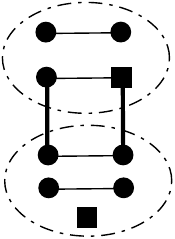} }}
+\vcenter{\hbox{\includegraphics[scale=0.5]{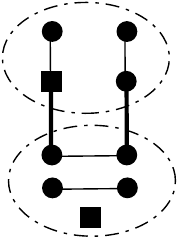} }} +\vcenter{\hbox{\includegraphics[scale=0.5]{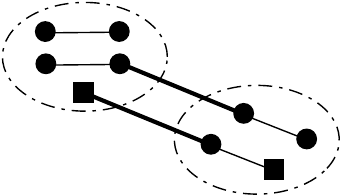} }} \Bigg\} \\\nonumber
&-\frac{10i u_5^{(2)} u_4^{(2)}}{N^3} \Bigg\{ \vcenter{\hbox{\includegraphics[scale=0.5]{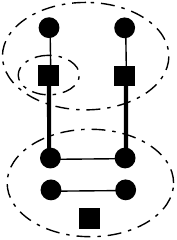} }} +\vcenter{\hbox{\includegraphics[scale=0.5]{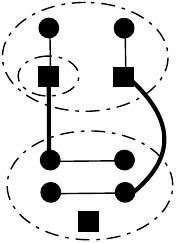} }} +\vcenter{\hbox{\includegraphics[scale=0.5]{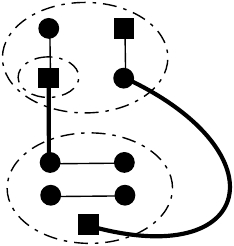} }}+\vcenter{\hbox{\includegraphics[scale=0.5]{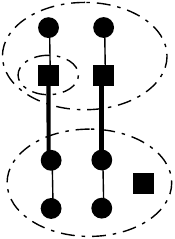} }}+\vcenter{\hbox{\includegraphics[scale=0.5]{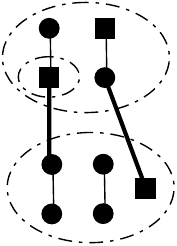} }} \Bigg\}\\\nonumber
&-\frac{10 u_5^{(3)} u_4^{(2)}}{N^4} \, \vcenter{\hbox{\includegraphics[scale=0.5]{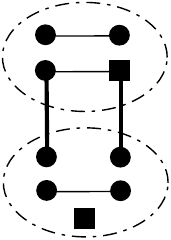} }}
+\frac{48i (u_4^{(1)})^2 u_3^{(1)}}{N^4} \Bigg\{ \vcenter{\hbox{\includegraphics[scale=0.5]{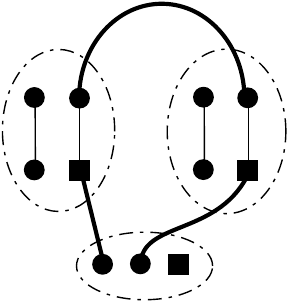} }}+\vcenter{\hbox{\includegraphics[scale=0.5]{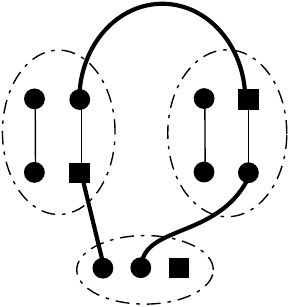} }} \Bigg\}\\\nonumber
&-\frac{96 u_4^{(1)}u_4^{(2)} u_3^{(1)}}{N^4} \, \vcenter{\hbox{\includegraphics[scale=0.5]{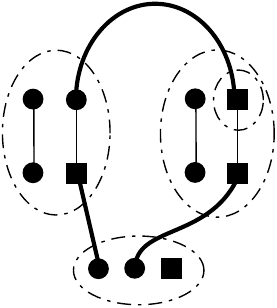} }} +\frac{48i (u_4^{(1)})^2 u_3^{(2)}}{N^3}\Bigg\{\vcenter{\hbox{\includegraphics[scale=0.5]{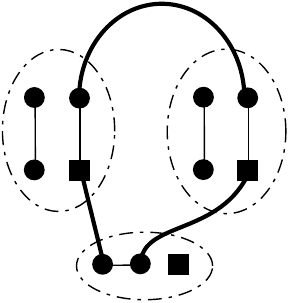} }} +\vcenter{\hbox{\includegraphics[scale=0.5]{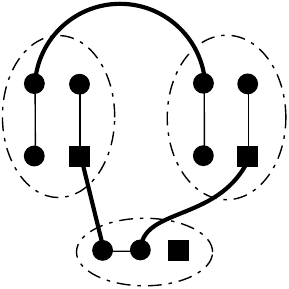} }}\\\nonumber
&+\vcenter{\hbox{\includegraphics[scale=0.5]{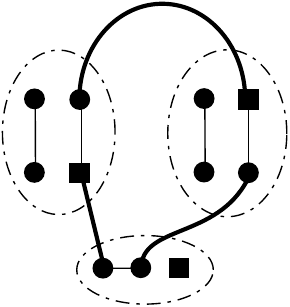} }} +\vcenter{\hbox{\includegraphics[scale=0.5]{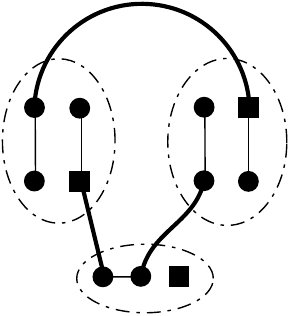} }}+\vcenter{\hbox{\includegraphics[scale=0.5]{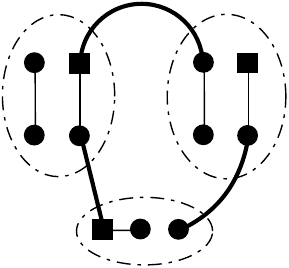} }} \Bigg\}-\frac{96 u_4^{(1)}u_4^{(2)} u_3^{(2)}}{N^3} \Bigg\{\vcenter{\hbox{\includegraphics[scale=0.5]{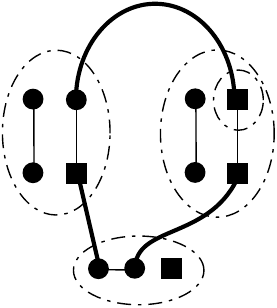} }}\\
& + \vcenter{\hbox{\includegraphics[scale=0.5]{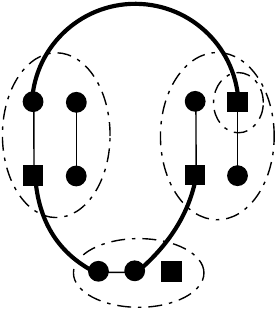} }}+\vcenter{\hbox{\includegraphics[scale=0.5]{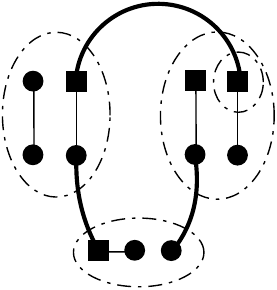} }} \Bigg\} \,,
\end{align}
\begin{align}
\nonumber \frac{i \dot{u}_5^{(3)}}{N^3} \vcenter{\hbox{\includegraphics[scale=0.6]{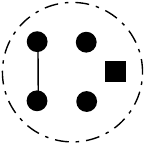} }}=&-\frac{5i u_6 u_3^{(1)}}{N^4} \Bigg\{ \vcenter{\hbox{\includegraphics[scale=0.5]{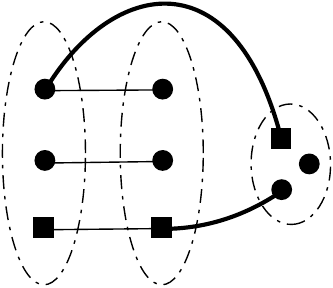} }} + \vcenter{\hbox{\includegraphics[scale=0.5]{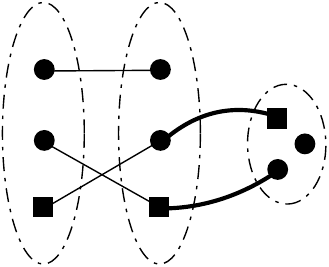} }} \Bigg\}
-\frac{10 u_5^{(3)} u_4^{(1)}}{N^4} \,\vcenter{\hbox{\includegraphics[scale=0.5]{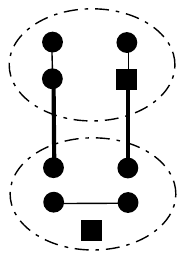} }}-\frac{10i u_5^{(1)} u_4^{(1)}}{N^4} \Bigg\{\\\nonumber
&\,\vcenter{\hbox{\includegraphics[scale=0.5]{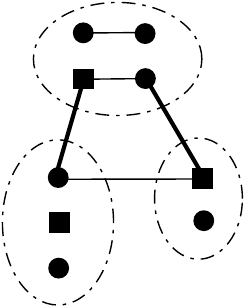} }}+\vcenter{\hbox{\includegraphics[scale=0.5]{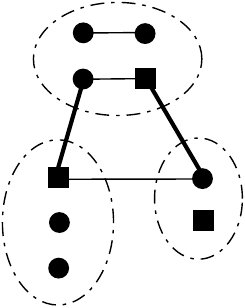} }}+\vcenter{\hbox{\includegraphics[scale=0.5]{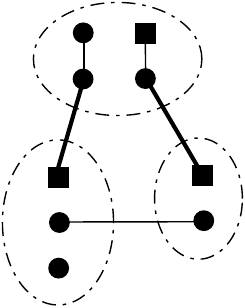} }} +\vcenter{\hbox{\includegraphics[scale=0.5]{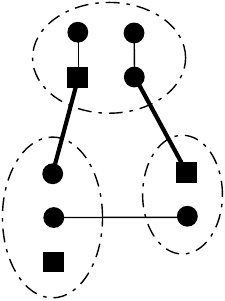} }}+\vcenter{\hbox{\includegraphics[scale=0.5]{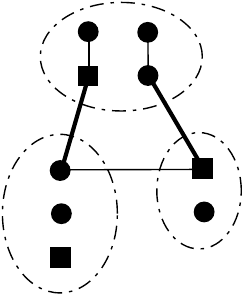} }} +\vcenter{\hbox{\includegraphics[scale=0.5]{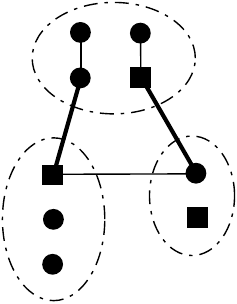} }} \Bigg\} \\\nonumber
&+\frac{48i (u_4^{(1)})^2 u_3^{(1)}}{N^4} \Bigg\{
\vcenter{\hbox{\includegraphics[scale=0.5]{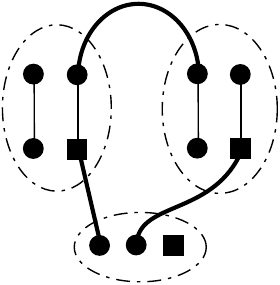} }}+\vcenter{\hbox{\includegraphics[scale=0.5]{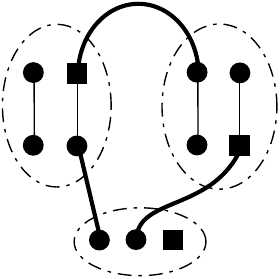} }} +\vcenter{\hbox{\includegraphics[scale=0.5]{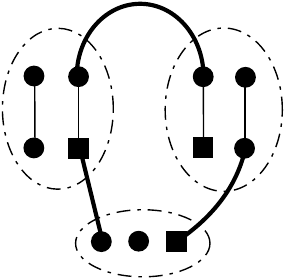} }} \Bigg\}-\frac{96u_4^{(1)} u_4^{(2)}u_3^{(1)}}{N^4} \Bigg\{ \vcenter{\hbox{\includegraphics[scale=0.5]{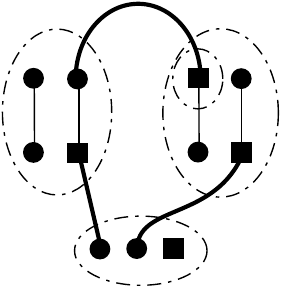} }} \\
&+\vcenter{\hbox{\includegraphics[scale=0.5]{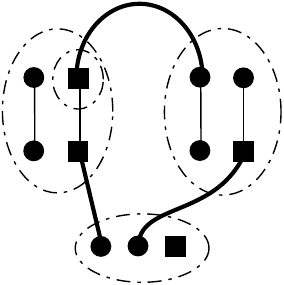} }} +\vcenter{\hbox{\includegraphics[scale=0.5]{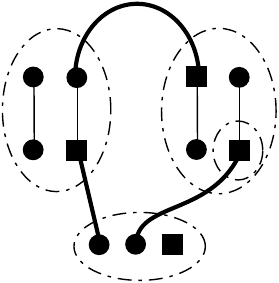} }} +\vcenter{\hbox{\includegraphics[scale=0.5]{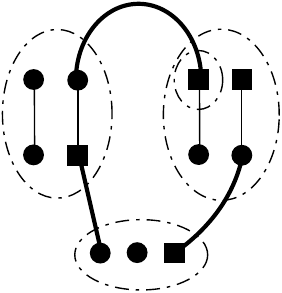} }}+\vcenter{\hbox{\includegraphics[scale=0.5]{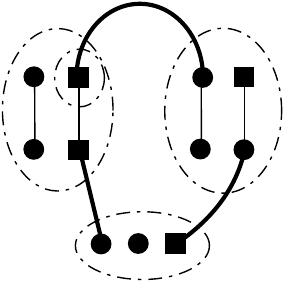} }}+\vcenter{\hbox{\includegraphics[scale=0.5]{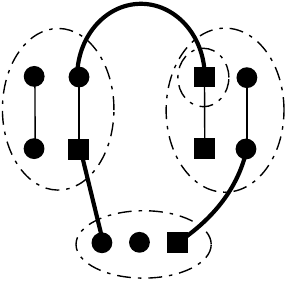} }}+\vcenter{\hbox{\includegraphics[scale=0.5]{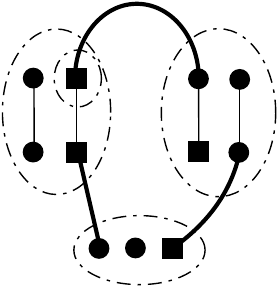} }} \Bigg\}\,,
\end{align}
\begin{align}
\nonumber \frac{\dot{u}_6}{N^2} \, \vcenter{\hbox{\includegraphics[scale=0.5]{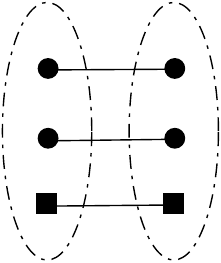} }}&= \frac{15 iu_6 u_4^{(1)}}{N^3}\Bigg\{ \vcenter{\hbox{\includegraphics[scale=0.5]{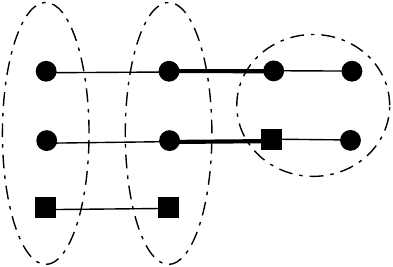}}}+\vcenter{\hbox{\includegraphics[scale=0.5]{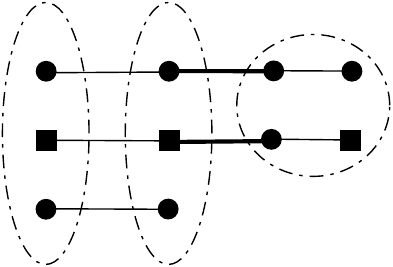}}}+\vcenter{\hbox{\includegraphics[scale=0.5]{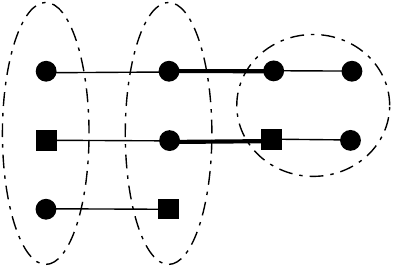}}}+\vcenter{\hbox{\includegraphics[scale=0.5]{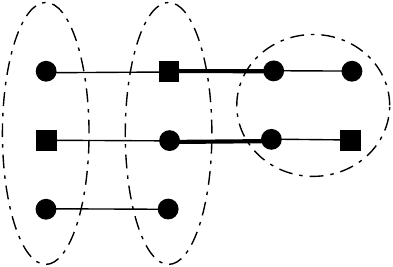}}}\Bigg\}\\\nonumber
&+ \frac{15 u_6 u_4^{(2)}}{N^3} \Bigg\{\vcenter{\hbox{\includegraphics[scale=0.5]{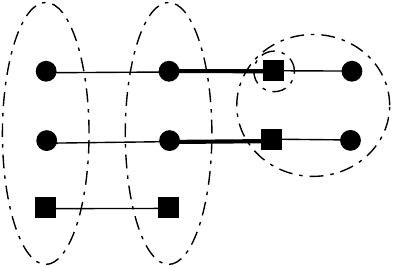}}}+\vcenter{\hbox{\includegraphics[scale=0.5]{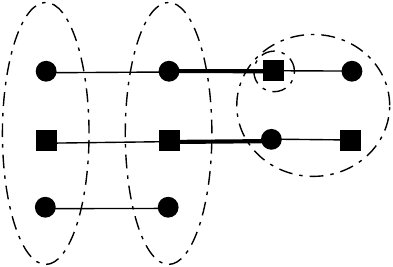}}}+\vcenter{\hbox{\includegraphics[scale=0.5]{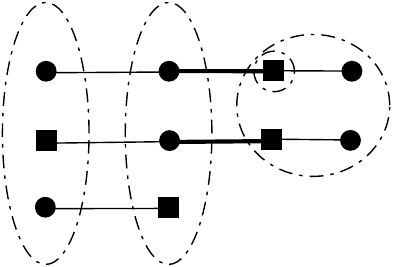}}} \Bigg\} \\
&+\frac{45 u_6 (u_3^{(2)})^2}{N^4}\Bigg\{\vcenter{\hbox{\includegraphics[scale=0.5]{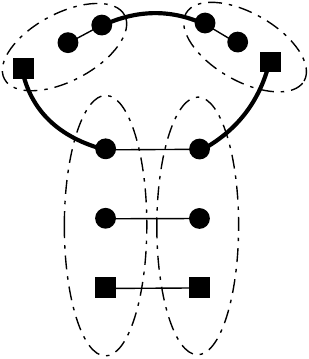}}}+\vcenter{\hbox{\includegraphics[scale=0.5]{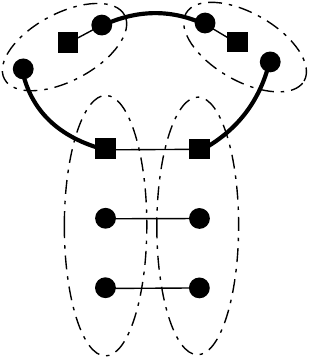}}} \, \Bigg\}\,.
\end{align}
 Translating them in formula following the same strategy as for $u_3^{(2)}$, we obtain:
\medskip
\begin{align}
\nonumber\dot{\bar{u}}_3^{(1)}=&-\frac{3}{2}(1+\eta_k){\bar{u}}_3^{(1)}-\frac{\alpha\bar{u}_5^{(1)}}{3N} (4+N)\frac{1+\eta_k}{(\bar{h}+\alpha)^2}-\frac{4\bar{u}_5^{(2)}}{3N}J_2^{(2)}-\frac{\bar{u}_5^{(3)}}{N}\Big[ (3+N)(\bar{I}_2+2\bar{J}_2^{(1)})\\\nonumber
&+8\bar{J}_2^{(2)}\Big]+\frac{108 (\bar{u}_3^{(1)})^3}{N}(\bar{I}_4+3\bar{J}_4^{(1)}+3\bar{J}_4^{(2)})+\frac{432(\bar{u}_3^{(1)})^2\bar{u}_3^{(2)}}{N}(\bar{I}_4+3\bar{J}_4^{(1)}+3\bar{J}_4^{(2)})\\\nonumber
&+\frac{324(\bar{u}_3^{(2)})^2\bar{u}_3^{(1)}}{N}(\bar{I}_4+3\bar{J}_4^{(1)}+3\bar{J}_4^{(2)})+\frac{(\bar{u}_3^{(2)})^3}{N}\Big[ 4(3N+5)(\bar{I}_4+3\bar{J}_4^{(1)})+64\bar{J}_4^{(2)} \Big]\\
&+\frac{72 \bar{u}_4^{(1)} \bar{u}_3^{(1)}}{N}(\bar{I}_3^{\prime}+3\bar{J}_3^{\prime (1)}+3\bar{J}_3^{\prime (2)})+\frac{18\alpha\bar{u}_4^{(2)}\bar{u}_3^{(1)}}{N}\frac{1+\eta_k}{(\bar{h}+\alpha)^3}\,,
\end{align}
\begin{align}
\nonumber\dot{\bar{u}}_4^{(1)}=&-2(1+\eta_k)\bar{u}_4^{(1)}-\frac{3\alpha\bar{u}_6}{2 N}\,(2+N) \frac{1+\eta_k}{(\bar{h}+\alpha)^2}+\frac{8 (\bar{u}_4^{(1)})^2}{N} \Big[(4+N) (\bar{I}_3^\prime+3\bar{J}_3^{\prime (1)})+2\bar{J}_3^{\prime (2)}\Big]\\\nonumber
&+\frac{8\alpha \bar{u}_4^{(1)} \bar{u}_4^{(2)}}{N} \frac{1+\eta_k}{(\bar{h}+\alpha)^3}+\frac{48 (\bar{u}_3^{(2)})^2\bar{u}_4^{(1)}}{N}\Big[6(\bar{I}_4+3\bar{J}_4^{(1)})+5\bar{J}_4^{(2)} \Big]\\
&-\frac{8(\bar{u}_3^{(2)})^4}{N}\Big[3(\bar{I}_5+3\bar{J}_5^{(1)})+2\bar{J}_5^{(2)} \Big]\,,
\end{align}
\begin{align}
\nonumber \dot{\bar{u}}_4^{(2)}=&-2(1+\eta_k)\bar{u}_4^{(2)}- \frac{12\bar{u}_6}{N}(\bar{I}_2+2\bar{J}_2^{(1)})-\frac{2\bar{u}_4^{(1)}\bar{u}_4^{(2)}}{N}\Big[(1+N)(\bar{I}_3^\prime+3\bar{J}_3^{\prime (1)})+3\bar{J}_3^{\prime (2)}\Big]\\
&-\frac{9\alpha(\bar{u}_4^{(2)})^2}{N}\frac{1+\eta_k}{(\bar{h}+\alpha)^3}+\frac{3(\bar{u}_3^{(2)})^2 \bar{u}_4^{(2)}}{4N}(11 (\bar{I}_4+3\bar{J}_4^{(1)})+12\bar{J}_4^{(2)})\,,
\end{align}
\begin{align}
\nonumber \dot{\bar{u}}_5^{(1)}=&-\frac{5}{2}(1+\eta_k){\bar{u}}_5^{(1)}+\frac{60\bar{u}_6\bar{u}_3^{(1)}}{N}(\bar{I}_3^{\prime}+3\bar{J}_3^{\prime (1)}+3\bar{J}_3^{\prime (2)})-\frac{40\bar{u}_5^{(1)} \bar{u}_4^{(1)}}{3N}\Big[7(\bar{I}_3^\prime+3\bar{J}_3^{\prime (1)})+12 \bar{J}_3^{\prime (2)} \Big]\\\nonumber
&-\frac{720 \bar{u}_5^{(1)} (\bar{u}_3^{(1)})^2}{N}(\bar{I}_4+3\bar{J}_4^{(1)}+3\bar{J}_4^{(2)})-\frac{80 \bar{u}_5^{(1)}\bar{u}_3^{(1)}\bar{u}_3^{(2)}}{N}\Big[7(\bar{I}_4+3\bar{J}_4^{(1)})+12\bar{J}_4^{(2)} \Big]\\
&-\frac{80\bar{u}_5^{(1)}(\bar{u}_3^{(2)})^2}{N}\Big[2(\bar{I}_4+3\bar{J}_4^{(1)})+3\bar{J}_4^{(2)} \Big]\,,
\end{align}
\begin{align}
\nonumber \dot{\bar{u}}_5^{(2)}=&-\frac{5}{2}(1+\eta_k){\bar{u}}_5^{(2)}+\frac{180 \alpha \bar{u}_6\bar{u}_3^{(1)}}{N} \frac{1+\eta_k}{(\bar{h}+\alpha)^3}+\frac{60 \alpha \bar{u}_6\bar{u}_3^{(2)}}{N}(N+3) \frac{1+\eta_k}{(\bar{h}+\alpha)^3}\\\nonumber
&+\frac{40 \bar{u}_5^{(2)} \bar{u}_4^{(1)}}{N}\Big[(4+N) (\bar{I}_3^\prime+3\bar{J}_3^{\prime (1)})+2\bar{J}_3^{\prime (2)} \Big]-\frac{190 \alpha \bar{u}_5^{(2)} \bar{u}_4^{(2)} }{3N}\frac{1+\eta_k}{(\bar{h}+\alpha)^3}\\\nonumber
&+\frac{40 \bar{u}_5^{(3)} \bar{u}_4^{(2)} }{3N^2}(\bar{I}_3^\prime+3\bar{J}_3^{\prime (1)}+3\bar{J}_3^{\prime (2)})+\frac{384 (\bar{u}_4^{(1)})^2 \bar{u}_3^{(1)}}{N}(\bar{I}_4+3\bar{J}_4^{(1)}+3\bar{J}_4^{(2)})\\\nonumber
&-\frac{384 \alpha \bar{u}_4^{(1)}\bar{u}_4^{(2)} \bar{u}_3^{(1)}}{N} \frac{1+\eta_k}{(\bar{h}+\alpha)^4}+\frac{64 (\bar{u}_4^{(1)})^2 \bar{u}_3^{(2)}}{N}\Big[ 41(\bar{I}_3^\prime+3\bar{J}_3^{\prime (1)})+18 \bar{J}_3^{\prime (2)} \Big]\\
&-\frac{128 \alpha \bar{u}_4^{(1)}\bar{u}_4^{(2)} \bar{u}_3^{(2)} }{N}(N+2) \frac{1+\eta_k}{(\bar{h}+\alpha)^4}\,,
\end{align}
\begin{align}
\nonumber \dot{\bar{u}}_5^{(3)}=&-\frac{5}{2}(1+\eta_k){\bar{u}}_5^{(3)}+\frac{90\alpha \bar{u}_6\bar{u}_3^{(1)}}{N^4} \frac{1+\eta_k}{(\bar{h}+\alpha)^3}+\frac{40 \bar{u}_5^{(3)} \bar{u}_4^{(1)}}{3N} (\bar{I}_3^\prime+3\bar{J}_3^{\prime (1)}+3\bar{J}_3^{\prime (2)})\\\nonumber
&+ \frac{40 \alpha \bar{u}_5^{(1)}\bar{u}_4^{(1)}}{9N} (17+3N)\frac{1+\eta_k}{(\bar{h}+\alpha)^3}+\frac{768(\bar{u}_4^{(1)})^2\bar{u}_3^{(1)}}{N}(\bar{I}_3^\prime+3\bar{J}_3^{\prime (1)}+3\bar{J}_3^{\prime (2)})\\
&-\frac{1792 \alpha \bar{u}_4^{(1)}\bar{u}_4^{(2)}\bar{u}_3^{(1)}}{N} \frac{1+\eta_k}{(\bar{h}+\alpha)^4}\,,
\end{align}
\begin{align}
\nonumber \dot{\bar{u}}_6=-3(1+\eta_k){\bar{u}}_6+&\frac{280 \bar{u}_6\bar{u}_4^{(1)}}{N}(\bar{I}_3+3\bar{J}_3^{(1)})-\frac{90 \alpha\bar{u}_6\bar{u}_4^{(2)}}{N} \frac{1+\eta_k}{(\bar{h}+\alpha)^3}\\
&-\frac{240 \alpha \bar{u}_6(\bar{u}_3^{(1)})^2}{N}\frac{1+\eta_k}{(\bar{h}+\alpha)^3}\frac{\bar{\Delta}^{(2)}}{(\bar{h}+\alpha)^2}\,. \label{equ6}
\end{align}
\end{appendices}

\bibliographystyle{plain}
\bibliography{biblio}

\begin{thebibliography}{100}

\bibitem{Crisanti1993}
H.~J.~Sommers A.~Crisanti, H.~Horner.
\newblock The sphericalp-spin interaction spin-glass model.
\newblock {\em Zeitschrift für Physik B}, 1993.

\bibitem{Sherrington2}
D.~Sherrington A.C.C.~Coolen, J.A.F.~Heimel.
\newblock Dynamics of the batch minority game with inhomogeneous decision
  noise.
\newblock {\em Physical Review E 65(1 Pt 2):016126}, January 2002.

\bibitem{kimanderson2008nonperturbative}
Johan Anderson and Eun-jin Kim.
\newblock Nonperturbative models of intermittency in edge turbulence.
\newblock {\em Physics of Plasmas}, 15(12):122303, 2008.

\bibitem{Aron_2010}
Camille Aron, Giulio Biroli, and Leticia~F Cugliandolo.
\newblock Symmetries of generating functionals of langevin processes with
  colored multiplicative noise.
\newblock {\em Journal of Statistical Mechanics: Theory and Experiment},
  2010(11):P11018, Nov 2010.

\bibitem{Kim2001}
A.~Latz B.~Kim.
\newblock The dynamics of the spherical p-spin model: from microscopic to
  asymptotics.
\newblock {\em EDP Sciences}, 2001.

\bibitem{Balog_2018}
Ivan Balog, Gilles Tarjus, and Matthieu Tissier.
\newblock Criticality of the random field ising model in and out of
  equilibrium: A nonperturbative functional renormalization group description.
\newblock {\em Physical Review B}, 97(9), Mar 2018.

\bibitem{Balog_2020}
Ivan Balog, Gilles Tarjus, and Matthieu Tissier.
\newblock Dimensional reduction breakdown and correction to scaling in the
  random-field ising model.
\newblock {\em Physical Review E}, 102(6), Dec 2020.

\bibitem{becchi1974abelian}
Carlo Becchi, Alain Rouet, and Raymond Stora.
\newblock The abelian higgs kibble model, unitarity of the s-operator.
\newblock {\em Physics Letters B}, 52(3):344--346, 1974.

\bibitem{Benaych1}
Florent Benaych-Georges and Antti Knowles.
\newblock Lectures on the local semicircle law for wigner matrices.
\newblock {\em arXiv:1601.04055}, 2018.

\bibitem{Berges_2002}
Jürgen Berges, Nikolaos Tetradis, and Christof Wetterich.
\newblock Non-perturbative renormalization flow in quantum field theory and
  statistical physics.
\newblock {\em Physics Reports}, 363(4-6):223–386, Jun 2002.

\bibitem{Neurobick2020understanding}
Christian Bick, Marc Goodfellow, Carlo~R Laing, and Erik~A Martens.
\newblock Understanding the dynamics of biological and neural oscillator
  networks through exact mean-field reductions: a review.
\newblock {\em The Journal of Mathematical Neuroscience}, 10(1):1--43, 2020.

\bibitem{Billoire_2005}
A~Billoire, L~Giomi, and E~Marinari.
\newblock The mean-field infinite range p = 3 spin glass: Equilibrium landscape
  and correlation time scales.
\newblock {\em Europhysics Letters (EPL)}, 71(5):824?830, Sep 2005.

\bibitem{Bouchaud2}
Jean~Philippe Bouchaud.
\newblock The (unfortunate) complexity of the economy.
\newblock {\em Physics World}, April 2009.

\bibitem{BryngelsonAl}
J~D Bryngelson and P~G Wolynes.
\newblock Spin glasses and the statistical mechanics of protein folding.
\newblock {\em Proceedings of the National Academy of Sciences of the United
  States of America}, Nov 1987.

\bibitem{B_ny_2018}
Cédric Bény.
\newblock Inferring relevant features: From qft to pca.
\newblock {\em International Journal of Quantum Information}, 16(08):1840012,
  Dec 2018.

\bibitem{BENY2015}
Cédric Bény and Tobias~J Osborne.
\newblock The renormalization group via statistical inference.
\newblock {\em New Journal of Physics}, 17(8):083005, Aug 2015.

\bibitem{Caiazzo1}
Antonio Caiazzo, Antonio Coniglio, and Mario Nicodemi.
\newblock Glass glass transition and new dynamical singularity points in an
  analytically solvable p-spin glass like model.
\newblock {\em Phys. Rev. Lett}, 2004.

\bibitem{Canet_2007}
Léonie Canet and Hugues Chaté.
\newblock A non-perturbative approach to critical dynamics.
\newblock {\em Journal of Physics A: Mathematical and Theoretical},
  40(9):1937–1949, Feb 2007.

\bibitem{Canet_2011}
Léonie Canet, Hugues Chaté, and Bertrand Delamotte.
\newblock General framework of the non-perturbative renormalization group for
  non-equilibrium steady states.
\newblock {\em Journal of Physics A: Mathematical and Theoretical},
  44(49):495001, Nov 2011.

\bibitem{carreras1996fluctuation}
BA~Carreras, C~Hidalgo, E~S{\'a}nchez, MA~Pedrosa, R~Balbin,
  I~Garc{\'\i}a-Cort{\'e}s, B~Van~Milligen, DE~Newman, and VE~Lynch.
\newblock Fluctuation-induced flux at the plasma edge in toroidal devices.
\newblock {\em Physics of Plasmas}, 3(7):2664--2672, 1996.

\bibitem{castellana2013renormalization}
Michele Castellana.
\newblock The renormalization group for disordered systems, 2013.

\bibitem{Castellana_20152}
Michele Castellana and Carlo Barbieri.
\newblock Hierarchical spin glasses in a magnetic field: A
  renormalization-group study.
\newblock {\em Physical Review B}, 91(2), Jan 2015.

\bibitem{Castellana_2015}
Michele Castellana and Giorgio Parisi.
\newblock Non-perturbative effects in spin glasses.
\newblock {\em Scientific Reports}, 5(1), Mar 2015.

\bibitem{Castellani1}
Tommaso Castellani and Andrea Cavagna.
\newblock Spin-glass theory for pedestrians.
\newblock {\em J. Stat. Mech. (2005) P05012}, june 2005.

\bibitem{Cavagna_2009}
Andrea Cavagna.
\newblock Supercooled liquids for pedestrians.
\newblock {\em Physics Reports}, 476(4-6):51?124, Jun 2009.

\bibitem{DeDominicisbook}
Irene~Giardina Cirano De~Dominicis.
\newblock {\em Random Fields and Spin Glasses: A Field Theory Approach}.
\newblock Cambridge University Press, 2009.

\bibitem{Contuccibook}
Cristian Contucci, Pierluigi;~Giardinà.
\newblock {\em Perspectives on Spin Glasses}.
\newblock Cambridge University Press, 2012.

\bibitem{Coolen1}
A.C.C. Coolen.
\newblock Non-equilibrium statistical mechanics of minority games.
\newblock {\em short review for Cergy conference proceedings}, 2002.

\bibitem{Cugliandolo1}
Leticia~F. Cugliandolo.
\newblock Dynamics of glassy systems.
\newblock {\em Lecture notes, Les Houches}, jul 2002.

\bibitem{Cugliandolo4}
Leticia~F Cugliandolo, D.~R. Grempel, and Constantino~A da~Silva~Santos.
\newblock The quantum spherical p-spin-glass model.
\newblock {\em Phys. Rev. B}, 2001.

\bibitem{Wett1}
C.Wetterich.
\newblock Average action and the renormalization group equations.
\newblock {\em Nuclear Physics B}, 1991.

\bibitem{Wett2}
C.Wetterich.
\newblock Exact evolution equation for the effective potential.
\newblock {\em Nuclear Physics B}, 1993.

\bibitem{Delamotte1}
Bertrand Delamotte.
\newblock An introduction to the nonperturbative renormalization group.
\newblock {\em Lecture Notes in Physics}, page 49–132, 2012.

\bibitem{DeDominicis1978}
C.~De Dominicis and L.~Peliti.
\newblock Field-theory renormalization and critical dynamics above tc: Helium,
  antiferromagnets, and liquid-gas systems.
\newblock {\em Phys. Rev. B}, 1978.

\bibitem{Duclut_2017}
Charlie Duclut and Bertrand Delamotte.
\newblock Frequency regulators for the nonperturbative renormalization group: A
  general study and the model a as a benchmark.
\newblock {\em Physical Review E}, 95(1), Jan 2017.

\bibitem{Dupuis_2021}
N.~Dupuis, L.~Canet, A.~Eichhorn, W.~Metzner, J.M. Pawlowski, M.~Tissier, and
  N.~Wschebor.
\newblock The nonperturbative functional renormalization group and its
  applications.
\newblock {\em Physics Reports}, Jan 2021.

\bibitem{erdmenger2021quantifying}
Johanna Erdmenger, Kevin~T. Grosvenor, and Ro~Jefferson.
\newblock Towards quantifying information flows: relative entropy in deep
  neural networks and the renormalization group, 2021.

\bibitem{FischerHertz}
K.~H. Fischer and J.~A. Hertz.
\newblock {\em Spin Glasses}.
\newblock Cambridge University Press, 1991.

\bibitem{Biolgarel1988mean}
T~Garel and H~Orland.
\newblock Mean-field model for protein folding.
\newblock {\em EPL (Europhysics Letters)}, 6(4):307, 1988.

\bibitem{Gredat_2014}
Damien Gredat, Hugues Chaté, Bertrand Delamotte, and Ivan Dornic.
\newblock Finite-scale singularity in the renormalization group flow of a
  reaction-diffusion system.
\newblock {\em Physical Review E}, 89(1), Jan 2014.

\bibitem{BenArous1}
Aukosh~Jagannath Gérard Ben~Arous, Reza~Gheissari.
\newblock Algorithmic thresholds for tensor pca.
\newblock {\em Ann. Probab. 48(4): 2052-2087}, July 2020.

\bibitem{BenArous2}
Aukosh~Jagannath Gérard Ben~Arous, Reza~Gheissari.
\newblock Bounding flows for spherical spin glass dynamics.
\newblock {\em Communications in Mathematical Physics volume 373,
  pages1011–1048}, 2020.

\bibitem{Takayama1986}
Koji~Nemoto Hajime~Takayama, Takayuki~Shirakura.
\newblock Dynamical mean field theory of spin glasses and their phase
  transitions in ac external fields.
\newblock {\em Progress of Theoretical Physics Supplement}, 1986.

\bibitem{hurst1951long}
Harold~Edwin Hurst.
\newblock Long-term storage capacity of reservoirs.
\newblock {\em Transactions of the American society of civil engineers},
  116(1):770--799, 1951.

\bibitem{Janssen1}
Hans-Karl Janssen.
\newblock On a lagrangean for classical field dynamics and renormalization
  group calculations of dynamical critical properties.
\newblock {\em Zeitschrift für Physik B}, 1976.

\bibitem{Bouchaud1}
Jorge Kurchan Marc~Mezard Jean Philippe~Bouchaud, Leticia F~Cugliandolo.
\newblock Out of equilibrium dynamics in spin-glasses and other glassy systems.
\newblock {\em Spin-glasses and random fields, A. P. Young Ed. (World
  Scientific)}, Feb 1997.

\bibitem{Kapoyannis_2000}
A.S. Kapoyannis and N.~Tetradis.
\newblock Quantum-mechanical tunnelling and the renormalization group.
\newblock {\em Physics Letters A}, 276(5-6):225–232, Nov 2000.

\bibitem{kaviraj2021random}
Apratim Kaviraj, Slava Rychkov, and Emilio Trevisani.
\newblock Random field ising model and parisi-sourlas supersymmetry ii.
  renormalization group, 2021.

\bibitem{kim2009probability}
Eun-jin Kim, Han-Li Liu, and Johan Anderson.
\newblock Probability distribution function for self-organization of shear
  flows.
\newblock {\em Physics of Plasmas}, 16(5):052304, 2009.

\bibitem{Kirkpatrick1987}
T.~R. Kirkpatrick and D.~Thirumalai.
\newblock p-spin-interaction spin-glass models: Connections with the structural
  glass problem.
\newblock {\em Phys. Rev. B}, 1987.

\bibitem{materkotliar2004strongly}
Gabriel Kotliar and Dieter Vollhardt.
\newblock Strongly correlated materials: Insights from dynamical mean-field
  theory.
\newblock {\em Physics today}, 57(3):53--60, 2004.

\bibitem{Krajewski1}
Thomas Krajewski.
\newblock A renormalisation group approach to the universality of wigner’s
  semicircle law for random matrices with dependent entries.
\newblock {\em Hal-01696480}, 2018.

\bibitem{Kurchan1992}
J.~Kurchan.
\newblock Supersymmetry in spin glass dynamics.
\newblock {\em EDP Sciences}, 1992.

\bibitem{Cugliandolo2}
J.~Kurchan L.~F.~Cugliandolo.
\newblock Analytical solution of the off-equilibrium dynamics of a long range
  spin-glass model.
\newblock {\em Phys. Rev. Lett.}, 1993.

\bibitem{lahoche2020field2}
Vincent Lahoche, Mohamed Ouerfeli, Dine~Ousmane Samary, and Mohamed
  Tamaazousti.
\newblock Field theoretical approach for signal detection in nearly continuous
  positive spectra ii: tensorial data.
\newblock {\em Entropy}, 2021.

\bibitem{lahoche2020field}
Vincent Lahoche, Dine~Ousmane Samary, and Mohamed Tamaazousti.
\newblock Field theoretical approach for signal detection in nearly continuous
  positive spectra i: Matricial data.
\newblock {\em Entropy}, 2020.

\bibitem{lahoche2021signal}
Vincent Lahoche, Dine~Ousmane Samary, and Mohamed Tamaazousti.
\newblock Signal detection in nearly continuous spectra and symmetry breaking.
\newblock {\em Symmetry}, 2021.

\bibitem{lahoche2022field}
Vincent Lahoche, Dine~Ousmane Samary, and Mohamed Tamaazousti.
\newblock Field theoretical approach for signal detection in nearly continuous
  positive spectra iii: Universal features.
\newblock {\em arXiv preprint arXiv:2201.04250}, 2022.

\bibitem{lahoche2020generalized}
Vincent Lahoche, Dine~Ousmane Samary, and Mohamed Tamaazousti.
\newblock Generalized scale behavior and renormalization group for principal
  component analysis.
\newblock {\em J. Stat. Mech}, 2022.

\bibitem{Cugliandolo3}
David S.~Dean Leticia F.~Cugliandolo.
\newblock Full dynamical solution for a spherical spin-glass model.
\newblock {\em J. Phys. A}, 1995.

\bibitem{MezardAl}
M.~Virasoro M.~Mezard, G.~Parisi.
\newblock {\em Spin glass theory and beyond}.
\newblock World scientific Publishing, 1987.

\bibitem{machmei2019mean}
Song Mei, Theodor Misiakiewicz, and Andrea Montanari.
\newblock Mean-field theory of two-layers neural networks: dimension-free
  bounds and kernel limit.
\newblock In {\em Conference on Learning Theory}, pages 2388--2464. PMLR, 2019.

\bibitem{Castellana3}
Giorgio~Parisi Michele~Castellana.
\newblock A renormalization group computation of the critical exponents of
  hierarchical spin glasses.
\newblock {\em Physical Review E}, 2010.

\bibitem{Morris_19942}
Tim~R. Morris.
\newblock Derivative expansion of the exact renormalization group.
\newblock {\em Physics Letters B}, 329(2-3):241–248, Jun 1994.

\bibitem{MORRIS_1994}
Tim~R. Morris.
\newblock The exact renormalization group and approximate solutions.
\newblock {\em International Journal of Modern Physics A}, 09(14):2411–2449,
  Jun 1994.

\bibitem{Moshe_2003}
Moshe Moshe and Jean Zinn-Justin.
\newblock Quantum field theory in the large n limit: a review.
\newblock {\em Physics Reports}, 385(3-6):69–228, Oct 2003.

\bibitem{Nishimori}
Hidetoshi Nishimori.
\newblock {\em Statistical Physics of Spin Glasses and Information Processing:
  An Introduction}.
\newblock Oxford Scholarship Online, 2010.

\bibitem{optimorland1985mean}
Henri Orland.
\newblock Mean-field theory for optimization problems.
\newblock {\em Journal de Physique Lettres}, 46(17):763--770, 1985.

\bibitem{DiFrancesco1}
J.~Zinn-Justin P.~Di~Francesco, P.~Ginsparg.
\newblock 2d gravity and random matrices.
\newblock {\em Phys.Rept}, 1995.

\bibitem{pawlowski2015physics}
Jan~M. Pawlowski, Michael~M. Scherer, Richard Schmidt, and Sebastian~J. Wetzel.
\newblock Physics and the choice of regulators in functional renormalisation
  group flows, 2015.

\bibitem{2018Entropie}
Pedro Pessoa and Ariel Caticha.
\newblock Exact renormalization groups as a form of entropic dynamics.
\newblock {\em Entropy}, 20(1):25, Jan 2018.

\bibitem{Pimentel_2002}
I.~R. Pimentel, T.~Temesvári, and C.~De~Dominicis.
\newblock Spin-glass transition in a magnetic field: A renormalization group
  study.
\newblock {\em Physical Review B}, 65(22), Jun 2002.

\bibitem{Pol1}
J.~Polchinski.
\newblock Renormalization and effective lagrangians.
\newblock {\em Nuclear Physics B}, 1984.

\bibitem{Potters1}
M~Potters and J~P Bouchaud.
\newblock A first course in random matrix theory (for physicists, engineers and
  data scientists).
\newblock {\em Cambridge University Press}, 2021.

\bibitem{Prokopec_2018}
Tomislav Prokopec and Gerasimos Rigopoulos.
\newblock Functional renormalization group for stochastic inflation.
\newblock {\em Journal of Cosmology and Astroparticle Physics},
  2018(08):013–013, Aug 2018.

\bibitem{Bausch1976}
H.~Wagner R.~Bausch, H. K.~Janssen.
\newblock Renormalized field theory of critical dynamics.
\newblock {\em Zeitschrift für Physik B}, 1976.

\bibitem{Riv2}
V.~Rivasseau.
\newblock {\em From Perturbative to Constructive Renormalization}.
\newblock Princeton series in physics, 1991.

\bibitem{Riv1}
V.~Rivasseau.
\newblock An introduction to renormalization.
\newblock {\em Séminaire Poincaré}, 2001.

\bibitem{Sherrington}
David Sherrington.
\newblock Physics and complexity.
\newblock {\em Phil. Trans. R. Soc. A}, 2010.

\bibitem{Sommers1}
H.~J. Sommers.
\newblock On the dynamic mean field theory of spin glasses.
\newblock {\em Zeitschrift für Physik B Condensed Matter}, 50(97-105), 1983.

\bibitem{Synatschke_2009}
Franziska Synatschke, Georg Bergner, Holger Gies, and Andreas Wipf.
\newblock Flow equation for supersymmetric quantum mechanics.
\newblock {\em Journal of High Energy Physics}, 2009(03):028–028, Mar 2009.

\bibitem{Tarjus_20082}
Gilles Tarjus and Matthieu Tissier.
\newblock Nonperturbative functional renormalization group for random field
  models and related disordered systems. i. effective average action formalism.
\newblock {\em Physical Review B}, 78(2), Jul 2008.

\bibitem{tarjus2015avalanches}
Gilles Tarjus and Matthieu Tissier.
\newblock Avalanches and perturbation theory in the random-field ising model,
  2015.

\bibitem{Tarjus_2020}
Gilles Tarjus and Matthieu Tissier.
\newblock Random-field ising and o(n) models: theoretical description through
  the functional renormalization group.
\newblock {\em The European Physical Journal B}, 93(3), Mar 2020.

\bibitem{thouless1977solution}
David~J Thouless, Philip~W Anderson, and Robert~G Palmer.
\newblock Solution of'solvable model of a spin glass'.
\newblock {\em Philosophical Magazine}, 35(3):593--601, 1977.

\bibitem{Tissier_2008}
Matthieu Tissier and Gilles Tarjus.
\newblock Nonperturbative functional renormalization group for random field
  models and related disordered systems. ii. results for the random
  fieldo(n)model.
\newblock {\em Physical Review B}, 78(2), Jul 2008.

\bibitem{Tissier_2011}
Matthieu Tissier and Gilles Tarjus.
\newblock Supersymmetry and its spontaneous breaking in the random field ising
  model.
\newblock {\em Physical Review Letters}, 107(4), Jul 2011.

\bibitem{Tissier_20122}
Matthieu Tissier and Gilles Tarjus.
\newblock Nonperturbative functional renormalization group for random field
  models and related disordered systems. iii. superfield formalism and
  ground-state dominance.
\newblock {\em Physical Review B}, 85(10), Mar 2012.

\bibitem{Tissier_2012}
Matthieu Tissier and Gilles Tarjus.
\newblock Nonperturbative functional renormalization group for random field
  models and related disordered systems. iv. supersymmetry and its spontaneous
  breaking.
\newblock {\em Physical Review B}, 85(10), Mar 2012.

\bibitem{vanDuijvendijk2010}
Kristina van Duijvendijk, Robert~L. Jack, and Fr{\'{e} }d{\'{e}}ric van
  Wijland.
\newblock Second-order dynamic transition in a $p=2$ spin-glass model.
\newblock {\em Physical Review E}, 81(1), jan 2010.

\bibitem{Wigner1}
Eugene~P. Wigner.
\newblock Random matrices in physics.
\newblock {\em SIAM REVIEW}, 1967.

\bibitem{wilkins2021functional}
Ashley Wilkins, Gerasimos Rigopoulos, and Enrico Masoero.
\newblock Functional renormalisation group for brownian motion i: The effective
  equations of motion, 2021.

\bibitem{wilkins2021functional2}
Ashley Wilkins, Gerasimos Rigopoulos, and Enrico Masoero.
\newblock Functional renormalisation group for brownian motion ii: Accelerated
  dynamics in and out of equilibrium, 2021.

\bibitem{Witten1981}
E.~Witten.
\newblock Dynamical breaking of supersymmetry.
\newblock {\em Nucl. Phys. B}, 1981.

\bibitem{Yeo_2020}
J.~Yeo and M.~A. Moore.
\newblock Possible instability of one-step replica symmetry breaking in p -spin
  ising models outside mean-field theory.
\newblock {\em Physical Review E}, 101(3), Mar 2020.

\bibitem{Zappal__2001}
D.~Zappalà.
\newblock Improving the renormalization group approach to the
  quantum-mechanical double well potential.
\newblock {\em Physics Letters A}, 290(1-2):35–40, Nov 2001.

\bibitem{ZinnJustin4}
J.~Zinn-Justin.
\newblock Vector models in the large n limit: a few applications.
\newblock {\em 11th Taiwan Spring School, Taipei}, 1997.

\bibitem{ZinnJustinBook1}
J.~Zinn-Justin.
\newblock {\em Quantum field theory and critical phenomena}.
\newblock Oxford Science Publication, 2002 (fourth edition).

\bibitem{ZinnJustinBook3}
J.~Zinn-Justin.
\newblock {\em Phase Transitions and Renormalization Group}.
\newblock Oxford Graduate Texts, 2007.

\bibitem{ZinnJustinBook2}
J.~Zinn-Justin.
\newblock {\em From random walks to random matrices}.
\newblock Oxford Graduate Texts, 2019.

\end{thebibliography}
\end{document}